\documentclass[aps,prb,twocolumn,groupedaddress]{revtex4-2}

\usepackage{graphicx}
\usepackage{amsmath}
\usepackage{mathtools, bm}
\usepackage{amssymb, bm}
\usepackage{times}
\usepackage{setspace}
\usepackage{hyperref}
\usepackage{amsthm}
\usepackage{diagbox}
\usepackage{float}
\usepackage{paralist}

\newcommand{\bra}[1]{\langle #1|}
\newcommand{\ket}[1]{| #1\rangle}

\begin{document}

\title{Superpositions of tree-tensor networks for single-reference ground states in the strong correlation regime}

\author{Dominic Bergeron}
\email[]{dominic.bergeron@usherbrooke.ca}
\affiliation{D\'epartement de physique, Universit\'e de Sherbrooke, Qu\'ebec, Canada}

\date{\today}

\begin{abstract}

The fermionic many-body problem in the strong correlation regime is notoriously difficult to tackle. It is inherently non-perturbative and the quantum Monte Carlo method is plagued by the fermionic sign problem when applied to such systems. In a previous work (Phys. Rev. B 101, 045109 (2020)), we have proposed to extend the single-reference coupled-cluster (SRCC) method to the strong correlation regime by truncating entanglement in the cluster operator, instead of truncating it with respect to the number or particle-hole excitations. The proposed extensions are applicable in practice if we can accurately represent the single-reference ground state using low-rank tensor decompositions (LRTD), such that the number of free parameters does not grow faster than a low-order polynomial in the number of excitations. To reach such a polynomially bounded parameterization, we have proposed a new type of LRTD called ``superpositions of tree-tensor networks'' (STTN), which use the same set of building blocs to define all the tensors involved in the CC equations, and combine different ``channels'', i.e. different types of pairing among excited particles and holes, in the decomposition of a given tensor. Those two principles are aimed at globally minimizing the total number of free parameters required to accurately represent the ground state. In this work, we show that STTN can indeed be compact and accurate representations of strongly correlated ground states by using them to express the CC cluster operator amplitudes and wave function coefficients of exact grounds states of small two-dimensional Hubbard clusters, at half-filling, up to three particle-hole excitations. We show the compactness of STTN by using a number of free parameters smaller than the number of equations in the CCSD approximation, i.e. much smaller than the number of fitted tensor elements. We find that, for the systems considered, the STTN are more accurate as the size of the system increases and that combining different channels in the decompositions of the most strongly correlated tensors is crucial to obtain good accuracy.

\end{abstract}

\maketitle

\section{Introduction}

The coupled-cluster (CC) method is arguably the most accurate method of quantum chemistry \cite{Coester_Kummel_1960, Cizek_1966, Paldus_Cizek_Shavitt_1972, Bartlett_Musial_2007}. Its success is due to the exponential ansatz for the wave function, which ensures size-extensivity, size-consistency, and makes it infinite-order from a diagrammatic point of view \cite{Bartlett_Musial_2007}. Single-reference CC (SRCC) is limited to weakly correlated systems however, and fails in the strong correlation regime \cite{Bulik_2015}. Because some of the most interesting physical systems are strongly correlated, e.g. high-$T_c$ cuprates, heavy fermions systems and Mott insulators, a strong correlation version of CC would be a great tool to have. So far, attempts to extend CC to the strong correlation regime do so by combining the method with either symmetry projection \cite{Bulik_2015, Qiu_2017,  Qiu_2018, Qiu_2019, Gomez_2019, Tsuchimochi_2019, Song_2022} or a correlation (Jastrow) operator \cite{Krotscheck_1980, Xian_2008, Neuscamman_2013}, but at the cost of introducing additional approximations on top of the CC main approximation (the truncation of the cluster operator). Multi-reference CC methods \cite{Jeziorski_Monkhorst_1981, Lyakh_2012} are also better adapted than SRCC to strong correlations, but not specifically designed for that purpose, and are considerably more complex to use than SRCC.

In Ref.\onlinecite{DBergeron_TCC_2020}, we have proposed extensions of SRCC to treat weak and strong correlations using a unique type of approximation. In the most straightforward of those extensions, instead of truncating the cluster operator $T$ with respect the the number of particle-hole excitations, as in conventional CC, it is the entanglement in $T$ that is truncated or, more specifically, the entanglement in the tensors of $T$. Here, truncating entanglement means that those tensors are represented using low-rank tensor decompositions, or tensor networks (TN), that can drastically reduce the number of free parameters used to define them. The main approximation in our proposed approach would therefore be in the definition of $T$, as in standard CC, with the important difference that, without the truncation of $T$ with respect to the number of excitations, the resulting SRCC wave function can represent both weakly and strongly correlated ground states. Our other SRCC extension proposal is based on CC-like equations that depend directly on the single-reference wave function coefficients and the TN are used to express those coefficients. Note that tensor decompositions have been used in the context of CC to speedup the calculations, but without extending the domain of validity of CC \cite{Hohenstein_2012, Benedikt_2013, Hohenstein_2013, Hohenstein_2013a, Parrish_2014, Schutski_2017, Tichai_2019, Parrish_2019, Lesiuk_2019, HohensteinFales_2022} and tensor network methods originally designed for spin systems have also been used in quantum chemistry, but not in CC \cite{Szalay_Pfeffer_2015}. One key requirement for those TN methods to be applicable is that the interactions are short range, a requirement which is not satisfied when delocalized orbitals are used, as is the CC method. 

Although our proposed extensions of SRCC can in principle be used with different types of TN representation, we have also proposed in Ref.\onlinecite{DBergeron_TCC_2020} a new type of TN representation designed specifically for the CC problem called ``superpositions of tree-tensor networks'' (STTN). The STTN are based on two principles aimed at minimizing in a global way the total number of free parameters required to express all the tensors involved in the CC equations: 1) all the decompositions for the different tensors share the same building blocs and 2) the decomposition of a given tensor is a sum of different binary tree-tensor networks (TTN) which are possible to construct for that tensor. We call those different TTN contributions ``channels'', in analogy with the channels in diagrammatic calculations. The different channels account for the different possible correlations among the excited particles and holes. An important aspect of the STTN representation is that, unlike in well-known TN approaches, no assumption of locality in the interactions between excited particles and holes is required, and thus one could keep using delocalized orbitals, for instance Hartree-Fock orbitals, as in conventional CC calculations.

In the present work, we illustrate how the STTN can accurately express either the $T$-amplitudes, or wave function coefficients, of exact strongly correlated ground states of small two-dimensional half-filled Hubbard clusters, using a number of free parameters much smaller than the number of tensor elements represented. We choose the Hubbard model because it is the simplest Hamiltonian that can model strongly correlated systems, which will help us interpret some of our results. The goal of those benchmarks is, on one hand, to verify the applicability of our proposed extensions of SRCC to interesting strongly correlated physical systems and, on the other hand, to show that STTN are a good choice of representation to implement those extensions. Therefore, they constitute an important step toward the development of those SRCC extensions.

The article is divided as follows: our methodology is described in section \ref{sec:Method}, the STTN construction is detailed in section \ref{sec:STTN}, numerical results are given in section \ref{sec:results}, which is followed by a discussion and a conclusion.

\section{Methodology}\label{sec:Method}

Let us assume that the ground state wave function of a fermionic many-body system is represented as a single-reference expansion of the form
\begin{equation}\label{eq:psi_sum}
\begin{split}
\ket{\psi}&=\ket{\phi}+\sum_{i,j} c_j^i \ket{\phi^i_j}+\sum_{\substack{\langle i_1,i_2\rangle \\\langle j_1,j_2\rangle}} c_{j_1j_2}^{i_1i_2}  \ket{\phi_{j_1j_2}^{i_1i_2}}+\ldots\\
\end{split}
\end{equation}
where $\langle \ldots \rangle$ means that the sum runs over all distinct combinations of indices,
\begin{equation}\label{eq:def_phi}
\ket{\phi}= a_{1}^\dagger a_{2}^\dagger \ldots a_{N}^\dagger \ket{0}\,,
\end{equation}
where $a_{j}^\dagger$ creates a particle in spin-orbital $j$, is the reference Slater determinant (SD), and
\begin{equation}\label{eq:phi_i_j}
\ket{\phi^{i_1,i_2,\ldots, i_k}_{j_1,j_2,\ldots,j_k}}=a_{i_k}^\dagger  a_{j_k} a_{i_{k-1}}^\dagger  a_{j_{k-1}} \ldots a_{i_1}^\dagger a_{j_1} \ket{\phi}\,,
\end{equation}
are excited SD's with $k$ particle-hole excitations with respect to the reference SD. Here, $j_l$ are occupied spin-orbitals, i.e. spin-orbitals among those defining the reference SD \eqref{eq:def_phi}, and $i_l$ are unoccupied ones.

We will also use the CC form of the wave function:
\begin{equation}\label{eq:CC_ansatz}
\ket{\psi}=e^{T}\ket{\phi}\,,
\end{equation}
where $T$ is the \textit{cluster operator} defined as
\begin{subequations}\label{eq:def_T_CC}
\begin{equation}\label{eq:def_T_CC_a}
T=T_1+T_2+\ldots+T_{n}
\end{equation}
where
\begin{equation}\label{eq:T_N_CC}
T_l=\sum_{\substack{\langle i_1i_2\ldots i_l\rangle,\\\langle j_1,j_2\ldots j_l\rangle}} d_{j_1,j_2\ldots j_l}^{i_1i_2\ldots i_l} a_{i_l}^\dagger a_{j_l} a_{i_{l-1}}^\dagger a_{j_{l-1}} \ldots a_{i_1}^\dagger a_{j_1}\,.
\end{equation}
\end{subequations}
By expanding the exponential in \eqref{eq:CC_ansatz} as a Taylor series, we obtain the form \eqref{eq:psi_sum} and can therefore express the coefficients $c_{j_1j_2\ldots}^{i_1i_2\ldots}$ in \eqref{eq:psi_sum} using the amplitudes $d_{j_1j_2\ldots}^{i_1i_2\ldots}$. We can then easily obtain the expressions of the amplitudes as functions of the coefficients which, up to three excitations, are
\begin{equation}\label{eq:wv_cfs_vs_T_ampl}
\begin{split}
d^{i_1}_{j_1}&=c^{i_1}_{j_1}\\
d^{i_1i_2}_{j_1j_2}&=c^{i_1i_2}_{j_1j_2}-d^{i_1}_{j_1}d^{i_2}_{j_2}+d^{i_2}_{j_1}d^{i_1}_{j_2}\\
d^{i_1i_2i_3}_{j_1j_2j_3}&=c^{i_1i_2i_3}_{j_1j_2j_3}-d^{i_1}_{j_1}d^{i_2i_3}_{j_2j_3}+d^{i_2}_{j_1}d^{i_1i_3}_{j_2j_3}+d^{i_3}_{j_1}d^{i_2i_1}_{j_2j_3}+d^{i_1}_{j_2}d^{i_2i_3}_{j_1j_3}\\
&-d^{i_2}_{j_2}d^{i_1i_3}_{j_1j_3}-d^{i_3}_{j_2}d^{i_2i_1}_{j_1j_3}+d^{i_1}_{j_3}d^{i_2i_3}_{j_2j_1}-d^{i_2}_{j_3}d^{i_1i_3}_{j_2j_1}\\
&-d^{i_3}_{j_3}d^{i_2i_1}_{j_2j_1}-d^{i_1}_{j_1}d^{i_2}_{j_2}d^{i_3}_{j_3}+d^{i_1}_{j_1}d^{i_3}_{j_2}d^{i_2}_{j_3}+d^{i_2}_{j_1}d^{i_1}_{j_2}d^{i_3}_{j_3}\\
&-d^{i_2}_{j_1}d^{i_3}_{j_2}d^{i_1}_{j_3}+d^{i_3}_{j_1}d^{i_2}_{j_2}d^{i_1}_{j_3}-d^{i_3}_{j_1}d^{i_1}_{j_2}d^{i_2}_{j_3}\,.
\end{split}
\end{equation}
We use those expressions to extract the exact cluster operator amplitudes from the exact ground state wave function coefficients. 

\begin{figure}
\includegraphics[width=0.51\columnwidth]{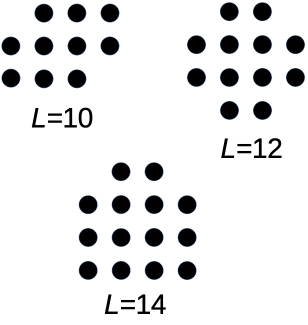}
\caption{\label{fig:clusters} Cluster geometries.}
\end{figure}

Now, our goal is to test the capacity of STTN to express the tensors $d_{j_1j_2\ldots}^{i_1i_2\ldots}$ and $c_{j_1j_2\ldots}^{i_1i_2\ldots}$ from a ground state $\ket{\psi}$ in the strong correlation regime, and this, using a number of free parameters much smaller than the number of tensor elements. To do so, we first compute the exact ground states of the Hubbard model
\begin{equation}
\hat{H}=-\sum_{ij} t_{ij} a^\dagger_i a_j + U \sum_i n_{i\uparrow}n_{i\downarrow}\,.
\end{equation}
where $n_{i\sigma}=a^\dagger_{i\sigma} a_{i\sigma}$, for the two-dimensional clusters of 10, 12 and 14 sites illustrated in Fig. \ref{fig:clusters}, at half filling, $U=10t$, where $t$ is the nearest neighbor hopping parameter, and next nearest neighbor hopping $t'=-0.3t$. Those Hamiltonian parameters are relevant to high-$T_c$ cuprates. We then optimize the STTN parameters to reproduce the tensors $d_{j_1j_2\ldots}^{i_1i_2\ldots}$ and $c_{j_1j_2\ldots}^{i_1i_2\ldots}$ up to three excitations, using a total number of free parameters smaller than the total number of single- and double-excitation SDs, i.e. smaller than the number of equations in the CCSD (CC single and double) approximation, and much smaller than the number of triple-excitation SDs. To assess the quality of the fits, we look at the agreement graphically between the STTN values and the exact amplitudes $d_{j_1j_2\ldots}^{i_1i_2\ldots}$ and coefficients $c_{j_1j_2\ldots}^{i_1i_2\ldots}$, and at the relative error on the energy obtained when using the STTN results in the CC energy expressions provided in appendix \ref{sec:DE_CC}.

The STTN fits are done with ground states written in unrestricted Hartree-Fock (UHF) basis. Since we had the freedom to choose which spin-orbital basis to use, we have also tried the STTN fits using natural spin-orbitals. However, in the strong correlation regime, natural orbitals produce a reference energy much higher than the ground state energy, namely a large correlation energy. Consequently, since it is the correlation energy that we actually compute in practice, they require an extremely small relative error on that energy to obtain a reasonably small relative error on the ground state energy. The natural orbitals are thus not a good choice. On the other hand, UHF orbitals, which yield the smallest possible correlation energy, require a much smaller relative error on the correlation energy to obtain a small relative error on the ground state energy. However, at large $U$, the many solutions to the HF equations produce many different energies. To obtain the true HF ground state, one can compute the solution to the HF equations many times and choose the solution with the smallest HF energy. Since we had access to many different solutions to the HF equations, we have tried the STTN fits using both the true HF ground state spin-orbitals and other spin-orbitals satisfying the HF equations, but producing a slightly larger reference energy, and found that the latter spin-orbitals produced better STTN fits and smaller errors on the energy. The results presented in section \ref{sec:results} are obtained using this type of spin-orbitals.

To obtain the clusters ground states, we use the Lanczos method. For the 10 and 12 sites clusters, the ground states in the UHF basis can be obtained directly by using the Hamiltonian matrix written in that basis. However, for the 14 sites cluster, the memory required to store the Hamiltonian matrix in a delocalized basis is too large to be practical. In that case, we first compute the ground state in the local orbital basis, in which the Hamiltonian matrix is highly sparse, and then transform that wave function to the UHF basis, which requires only a small amount of memory. However, because of the large size of the Hilbert space (11778624), this is still a heavy calculation that requires the use of a graphics processing unit (GPU). Fortunately, we only need the coefficients of single-, double- and triple-excitation determinants, so that the transformation can be performed in less than a day on a NVidia A100SXM4 GPU. To optimize the STTN free parameters, we minimize the quadratic distance ($\chi^2$) between the exact and the STTN $T$-amplitudes, or wave function coefficients, using the AMSGrad algorithm \cite{reddi2019}. 

\section{Superpositions of tree-tensor networks}\label{sec:STTN}

The STTN are based on two principles:
\begin{compactenum}[i)]
\item All the tensors are decomposed using the same building blocs.
\item The decomposition of a given tensor is a sum of the different binary tree-tensor networks that can be constructed for this tensor.
\end{compactenum}
We will call ``channels'' the different contributions to a given tensor. The STTN structure is described in details in Ref. \onlinecite{DBergeron_TCC_2020}. However, the decompositions used in this work slightly differ from those of Ref. \onlinecite{DBergeron_TCC_2020}. We therefore provide their definition below. We describe them as if they were used to represent $T$-amplitudes $d_{j_1j_2\ldots}^{i_1i_2\ldots}$, but we also use the same decompositions to represent wave function coefficients $c_{j_1j_2\ldots}^{i_1i_2\ldots}$.

\begin{figure}
\includegraphics[width=\columnwidth]{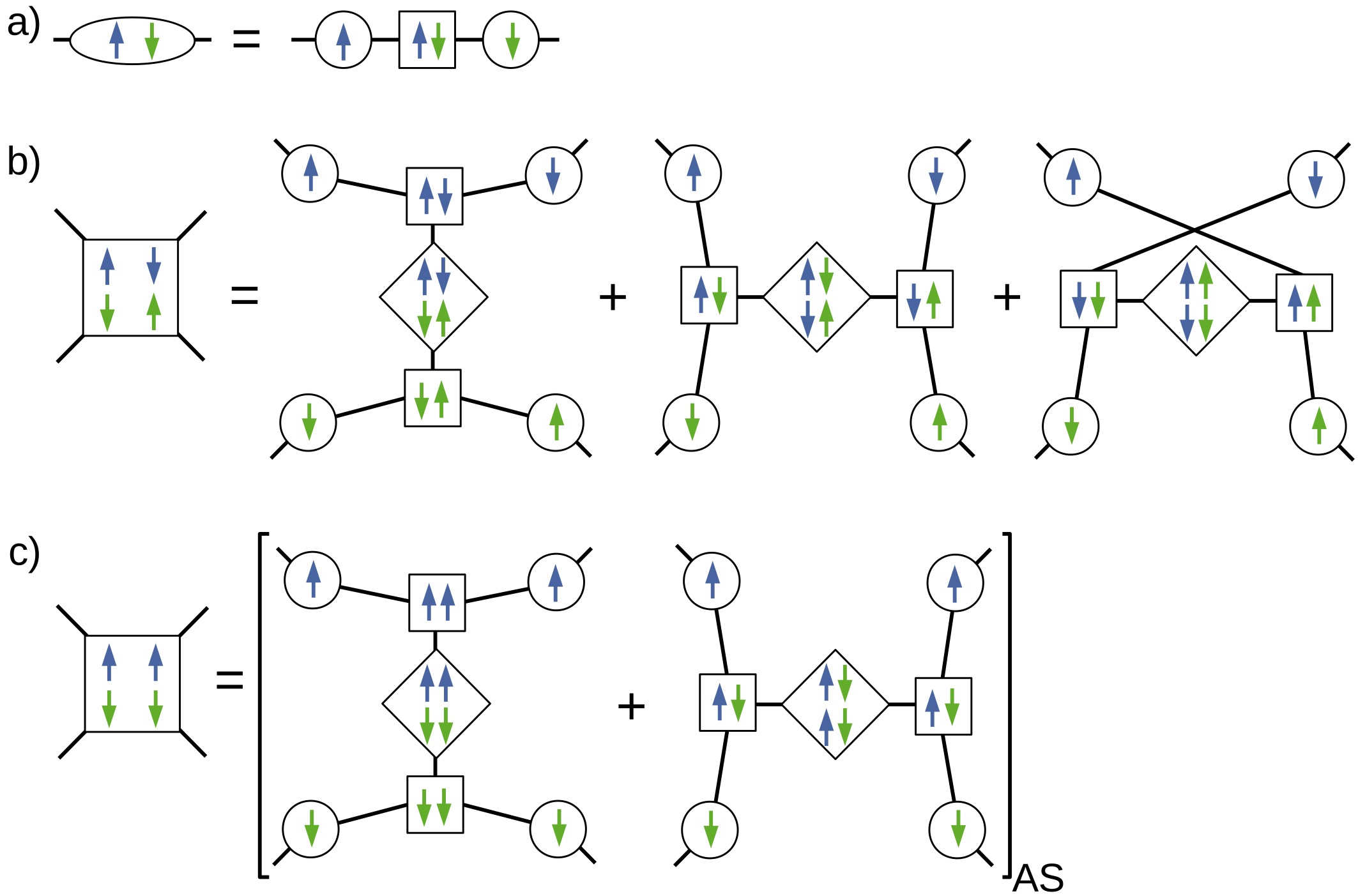}
\caption{\label{fig:decomp_c2} Graphical representation of the STTN for single- and double-excitation coefficients. The blue and green arrows represent particles and holes, respectively, and their spin orientations. Figure (a) is a representation of Eq. \eqref{eq:c1_SVD}, (b) corresponds to Eq. \eqref{eq:c2_ud_sum} and (c), Eq. \eqref{eq:c2_uu_sum}. The circles represent the $u$ and $v$ matrices, the squares with two arrows correspond to the $\kappa$ tensors and the diamonds with four arrows, the $\lambda$ tensors. The brackets with subscript ``AS'' in (c) indicates that the decompositions are antisymmetrized with respect to exchange of same spin particle or hole indices.}
\end{figure}

For the single-excitation amplitudes, the decomposition is
\begin{equation}\label{eq:c1_SVD}
d^{i\uparrow}_{j\downarrow}=\kappa_{0}^{p\bar{h}}\sum_{k=1}^{s_{p}}\sum_{l=1}^{s_{h}} \kappa_{kl1}^{p\bar{h}} u_{i k}^{\uparrow} v_{j l}^{\downarrow}\,,
\end{equation}
where matrices $u^{\uparrow}$ and $v^{\downarrow}$ are orthogonal and defined as Cayley transforms of skew-symmetric matrices. $d^{i\downarrow}_{j\uparrow}$ is defined similarly. Here $\kappa_{kl1}$ is the first matrix slice of the third order tensor $\kappa$, which is also used in decompositions for higher excitations. The prefactor $\kappa_{0}^{p\bar{h}}$ is required because the matrix slices of $\kappa$ are normalized. They are also orthogonal to each other since their elements are defined by a column in an orthogonal matrix, which is itself defined as the Cayley transform of a skew-symmetric matrix. Expression \eqref{eq:c1_SVD} is represented graphically in Fig. \ref{fig:decomp_c2}a.

In the following, we label tensors using combinations of $p$, $\bar{p}$, $h$ and $\bar{h}$ as superscripts, corresponding respectively to spin up and spin down particle and spin up and spin down hole. The order in which they appear identifies the channel. 

For two particle-hole excitations with opposite spins, there are three different channels if we use pairwise entanglement:
\begin{equation}\label{eq:decomp_c2_ud_pp_s}
\begin{split}
\left(d^{i_1\uparrow i_2\downarrow}_{j_1\downarrow j_2\uparrow}\right)_{p\bar{p}\bar{h}h}=&\lambda_{0}^{p\bar{p}\bar{h}h}\sum_{k=1}^{s_{p\bar{p}}}\sum_{l=1}^{s_{h\bar{h}}}\sum_{m,n=1}^{s_{p}}\sum_{q,r=1}^{s_{h}} \lambda_{kl1}^{p\bar{p}\bar{h}h} \kappa_{mnk}^{p\bar{p}} \kappa_{qrl}^{\bar{h}h}\\
&\times u^{\uparrow}_{i_1m}u^{\downarrow}_{i_2n} v^{\downarrow}_{j_1q}v^{\uparrow}_{j_2r}\,,
\end{split}
\end{equation}
\begin{equation}\label{eq:decomp_c2_ud_ph_s}
\begin{split}
\left(d^{i_1\uparrow i_2\downarrow}_{j_1\downarrow j_2\uparrow}\right)_{p\bar{h}\bar{p}h}=&\lambda_{0}^{p\bar{h}\bar{p}h}\sum_{k,l=1}^{s_{p\bar{h}}}\sum_{m,q=1}^{s_{p}}\sum_{n,r=1}^{s_{h}} \lambda_{kl1}^{p\bar{h}\bar{p}h} \kappa_{mnk}^{p\bar{h}} \kappa_{qrl}^{\bar{p}h}\\
&\times u^{\uparrow}_{i_1m}v^{\downarrow}_{j_1n}u^{\downarrow}_{i_2q} v^{\uparrow}_{j_2r}\,
\end{split}
\end{equation}
and
\begin{equation}\label{eq:decomp_c2_ud_ph_t}
\begin{split}
\left(d^{i_1\uparrow i_2\downarrow}_{j_1\downarrow j_2\uparrow}\right)_{ph\bar{p}\bar{h}}=& \lambda_{0}^{ph\bar{p}\bar{h}}\sum_{k,l=1}^{s_{ph}}\sum_{m,q=1}^{s_{p}}\sum_{n,r=1}^{s_{h}} \lambda_{kl1}^{ph\bar{p}\bar{h}} \kappa_{mnk}^{ph} \kappa_{qrl}^{\bar{p}\bar{h}}\\
&\times u^{\uparrow}_{i_1m}v^{\uparrow}_{j_2n} u^{\downarrow}_{i_2q}v^{\downarrow}_{j_1r} \,,
\end{split}
\end{equation}
where $u^{\sigma}$ and $v^{\sigma}$ and the first matrix slice of $\kappa^{p\bar{h}}$ and $\kappa^{\bar{p}h}$ are the same as in \eqref{eq:c1_SVD} and the decomposition for $d^{i\downarrow}_{j\uparrow}$, respectively, and we have assumed $s_{\bar{p}}=s_{p}$, $s_{\bar{h}}=s_{h}$, $s_{\bar{p}h}=s_{p\bar{h}}$ and $s_{\bar{p}\bar{h}}=s_{ph}$. 

As for the $\kappa_0$ prefactors in the single-excitations definitions, the $\lambda_{0}$ prefactors are required because the $\lambda_{kl1}$ matrices are the first matrix slices of the $\lambda$ third order tensors, which are also used to define higher excitation decompositions, and are defined the same way as the $\kappa$ tensors, using Cayley transforms of skew-symmetric matrices, so that the matrix slices are normalized and orthogonal to each other. 

Now, we define the opposite spins double-excitation tensor as a combination of \eqref{eq:decomp_c2_ud_pp_s}, \eqref{eq:decomp_c2_ud_ph_s} and \eqref{eq:decomp_c2_ud_ph_t}:
\begin{equation}\label{eq:c2_ud_sum}
d^{i_1\uparrow i_2\downarrow}_{j_1\downarrow j_2\uparrow}=\left(d^{i_1\uparrow i_2\downarrow}_{j_1\downarrow j_2\uparrow}\right)_{p\bar{p}\bar{h}h}+\left(d^{i_1\uparrow i_2\downarrow}_{j_1\downarrow j_2\uparrow}\right)_{p\bar{h}\bar{p}h}+\left(d^{i_1\uparrow i_2\downarrow}_{j_1\downarrow j_2\uparrow}\right)_{ph\bar{p}\bar{h}}\,.
\end{equation}
The graphical representation of Eq. \eqref{eq:c2_ud_sum} is shown in Fig. \ref{fig:decomp_c2}b.

For the same spin double-excitation tensor, there are two different channels based on pairwise entanglement, which must respect the Pauli exclusion principle:
\begin{equation}
\begin{split}
\left(d^{i_1\uparrow i_2\uparrow}_{j_1\downarrow j_2\downarrow}\right)_{pp\bar{h}\bar{h}}=&\lambda_{0}^{pp\bar{h}\bar{h}}\sum_{k=1}^{s_{pp}}\sum_{l=1}^{s_{hh}}\sum_{m,n=1}^{s_{p}} \sum_{q,r=1}^{s_{h}} \lambda_{kl1}^{pp\bar{h}\bar{h}} \kappa_{mnk}^{pp} \kappa_{qrl}^{\bar{h}\bar{h}}\\
&\times 
\begin{vmatrix}
u^{\uparrow}_{i_1m} & u^{\uparrow}_{i_1n}\\
u^{\uparrow}_{i_2m} & u^{\uparrow}_{i_2n}
\end{vmatrix}
\begin{vmatrix}
v^{\downarrow}_{j_1q}& v^{\downarrow}_{j_1r}\\ 
v^{\downarrow}_{j_2q} & v^{\downarrow}_{j_2r}
\end{vmatrix}\,,
\end{split}
\end{equation}
\begin{equation}\label{eq:decomp_c2_uu_ph}
\begin{split}
\left(d^{i_1\uparrow i_2\uparrow}_{j_1\downarrow j_2\downarrow}\right)_{p\bar{h}p\bar{h}}=&\lambda_{0}^{p\bar{h}p\bar{h}}\sum_{k,l=1}^{s_{p\bar{h}}}\sum_{m,q=1}^{s_{p}}\sum_{n,r=1}^{s_{h}} \lambda_{kl1}^{p\bar{h}p\bar{h}} \kappa_{mnk}^{p\bar{h}} \kappa_{qrl}^{p\bar{h}}\\
&\times
\begin{vmatrix}
 u^{\uparrow}_{i_1m} &u^{\uparrow}_{i_1q}\\
 u^{\uparrow}_{i_2m} &u^{\uparrow}_{i_2q}
\end{vmatrix}
\begin{vmatrix}
v^{\downarrow}_{j_1n}& v^{\downarrow}_{j_1r}\\ 
v^{\downarrow}_{j_2n} & v^{\downarrow}_{j_2r}
\end{vmatrix}\,,
\end{split}
\end{equation}
where the matrix $\lambda_{kl1}^{p\bar{h}p\bar{h}}$ is symmetric, the notation $|\ldots|$ indicates the determinant, and we use
\begin{equation}\label{eq:c2_uu_sum}
d^{i_1\uparrow i_2\uparrow}_{j_1\downarrow j_2\downarrow}=\left(d^{i_1\uparrow i_2\uparrow}_{j_1\downarrow j_2\downarrow}\right)_{pp\bar{h}\bar{h}}+\left(d^{i_1\uparrow i_2\uparrow}_{j_1\downarrow j_2\downarrow}\right)_{p\bar{h}p\bar{h}}\,.
\end{equation}
Equation \eqref{eq:c2_uu_sum} is represented in Fig. \ref{fig:decomp_c2}c.

\begin{figure}
\includegraphics[width=\columnwidth]{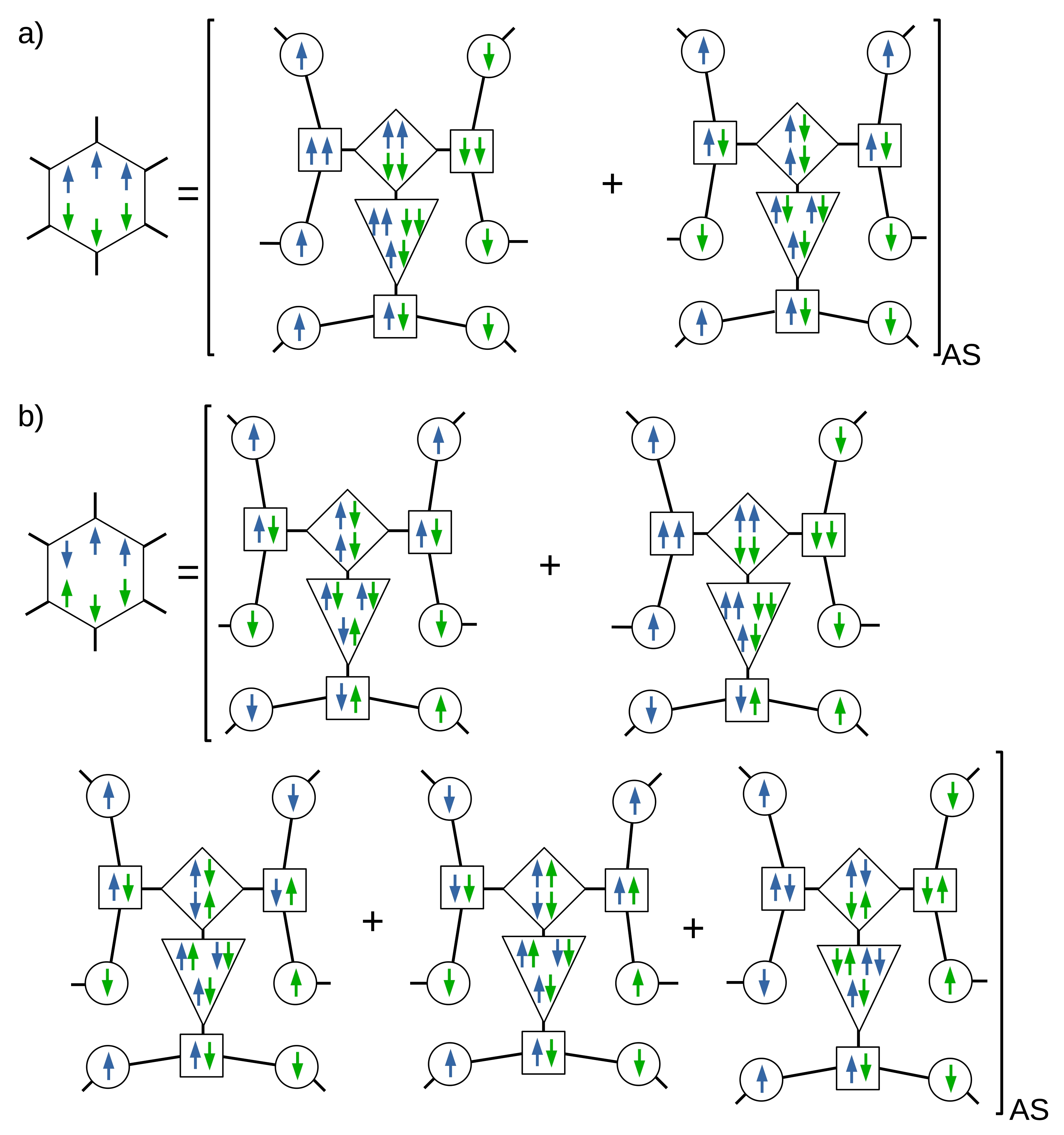}
\caption{\label{fig:decomp_c3} Graphical representation of the STTN for triple-excitation coefficients. Figure (a) is a representation of Eq. \eqref{eq:c3_u_sum} and (b) corresponds to Eq. \eqref{eq:c3_duu_sum}. Here, the triangles represent the $\mu$ matrices. See Fig. \ref{fig:decomp_c2} for other details on the notation. Note that, to avoid line crossings, the relative position of the circles representing the $u$ and $v$ matrices is different for different decompositions, unlike in Fig. \ref{fig:decomp_c2}, which is of no consequence since the antisymmetrization indicated by the brackets implies a summation with permutation of identical quasiparticle indices.}
\end{figure}

At three excitations and more, if we allow any type of pairwise entanglement in the channels, the number of possible channels becomes quite large, therefore, to limit their number, we construct the channels from the same building blocs as the single- and double-excitation tensors. For the parallel spin triple-excitation tensor, we obtain two different channels. The first is
\begin{equation}\label{eq:decomp_c3_uuu_ph}
\begin{split}
&\left(d^{i_1\uparrow i_2\uparrow i_3 \uparrow}_{j_1\downarrow j_2\downarrow j_3 \downarrow}\right)_{p\bar{h}p\bar{h}p\bar{h}}=\sum_{m=1}^{s_{p\bar{h}p\bar{h}}}\sum_{l_1,l_2,l_3=1}^{s_{p\bar{h}}} \sum_{k_1,k_3,k_5=1}^{s_{p}}\sum_{k_2,k_4,k_6=1}^{s_{h}}\\
&\sum_{\pi,\chi \in S_3}\left(\mu_{l_1m}^{p\bar{h}p\bar{h}p\bar{h}} \lambda_{l_2l_3m}^{p\bar{h}p\bar{h}}+\mu_{l_2m}^{p\bar{h}p\bar{h}p\bar{h}} \lambda_{l_1l_3m}^{p\bar{h}p\bar{h}}+\mu_{l_3m}^{p\bar{h}p\bar{h}p\bar{h}} \lambda_{l_1l_2m}^{p\bar{h}p\bar{h}}\right)\\
&\qquad\times \kappa_{k_1k_2l_1}^{p\bar{h}}\kappa_{k_3k_4l_2}^{p\bar{h}}\kappa_{k_5k_6l_3}^{p\bar{h}}\epsilon_{\pi_1\pi_2\pi_3}\epsilon_{\chi_1\chi_2\chi_3}\\
&\qquad\times u_{i_{\pi_1}k_1}^{\uparrow}u_{i_{\pi_2}k_3}^{\uparrow}u_{i_{\pi_3}k_5}^{\uparrow}v_{j_{\chi_1}k_2}^{\downarrow}v_{j_{\chi_2}k_4}^{\downarrow}v_{j_{\chi_3}k_6}^{\downarrow}\,,
\end{split}
\end{equation}
where $\epsilon_{\pi_1\pi_2\pi_3}$ is the Levi-Civita symbol, $S_3$ is the permutation group for the set $\{1,2,3\}$, and the symmetrization of the product $\mu\lambda$ is necessary to ensure the antisymmetry of the coefficient with respect to exchange of particle indices or of hole indices. The second channel is
\begin{equation}\label{eq:decomp_c3_uuu_phpphh}
\begin{split}
&\left(d^{i_1\uparrow i_2\uparrow i_3 \uparrow}_{j_1\downarrow j_2\downarrow j_3 \downarrow}\right)_{p\bar{h},pp\bar{h}\bar{h}}=\sum_{m=1}^{s_{pp\bar{h}\bar{h}}}\sum_{l_1=1}^{s_{p\bar{h}}}\sum_{l_2=1}^{s_{pp}}\sum_{l_3=1}^{s_{hh}}\\
&\sum_{k_1,k_3,k_4=1}^{s_{p}} \sum_{k_2,k_5,k_6=1}^{s_{h}} \sum_{\pi,\chi\in S_3}\mu_{l_1m}^{p\bar{h},pp\bar{h}\bar{h}} \lambda_{l_2l_3m}^{pp\bar{h}\bar{h}}\\
&\qquad \times\kappa_{k_1k_2l_1}^{p\bar{h}}\kappa_{k_3k_4l_2}^{pp}\kappa_{k_5k_6l_3}^{\bar{h}\bar{h}} \epsilon_{\pi_1\pi_2\pi_3}\epsilon_{\chi_1\chi_2\chi_3}\\
&\qquad\times u_{i_{\pi_1}k_1}^{\uparrow}u_{i_{\pi_2}k_3}^{\uparrow}u_{i_{\pi_1}k_4}^{\uparrow}v_{j_{\chi_1}k_2}^{\downarrow}v_{j_{\chi_2}k_5}^{\downarrow}v_{j_{\chi_3}k_6}^{\downarrow}\,.
\end{split}
\end{equation}
and the parallel spin triple-excitation tensor is expressed as
\begin{equation}\label{eq:c3_u_sum}
d^{i_1\uparrow i_2\uparrow i_3 \uparrow}_{j_1\downarrow j_2\downarrow j_3 \downarrow}=\left(d^{i_1\uparrow i_2\uparrow i_3 \uparrow}_{j_1\downarrow j_2\downarrow j_3 \downarrow}\right)_{p\bar{h}p\bar{h}p\bar{h}}+\left(d^{i_1\uparrow i_2\uparrow i_3 \uparrow}_{j_1\downarrow j_2\downarrow j_3 \downarrow}\right)_{p\bar{h},pp\bar{h}\bar{h}}\,,
\end{equation}
and is represented graphically in Fig. \ref{fig:decomp_c3}a. 

When one spin is different at three excitations, there are five channels that can be constructed with the building blocs used in the double-excitation tensors:
\begin{equation}\label{eq:c3_pdhupuhdpuhd}
\begin{split}
&\left(d^{i_1\downarrow i_2\uparrow i_3 \uparrow}_{j_1\uparrow j_2\downarrow j_3 \downarrow}\right)_{\bar{p}h,p\bar{h}p\bar{h}}=\sum_{m=1}^{s_{p\bar{h}p\bar{h}}}\sum_{l_1,l_2,l_3=1}^{s_{p\bar{h}}}\sum_{k_1,k_3,k_5=1}^{s_{p}}\sum_{k_2,k_4,k_6=1}^{s_{h}}\\
&\qquad\mu_{l_1m}^{\bar{p}h,p\bar{h}p\bar{h}} \lambda_{l_2l_3m}^{p\bar{h}p\bar{h}} \kappa_{k_1k_2l_1}^{\bar{p}h}\kappa_{k_3k_4l_2}^{p\bar{h}}\kappa_{k_5k_6l_3}^{p\bar{h}}\\
&\qquad\times u_{i_1k_1}^{\downarrow}v_{j_1k_2}^{\uparrow}
\begin{vmatrix}
u^{\uparrow}_{i_2k_3} &u^{\uparrow}_{i_2k_5}\\
u^{\uparrow}_{i_3k_3} & u^{\uparrow}_{i_3k_5}
\end{vmatrix}
\begin{vmatrix}
v^{\downarrow}_{j_2k_4}& v^{\downarrow}_{j_2k_6}\\ 
v^{\downarrow}_{j_3k_4} & v^{\downarrow}_{j_3k_6}
\end{vmatrix}\,,
\end{split}
\end{equation}
\begin{equation}\label{eq:c3_pdhupupuhdhd}
\begin{split}
&\left(d^{i_1\downarrow i_2\uparrow i_3 \uparrow}_{j_1\uparrow j_2\downarrow j_3 \downarrow}\right)_{\bar{p}h,pp\bar{h}\bar{h}}=\sum_{m=1}^{s_{pp\bar{h}\bar{h}}}\sum_{l_1=1}^{s_{p\bar{h}}}\sum_{l_2=1}^{s_{pp}}\sum_{l_3=1}^{s_{hh}} \sum_{k_1,k_3,k_4=1}^{s_{p}}\\
&\quad\sum_{k_2,k_5,k_6=1}^{s_{h}}\mu_{l_1m}^{\bar{p}h,pp\bar{h}\bar{h}} \lambda_{l_2l_3m}^{pp\bar{h}\bar{h}} \kappa_{k_1k_2l_1}^{\bar{p}h}\kappa_{k_3k_4l_2}^{pp}\kappa_{k_5k_6l_3}^{\bar{h}\bar{h}}\\
&\qquad\times u_{i_1k_1}^{\downarrow}v_{j_1k_2}^{\uparrow}
\begin{vmatrix}
u^{\uparrow}_{i_2k_3} & u^{\uparrow}_{i_2k_4}\\
u^{\uparrow}_{i_3k_3} & u^{\uparrow}_{i_3k_4}
\end{vmatrix}
\begin{vmatrix}
v^{\downarrow}_{j_2k_5}& v^{\downarrow}_{j_2k_6}\\ 
v^{\downarrow}_{j_3k_5} & v^{\downarrow}_{j_3k_6}
\end{vmatrix}\,,
\end{split}
\end{equation}
\begin{equation}\label{eq:c3_puhdpuhdpdhu}
\begin{split}
&\left(d^{i_1\downarrow i_2\uparrow i_3 \uparrow}_{j_1\uparrow j_2\downarrow j_3 \downarrow}\right)_{p\bar{h},p\bar{h}\bar{p}h}=\sum_{m=1}^{s_{p\bar{h}\bar{p}h}}\sum_{l_1,l_2,l_3=1}^{s_{p\bar{h}}}\sum_{k_1,k_3,k_5=1}^{s_{p}}\sum_{k_2,k_4,k_6=1}^{s_{h}}\\
&\qquad\left(\mu_{l_1m}^{p\bar{h},p\bar{h}\bar{p}h} \lambda_{l_2l_3m}^{p\bar{h}\bar{p}h}+\mu_{l_2m}^{p\bar{h},p\bar{h}\bar{p}h} \lambda_{l_1l_3m}^{p\bar{h}\bar{p}h}\right) \kappa_{k_1k_2l_1}^{p\bar{h}}\kappa_{k_3k_4l_2}^{p\bar{h}}\kappa_{k_5k_6l_3}^{\bar{p}h}\\
&\qquad\times u_{i_1k_5}^{\downarrow}v_{j_1k_6}^{\uparrow}
\begin{vmatrix}
u^{\uparrow}_{i_2k_1} & u^{\uparrow}_{i_2k_3}\\
u^{\uparrow}_{i_3k_1} & u^{\uparrow}_{i_3k_3}
\end{vmatrix}
\begin{vmatrix}
v^{\downarrow}_{j_2k_2}& v^{\downarrow}_{j_2k_4}\\ 
v^{\downarrow}_{j_3k_2} & v^{\downarrow}_{j_3k_4}
\end{vmatrix}\,,
\end{split}
\end{equation}
\begin{equation}\label{eq:c3_puhdpuhupdhd}
\begin{split}
&\left(d^{i_1\downarrow i_2\uparrow i_3 \uparrow}_{j_1\uparrow j_2\downarrow j_3 \downarrow}\right)_{p\bar{h},\bar{p}\bar{h}ph}=\sum_{m=1}^{s_{\bar{p}\bar{h}ph}}\sum_{l_1=1}^{s_{p\bar{h}}}\sum_{l_2,l_3=1}^{s_{ph}}\sum_{k_1,k_3,k_5=1}^{s_{p}}\sum_{k_2,k_4,k_6=1}^{s_{h}}\\
&\qquad\mu_{l_1m}^{p\bar{h},\bar{p}\bar{h}ph} \lambda_{l_2l_3m}^{\bar{p}\bar{h}ph} \kappa_{k_1k_2l_1}^{p\bar{h}}\kappa_{k_3k_4l_2}^{\bar{p}\bar{h}}\kappa_{k_5k_6l_3}^{ph}\\
&\qquad\times u_{i_1k_3}^{\downarrow}v_{j_1k_6}^{\uparrow}
\begin{vmatrix}
u^{\uparrow}_{i_2k_5} & u^{\uparrow}_{i_2k_1}\\
u^{\uparrow}_{i_3k_5} & u^{\uparrow}_{i_3k_1}
\end{vmatrix}
\begin{vmatrix}
v^{\downarrow}_{j_2k_4}& v^{\downarrow}_{j_2k_2}\\ 
v^{\downarrow}_{j_3k_4} & v^{\downarrow}_{j_3k_2}
\end{vmatrix}\,,
\end{split}
\end{equation}
\begin{equation}\label{eq:c3_puhdpupdhdhu}
\begin{split}
&\left(d^{i_1\downarrow i_2\uparrow i_3 \uparrow}_{j_1\uparrow j_2\downarrow j_3 \downarrow}\right)_{p\bar{h},\bar{p}ph\bar{h}}=\sum_{m=1}^{s_{\bar{p}ph\bar{h}}}\sum_{l_1=1}^{s_{p\bar{h}}}\sum_{l_2=1}^{s_{p\bar{p}}}\sum_{l_3=1}^{s_{h\bar{h}}} \sum_{k_1,k_3,k_4=1}^{s_{p}}\\
&\quad\sum_{k_2,k_5,k_6=1}^{s_{h}}\mu_{l_1m}^{p\bar{h},\bar{p}ph\bar{h}} \lambda_{l_2l_3m}^{\bar{p}ph\bar{h}} \kappa_{k_1k_2l_1}^{p\bar{h}}\kappa_{k_3k_4l_2}^{\bar{p}p}\kappa_{k_5k_6l_3}^{h\bar{h}}\\
&\qquad\times u_{i_1k_3}^{\downarrow}v_{j_1k_5}^{\uparrow}
\begin{vmatrix}
u^{\uparrow}_{i_2k_4} & u^{\uparrow}_{i_2k_1}\\
u^{\uparrow}_{i_3k_4} & u^{\uparrow}_{i_3k_1}
\end{vmatrix}
\begin{vmatrix}
v^{\downarrow}_{j_2k_6}& v^{\downarrow}_{j_2k_2}\\ 
v^{\downarrow}_{j_3k_6} & v^{\downarrow}_{j_3k_2}
\end{vmatrix}\,,
\end{split}
\end{equation}
and we use
\begin{equation}\label{eq:c3_duu_sum}
\begin{split}
&d^{i_1\downarrow i_2\uparrow i_3 \uparrow}_{j_1\uparrow j_2\downarrow j_3 \downarrow}=\left(d^{i_1\downarrow i_2\uparrow i_3 \uparrow}_{j_1\uparrow j_2\downarrow j_3 \downarrow}\right)_{\bar{p}h,p\bar{h}p\bar{h}}+\left(d^{i_1\downarrow i_2\uparrow i_3 \uparrow}_{j_1\uparrow j_2\downarrow j_3 \downarrow}\right)_{\bar{p}h,pp\bar{h}\bar{h}}\\
&+\left(d^{i_1\downarrow i_2\uparrow i_3 \uparrow}_{j_1\uparrow j_2\downarrow j_3 \downarrow}\right)_{p\bar{h},p\bar{h}\bar{p}h}+\left(d^{i_1\downarrow i_2\uparrow i_3 \uparrow}_{j_1\uparrow j_2\downarrow j_3 \downarrow}\right)_{p\bar{h},\bar{p}\bar{h}ph}\\&+\left(d^{i_1\downarrow i_2\uparrow i_3 \uparrow}_{j_1\uparrow j_2\downarrow j_3 \downarrow}\right)_{p\bar{h},\bar{p}ph\bar{h}}\,,
\end{split}
\end{equation}
which is depicted in Fig. \ref{fig:decomp_c3}b. Note that no prefactors are required in expressions \eqref{eq:decomp_c3_uuu_ph}, \eqref{eq:decomp_c3_uuu_phpphh}, and \eqref{eq:c3_pdhupuhdpuhd} to \eqref{eq:c3_puhdpupdhdhu} since the $\mu$ matrices are not normalized. The $u$ and $v$ matrices, the $\kappa$ tensors, and the first matrix slices of the $\lambda$ tensors are the same as in the single- and double-excitation decompositions.

We have mentioned that the $u$ and $v$ matrices and the $\kappa$ and $\lambda$ tensors are all computed using Cayley transforms of skew-symmetric matrices. This transform involves the inversion of a matrix which could be computationally expensive for large matrices. However, for the $\kappa$ and $\lambda$ tensors, those skew-symmetric matrices are actually band matrices, with bandwidth equal to the third dimension of the tensors, and which is small compared to the matrix dimension, so that the inversion of those band matrices is fast. On the other hand $u$ and $v$ are square matrices in the cases shown in section \ref{sec:results} and thus their corresponding skew-symmetric matrices are dense, but because the system sizes considered here are small, so are the sizes $s_p=s_h=N_\sigma$ of those matrices, where $N_\sigma$ is the number of spin $\sigma$ occupied spin-orbitals. For large systems however, $s_h$ and $s_p$ should be kept small enough so that the inversion of their corresponding skew-symmetric matrices is not computationally too heavy.

\begin{figure}
\includegraphics[width=0.9\columnwidth]{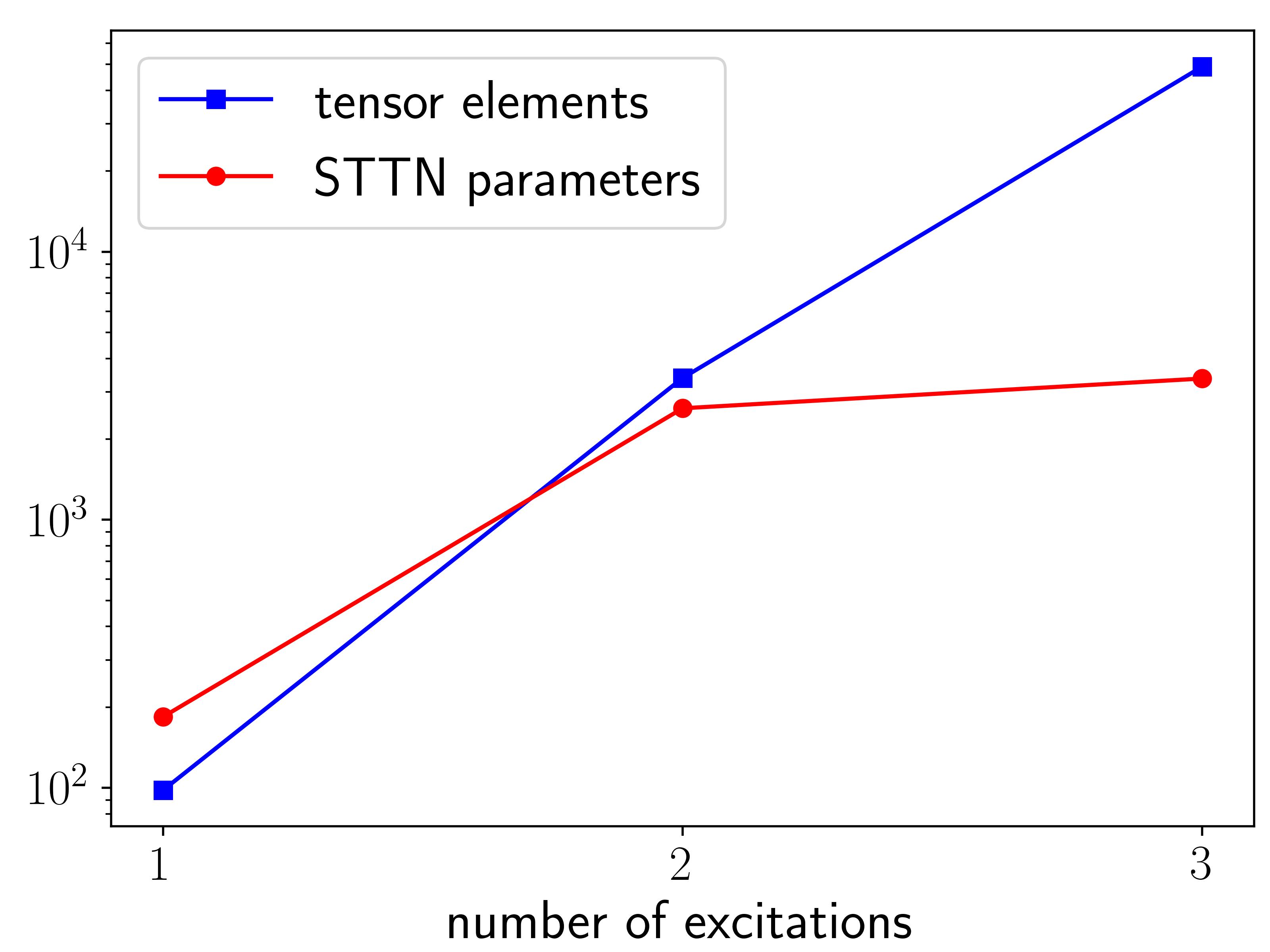}
\caption{\label{fig:Npar_vs_Nx} Cumulative total number of tensor elements (Slater determinants) and STTN free parameters as functions of excitation number for the case shown in Fig. \ref{fig:14_sites_HF_T_ampl}.}
\end{figure}
Figure \ref{fig:Npar_vs_Nx} illustrates how the total number of STTN free parameters increases with the number of excitations compared with the total number of tensor elements for the 14 sites cluster example shown section \ref{sec:results}. Note how the additional number of free parameters required to define the triple-excitation tensors is smaller than the total number of free parameters used to define single- and double-excitation decompositions. In that case it is less than $30\%$ of that number, while the number of triple-excitation tensor elements is much larger. If we were to define the decompositions for quadruple-excitations tensors, because the same building blocs as for the lower order decompositions would be used, we would only need to define additional matrices to connect $\lambda$ tensors together, and thus the number of additional parameters would be quite small relatively to the total number of free parameters used to define the single-, double- and triple-excitations decompositions, for which 10 $\kappa$ and 7 $\lambda$ third order tensors have to be defined. This is what makes the STTN representation so compact in terms of number of free parameters.

\section{Results}\label{sec:results}

In the present section, we provide the results of the fits to the $T$-amplitudes $d_{j_1j_2\ldots}^{i_1i_2\ldots}$. The results of the fits to the wave function coefficients $c_{j_1j_2\ldots}^{i_1i_2\ldots}$ are in appendix \ref{sec:results_STTN_wv_cfs}.  For each case presented in this section, we have tried many different sets of tensor parameters and chosen the ones yielding the smallest $\chi^2$. There is therefore no bias in our optimization process toward the exact energy, so that the error on the energy is a good metric to assert the quality of the fits. The STTN tensor dimensions used in the fits are listed in tables \ref{tab:tensor_dim_T} and \ref{tab:tensor_dim_T_SC} of appendix \ref{sec:tensor_dim}. Although the energy obtained using the STTN $T$-amplitudes or wave function coefficients is not variational, we have found that, for the systems considered, it is always larger than the exact energy. In other non-half-filed systems not shown here however, it is possible to obtain an energy smaller than the exact one.
\begin{figure}[h]
\includegraphics[width=0.51\columnwidth]{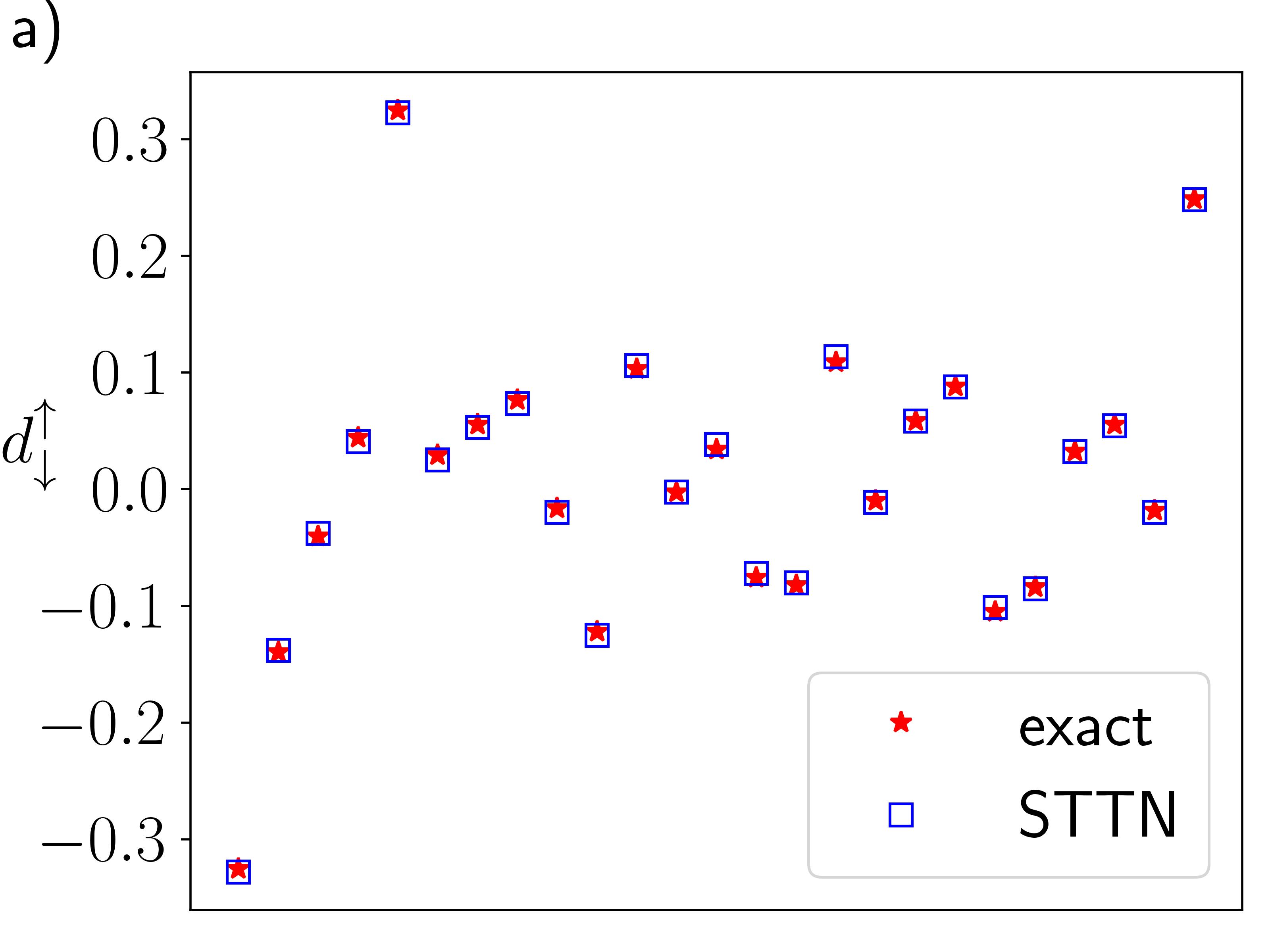}\includegraphics[width=0.51\columnwidth]{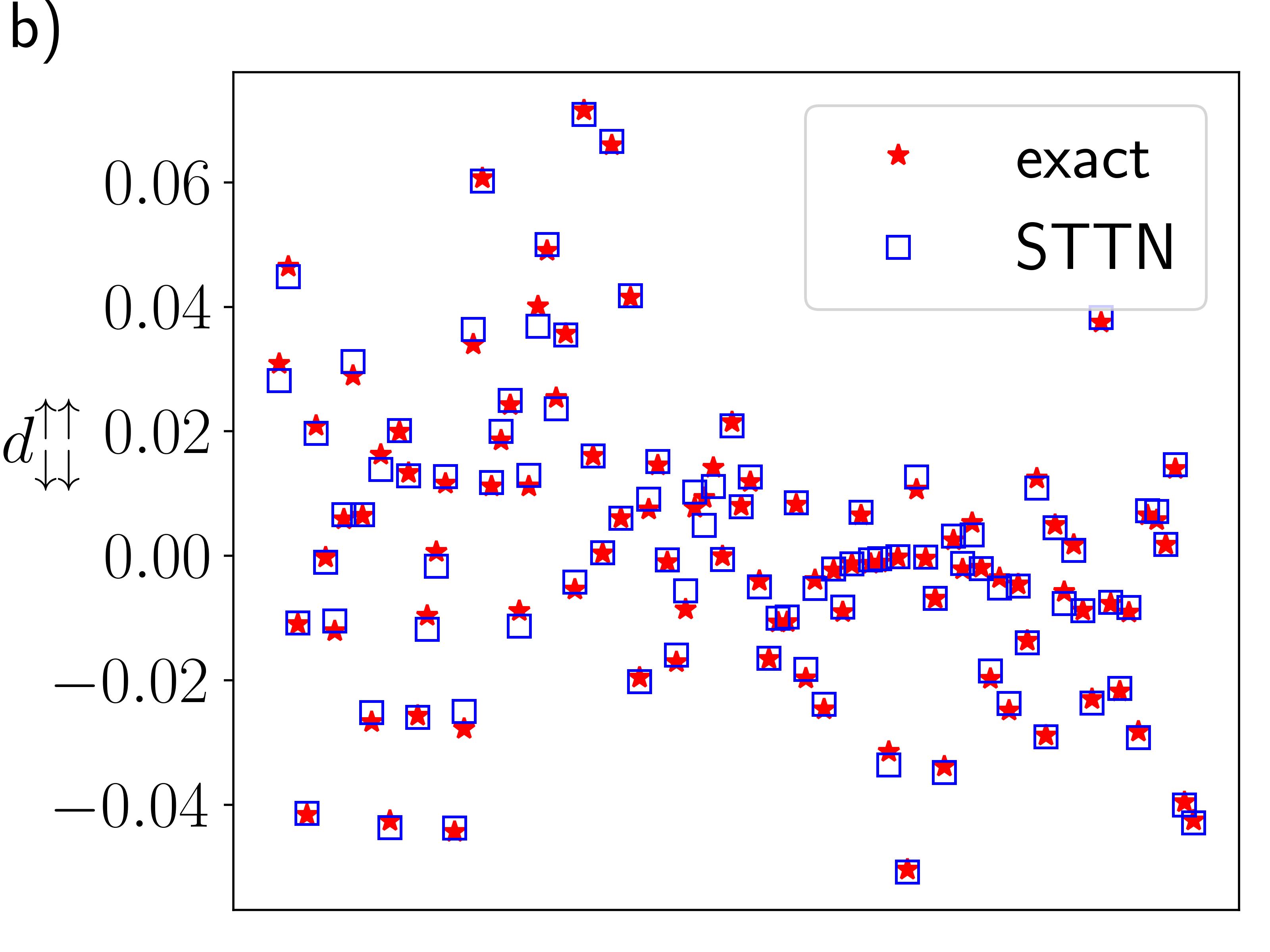}\\
\includegraphics[width=0.51\columnwidth]{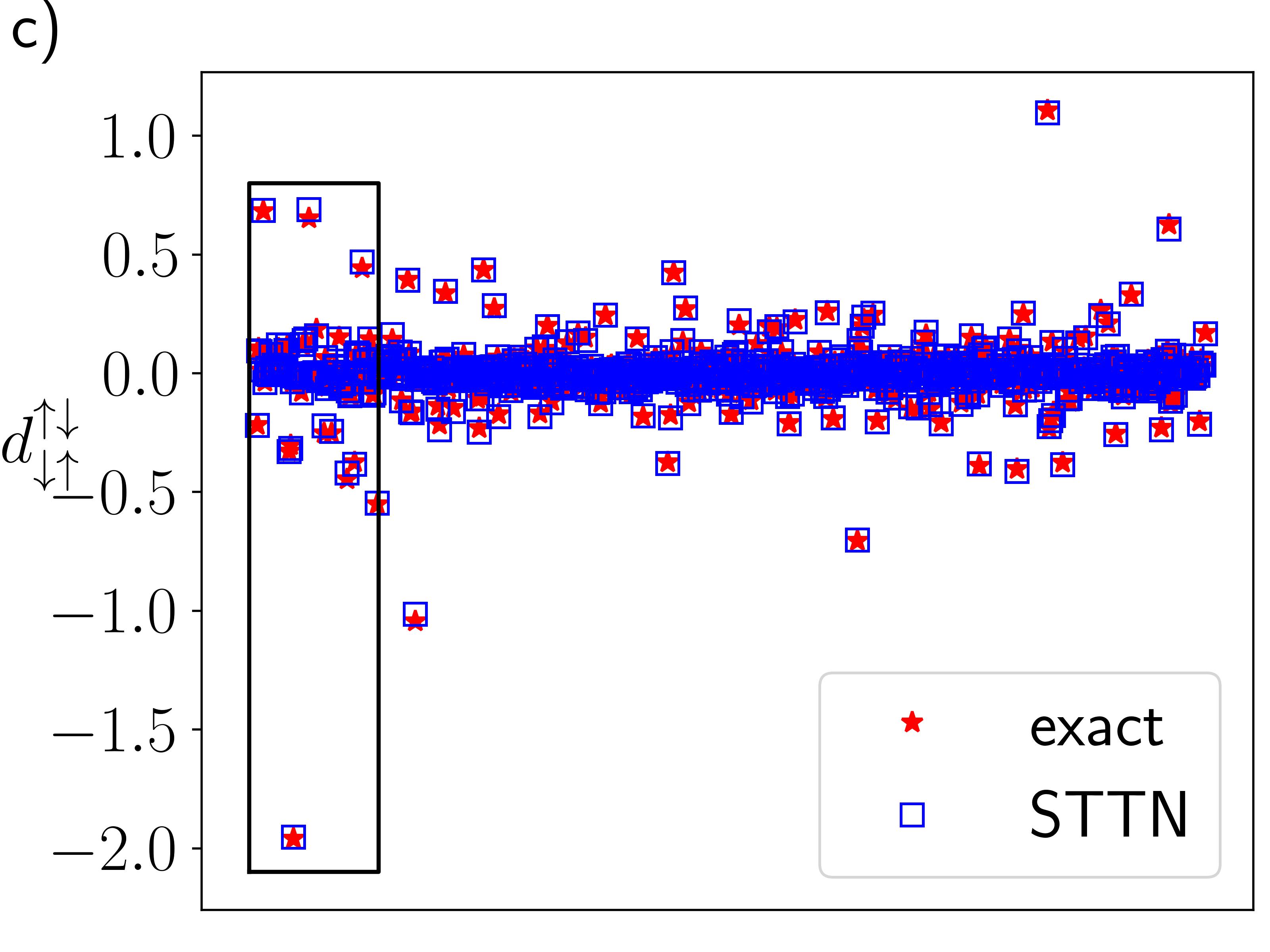}\includegraphics[width=0.51\columnwidth]{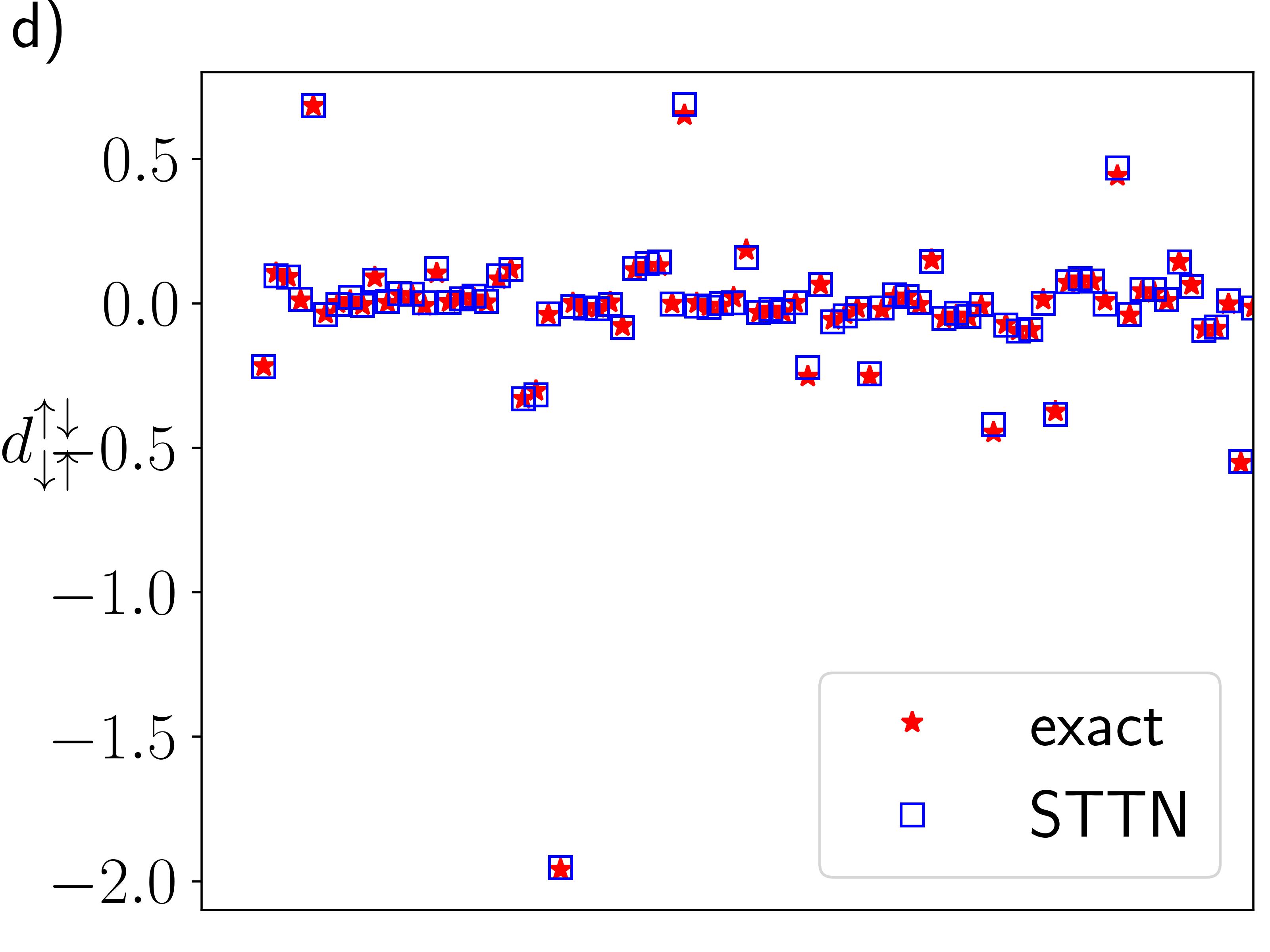}\\
\includegraphics[width=0.51\columnwidth]{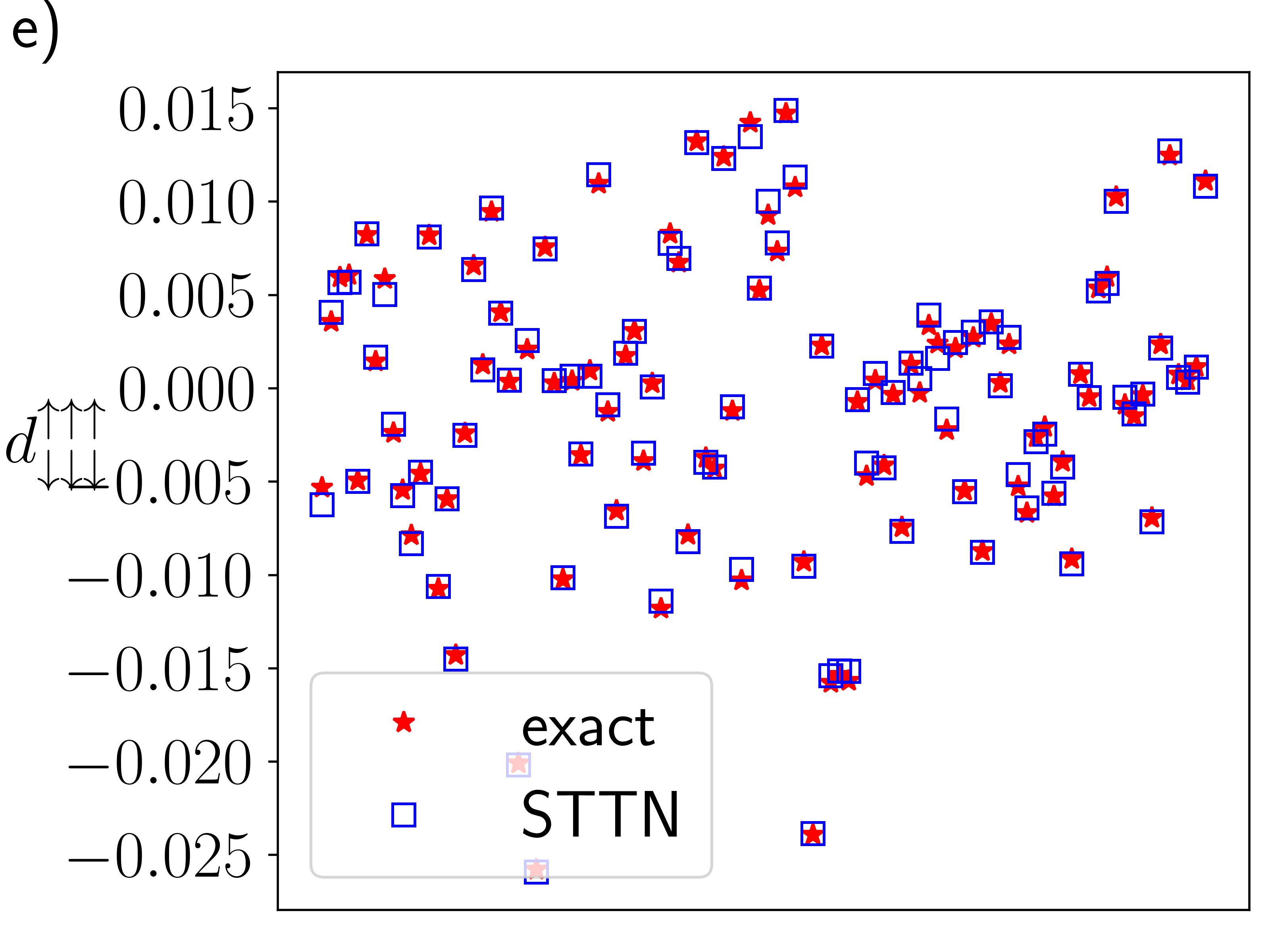}\includegraphics[width=0.51\columnwidth]{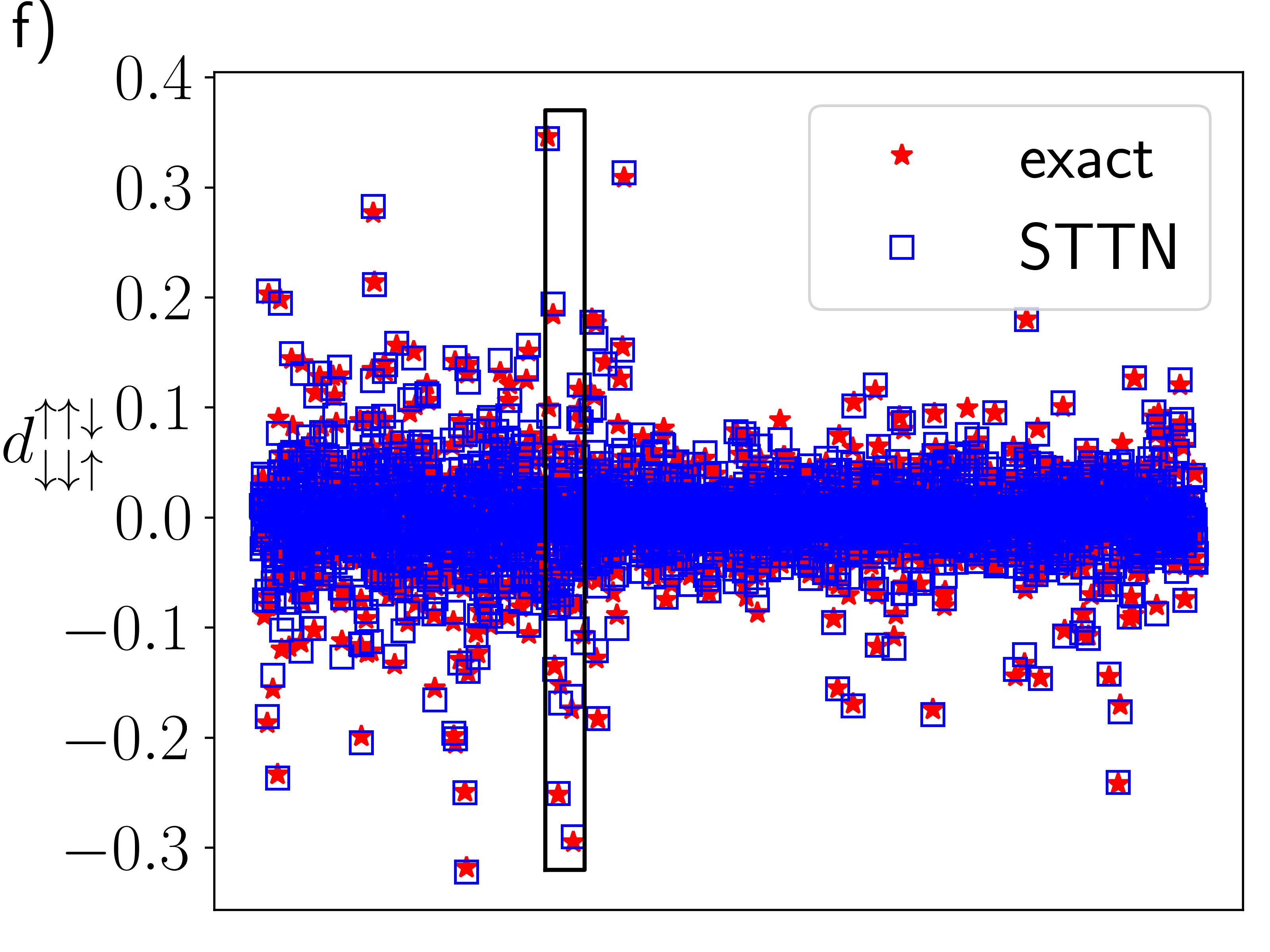}\\
\includegraphics[width=0.51\columnwidth]{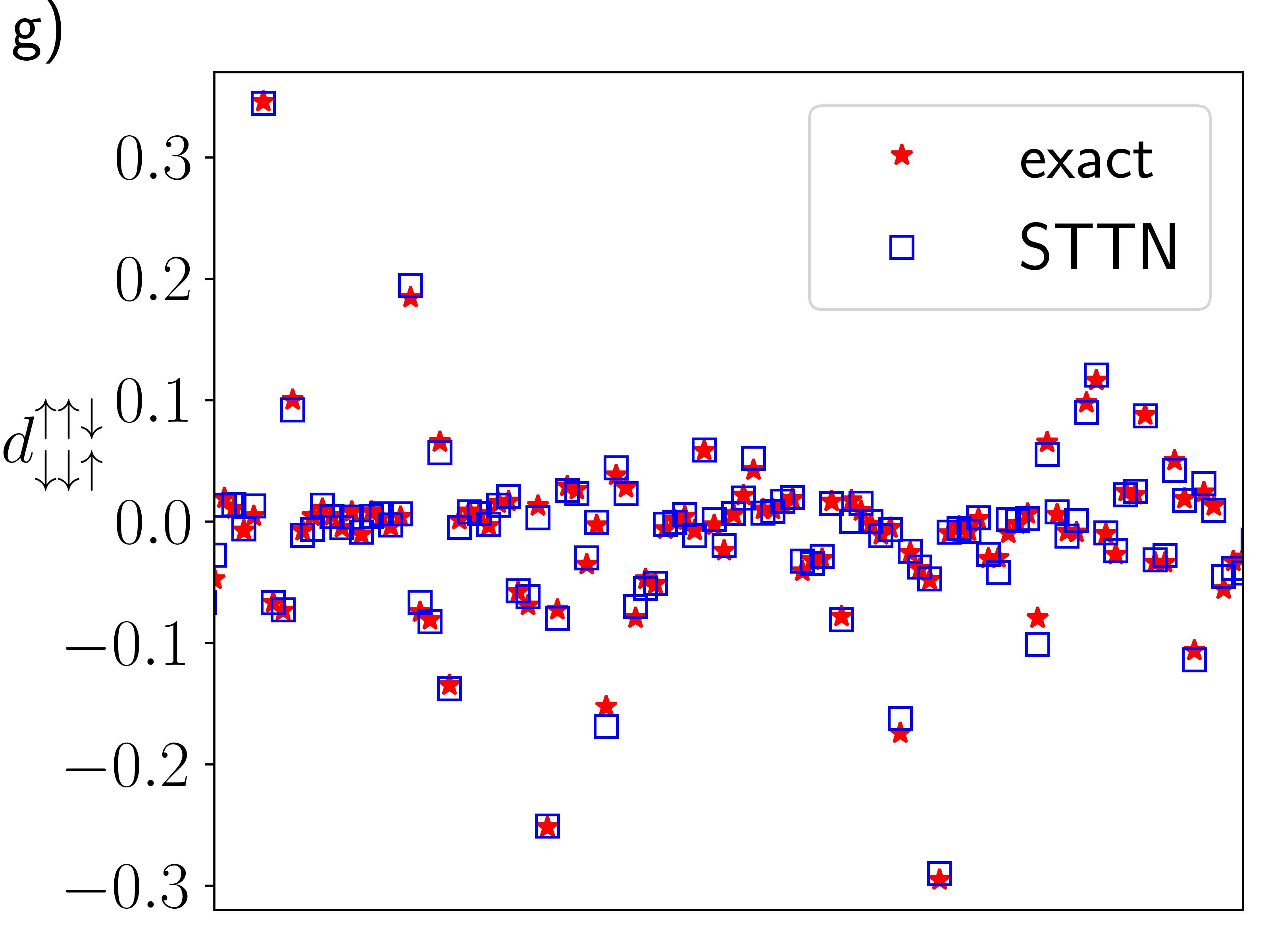}
\caption{\label{fig:10_sites_HF_T_ampl} For the 10 sites cluster, exact $T$-amplitudes and STTN fit for a) the spin-up single-excitation tensor, b) the up-up double-excitation tensor, c) the up-down double-excitation tensor, d) the region delimited by the black rectangle in c), e) the up-up-up triple-excitation tensor, f) the up-up-down triple-excitation tensor and g) the region delimited by the black rectangle in f). The horizontal axes correspond to arbitrary tensor indices. The tensor dimensions used in the STTN are listed in table \ref{tab:tensor_dim_T}.}
\end{figure}
\begin{figure}[h]
\includegraphics[width=0.51\columnwidth]{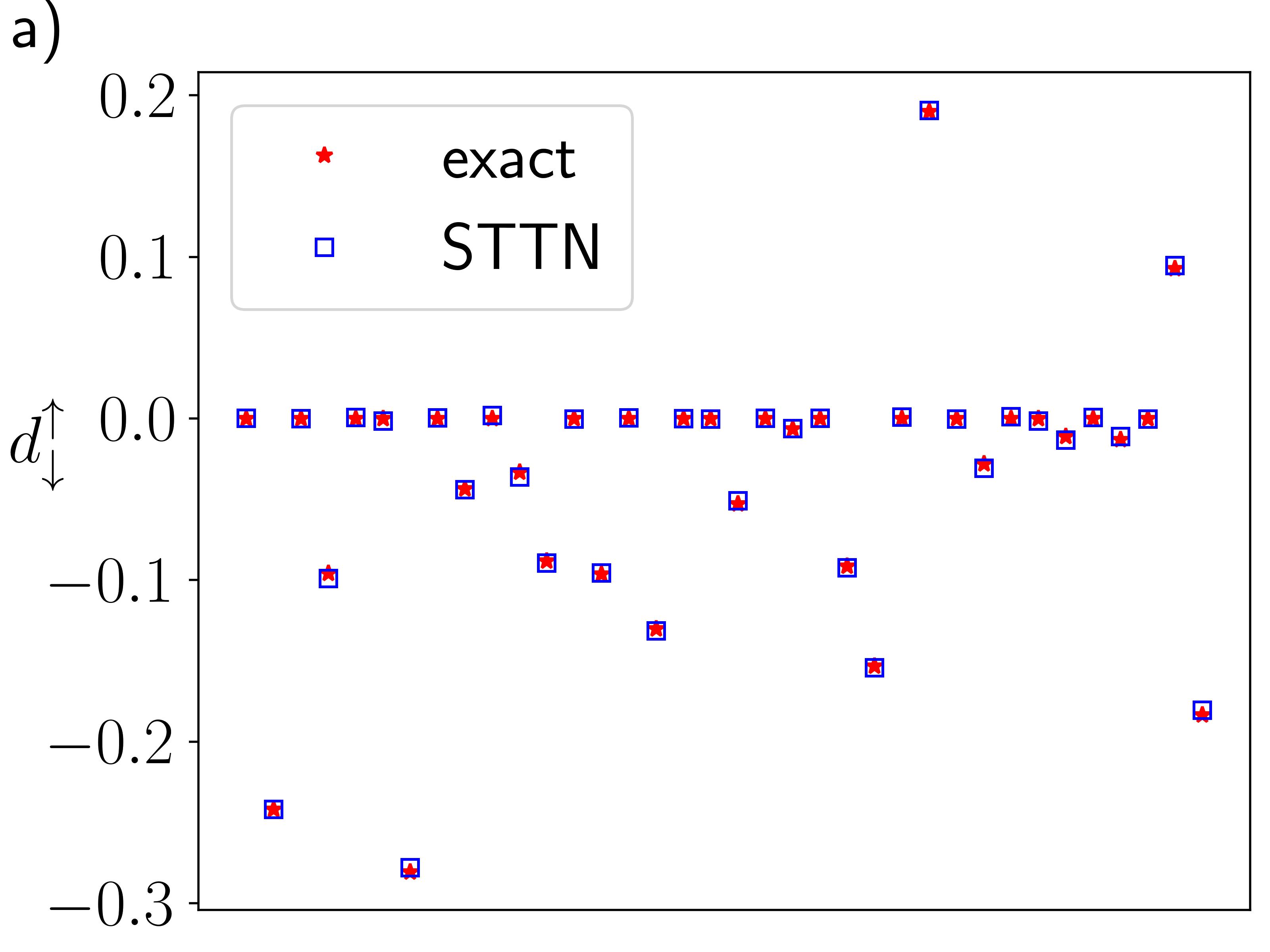}\includegraphics[width=0.51\columnwidth]{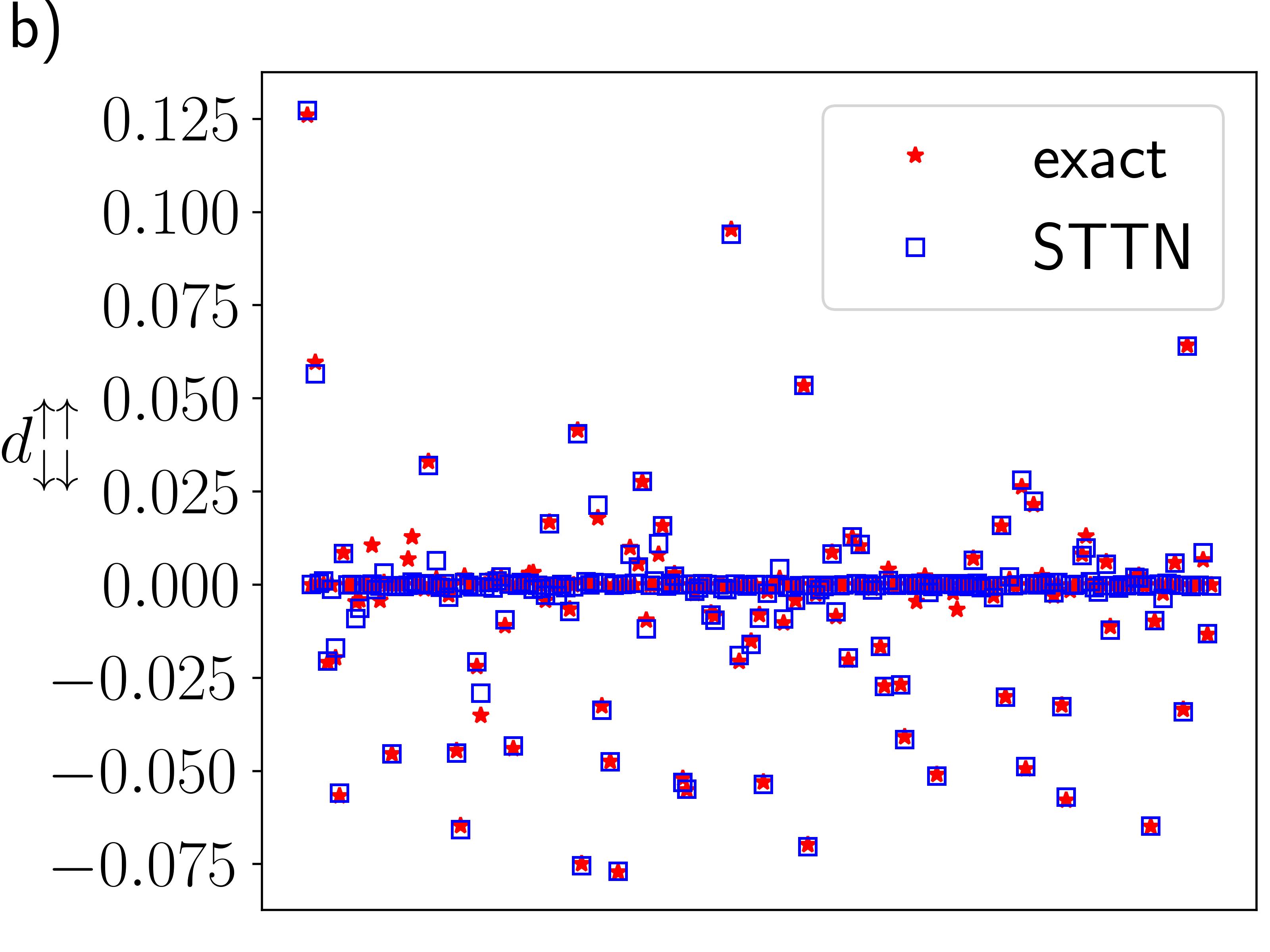}\\
\includegraphics[width=0.51\columnwidth]{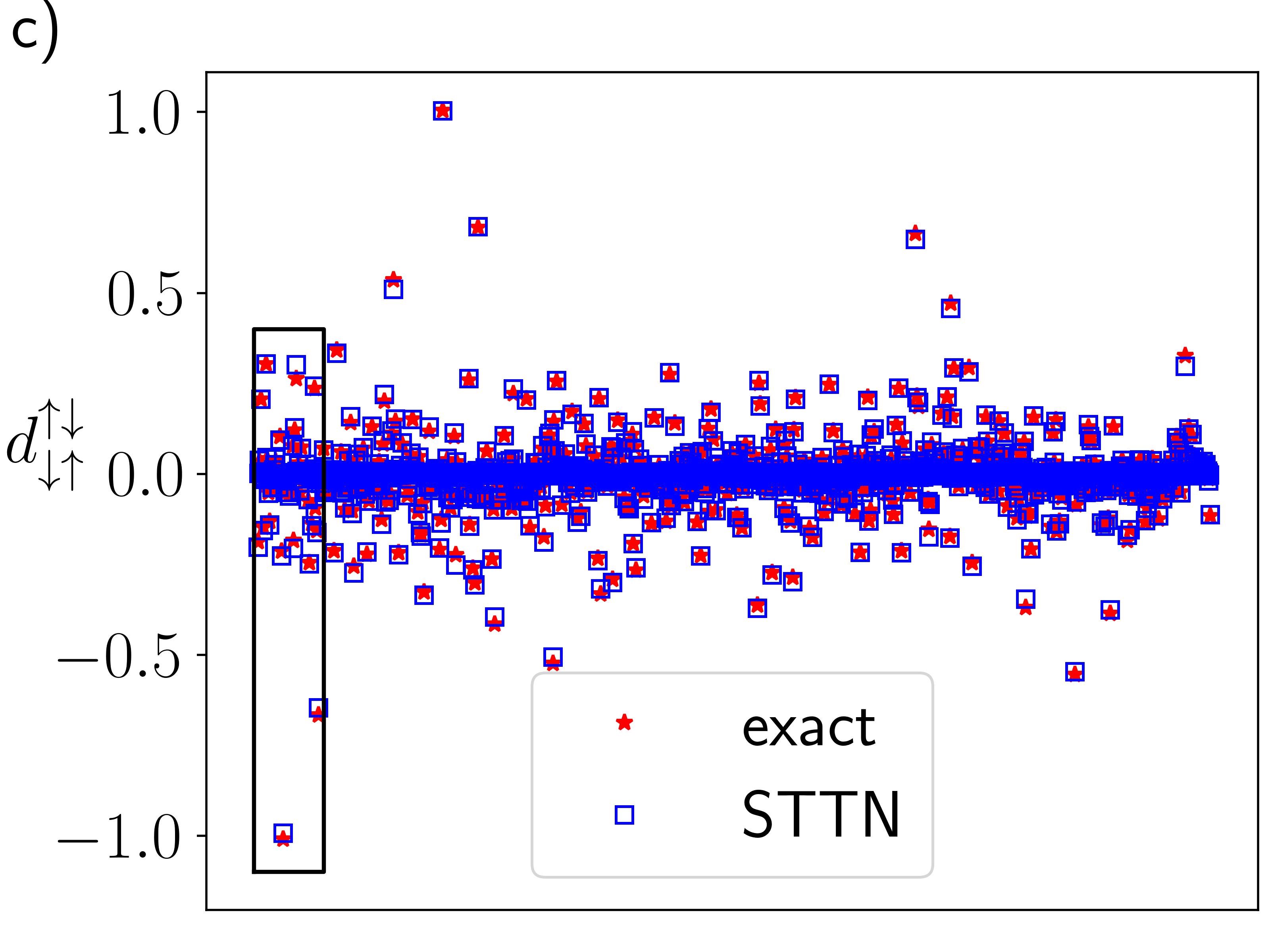}\includegraphics[width=0.51\columnwidth]{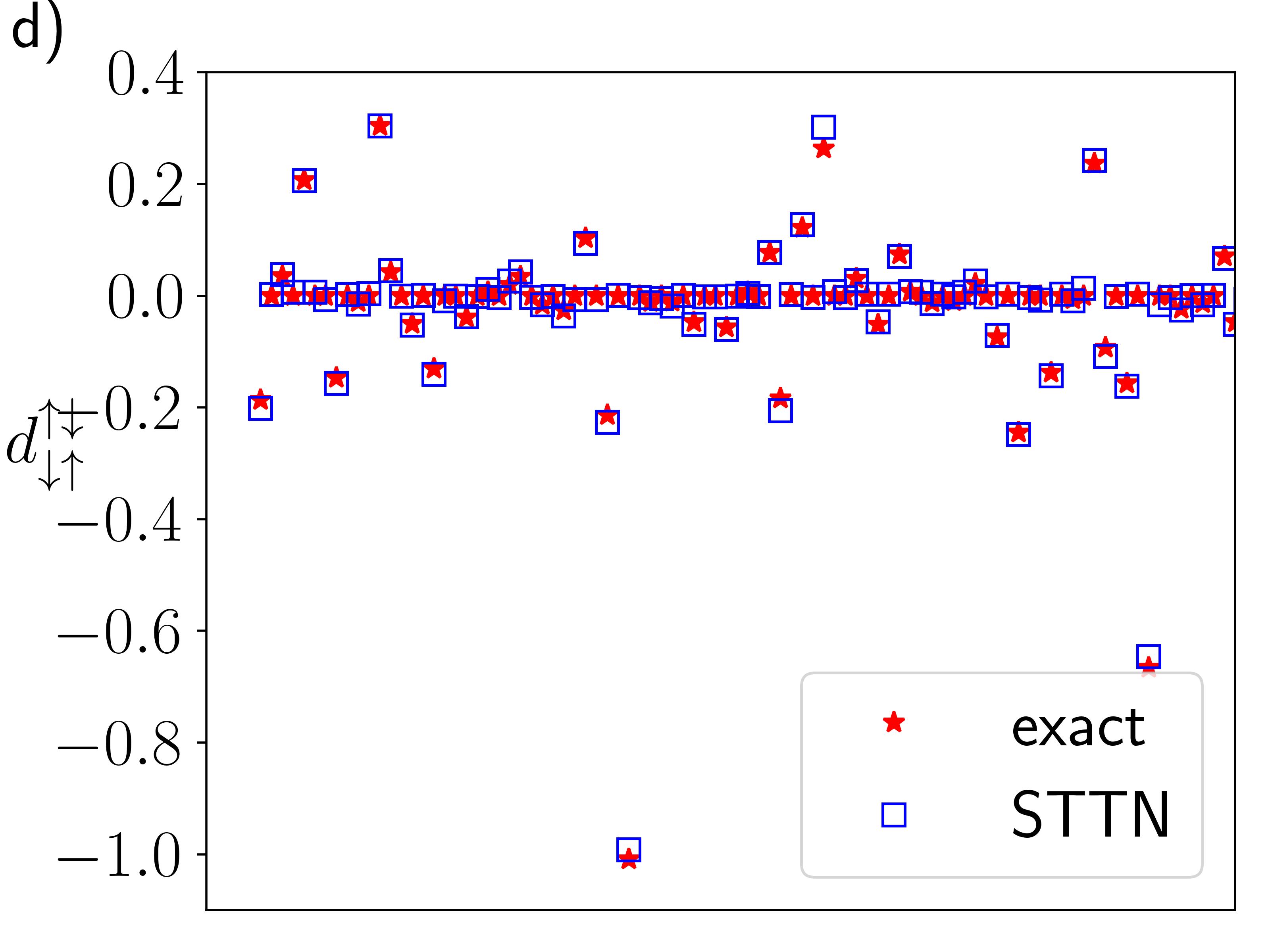}\\
\includegraphics[width=0.51\columnwidth]{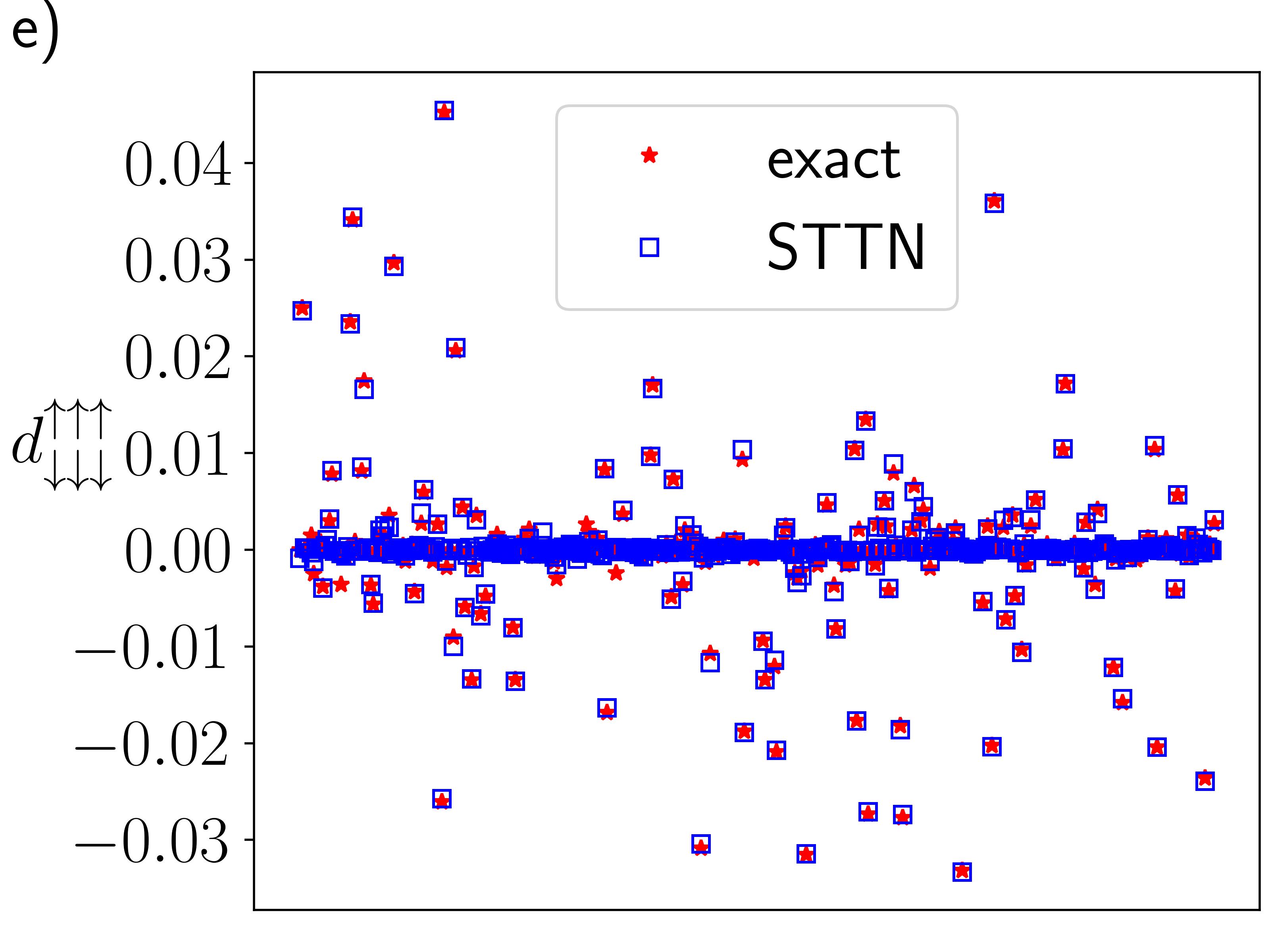}\includegraphics[width=0.51\columnwidth]{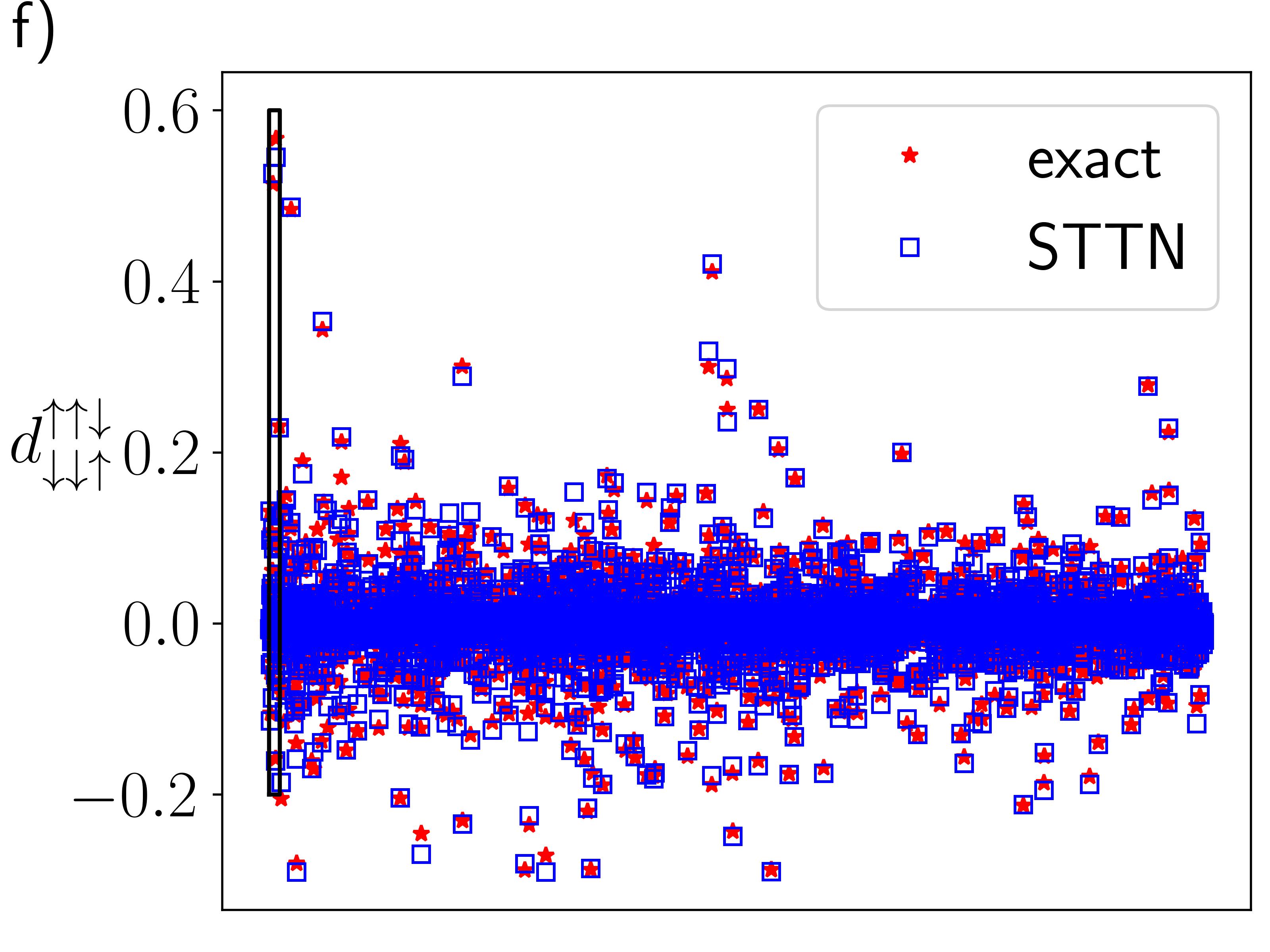}\\
\includegraphics[width=0.51\columnwidth]{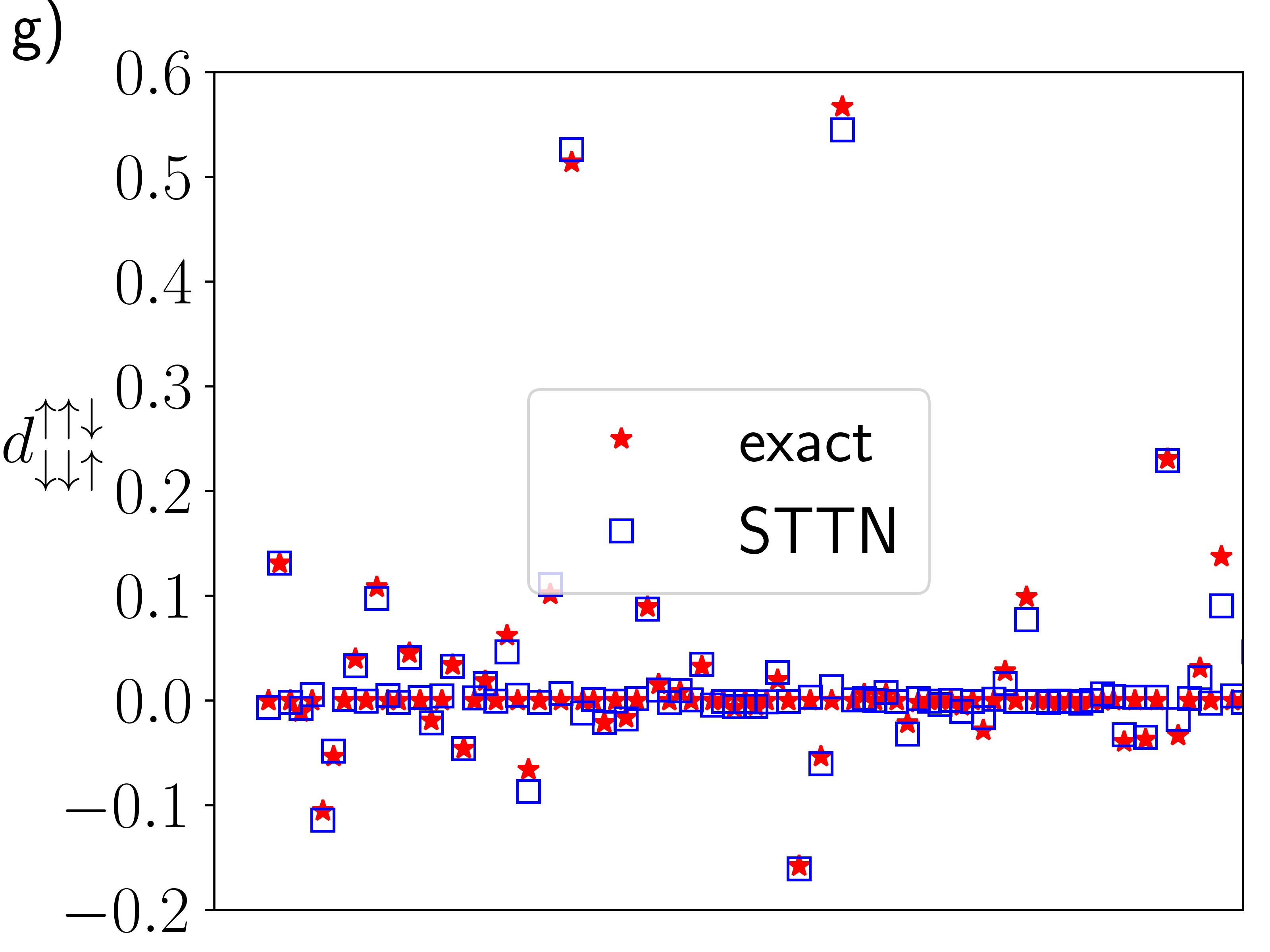}
\caption{\label{fig:12_sites_HF_T_ampl} For the 12 sites cluster, exact $T$-amplitudes and STTN fit for a) the single-excitation tensor, b) the up-up double-excitation tensor, c) the up-down double-excitation tensor, d) the region delimited by the black rectangle in c), e) the up-up-up triple-excitation tensor, f) the up-up-down triple-excitation tensor and g) the region delimited by the black rectangle in f). The horizontal axes correspond to arbitrary tensor indices. The tensor dimensions used in the STTN are listed in table \ref{tab:tensor_dim_T}.}
\end{figure}
\begin{figure}[h]
\includegraphics[width=0.51\columnwidth]{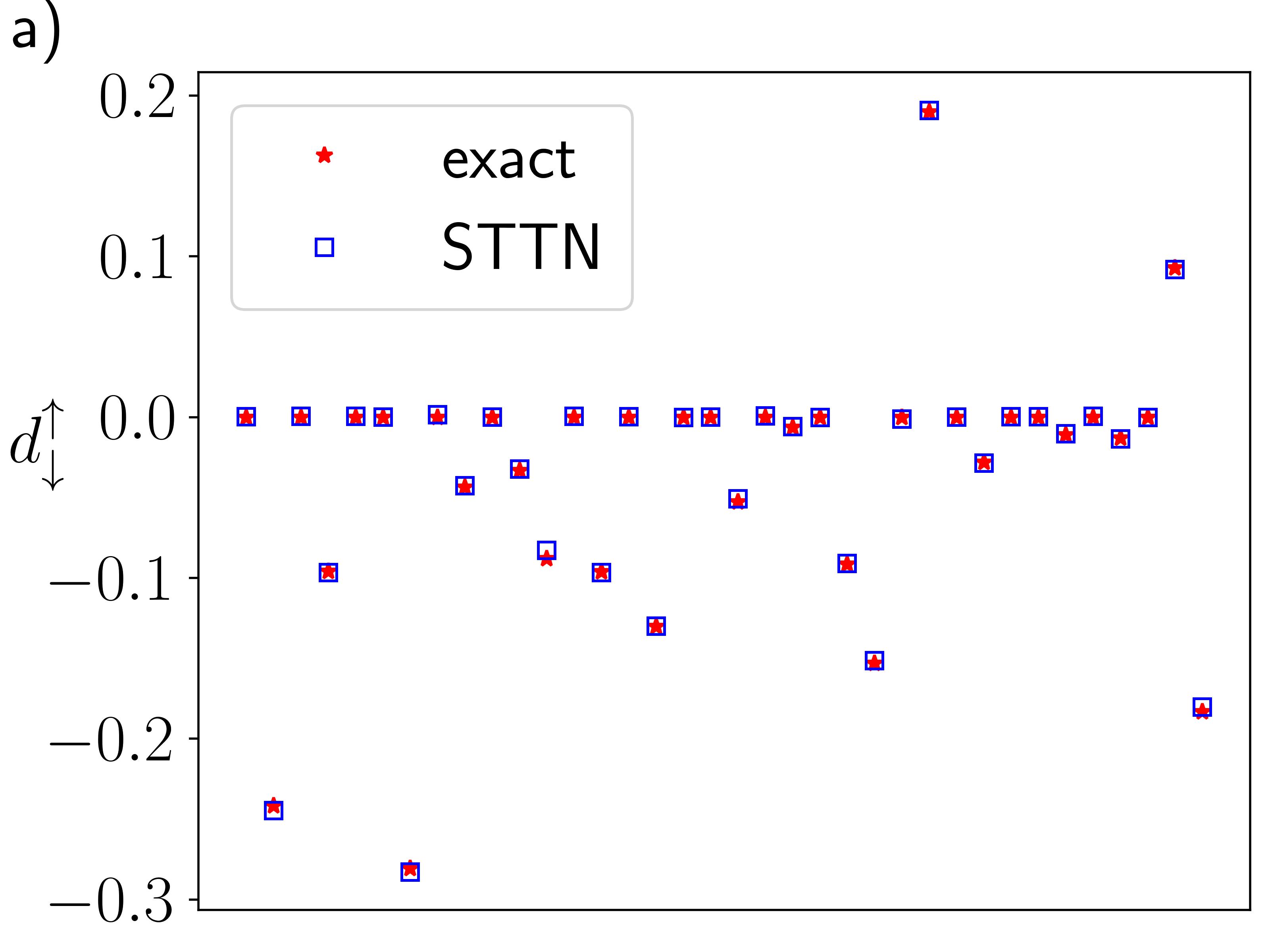}\includegraphics[width=0.51\columnwidth]{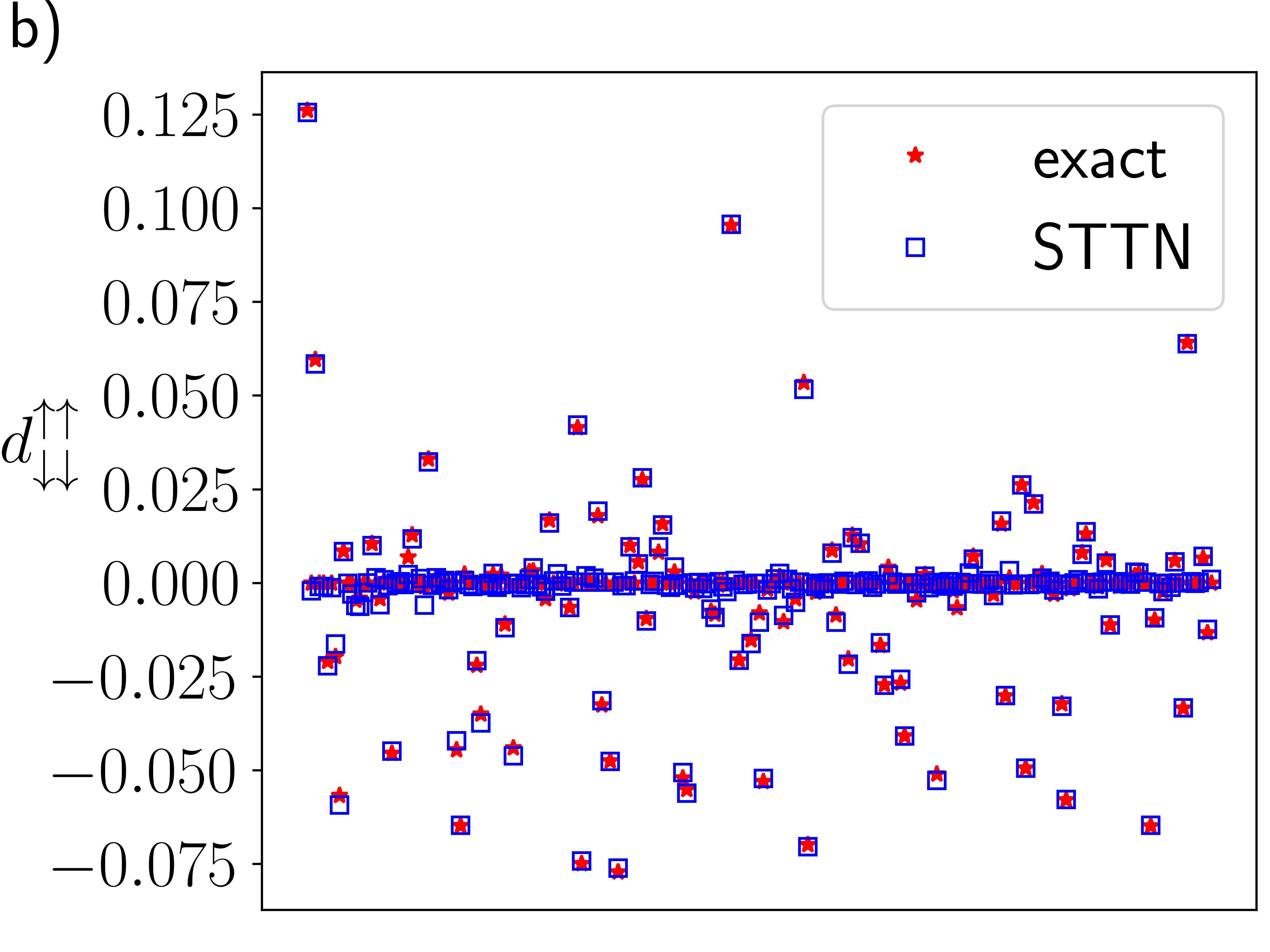}\\
\includegraphics[width=0.51\columnwidth]{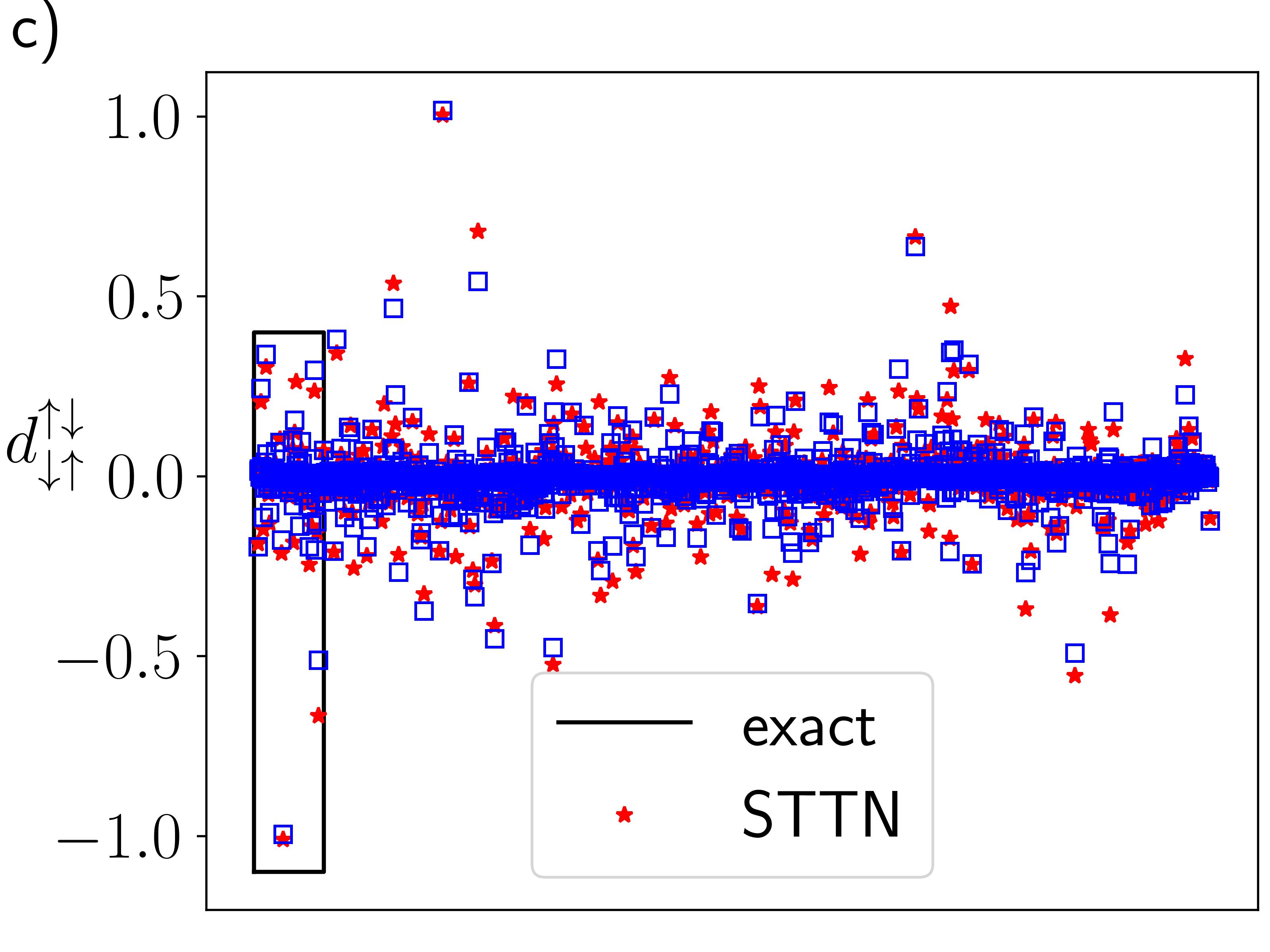}\includegraphics[width=0.51\columnwidth]{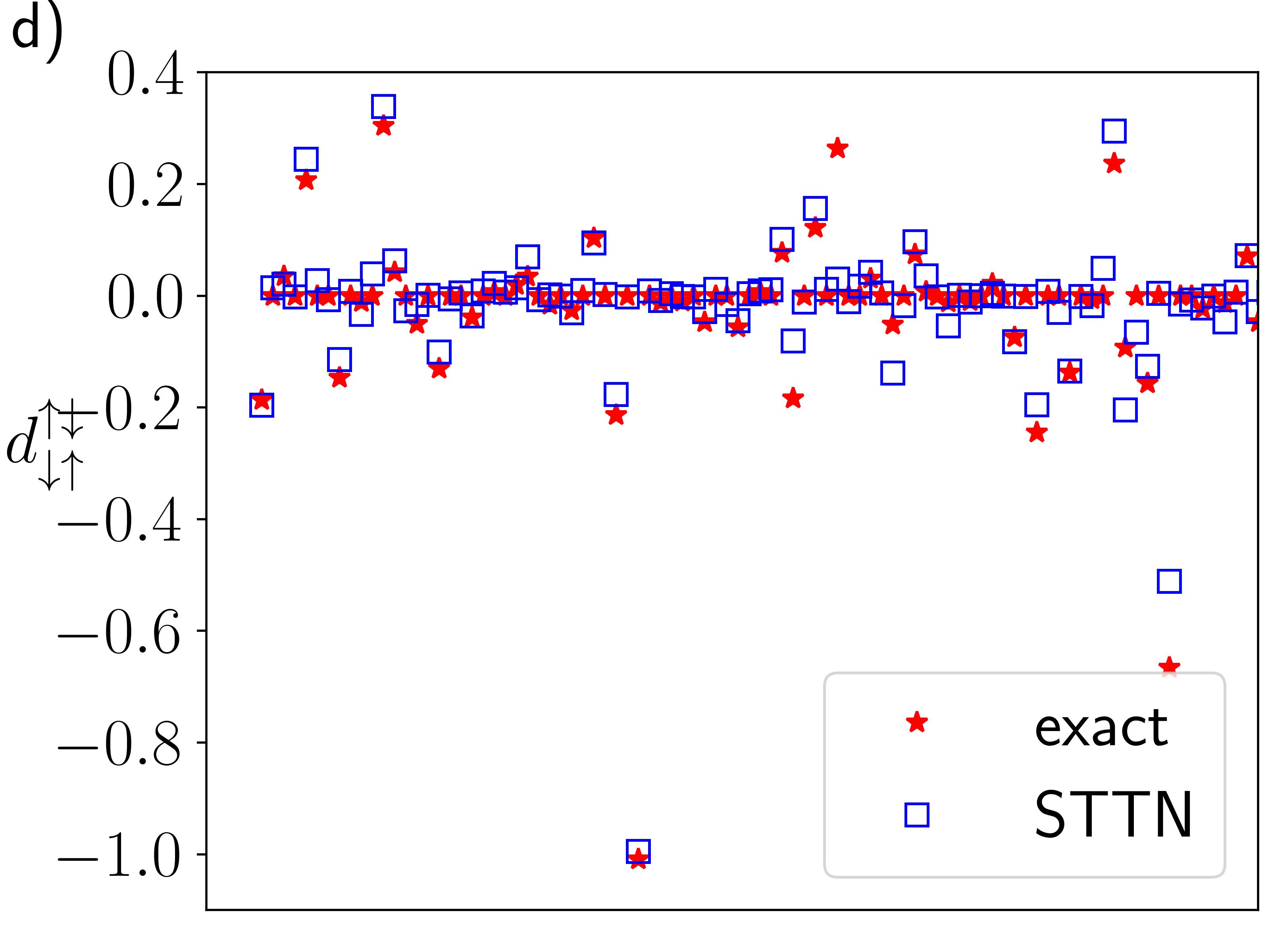}\\
\includegraphics[width=0.51\columnwidth]{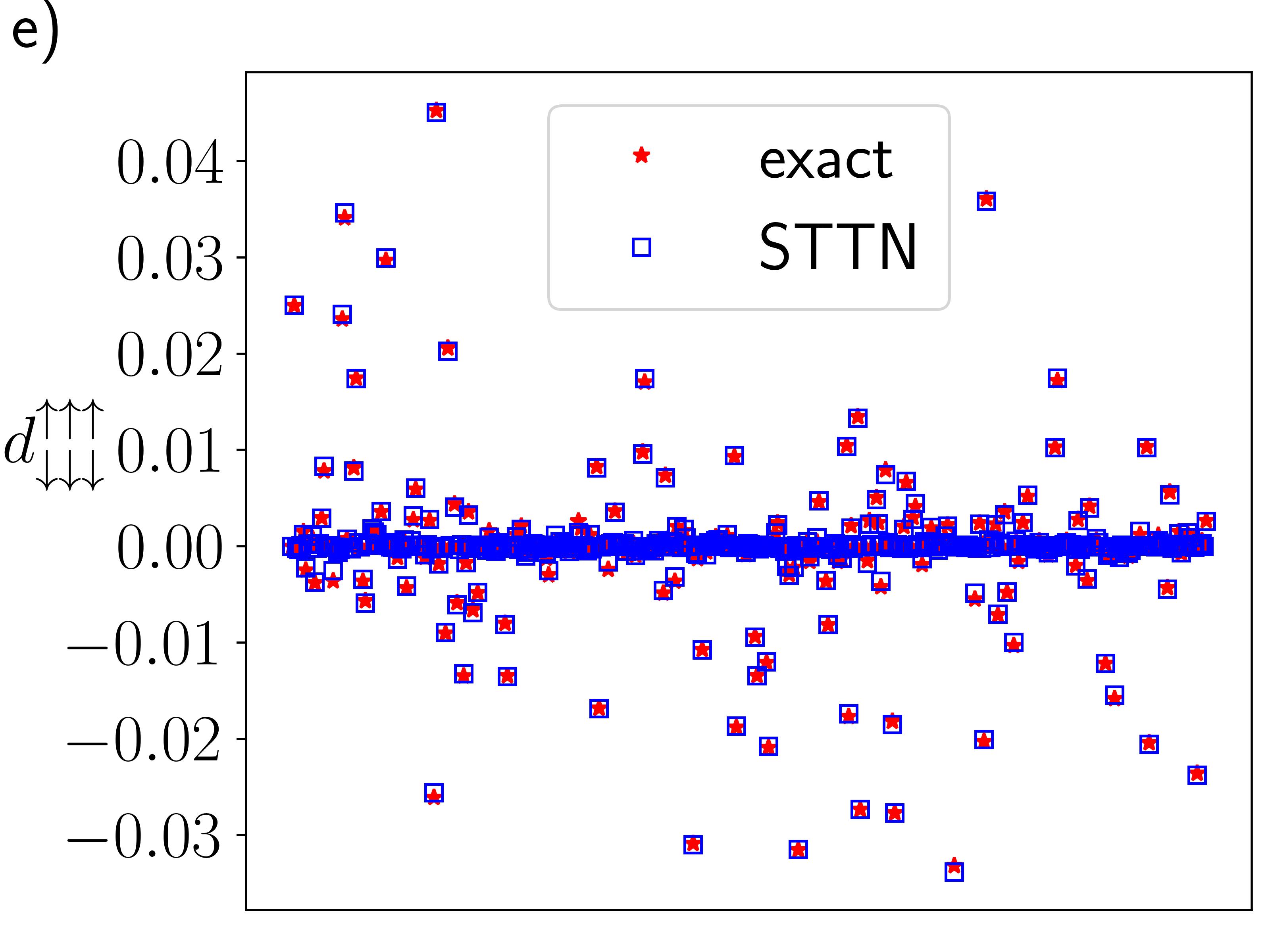}\includegraphics[width=0.51\columnwidth]{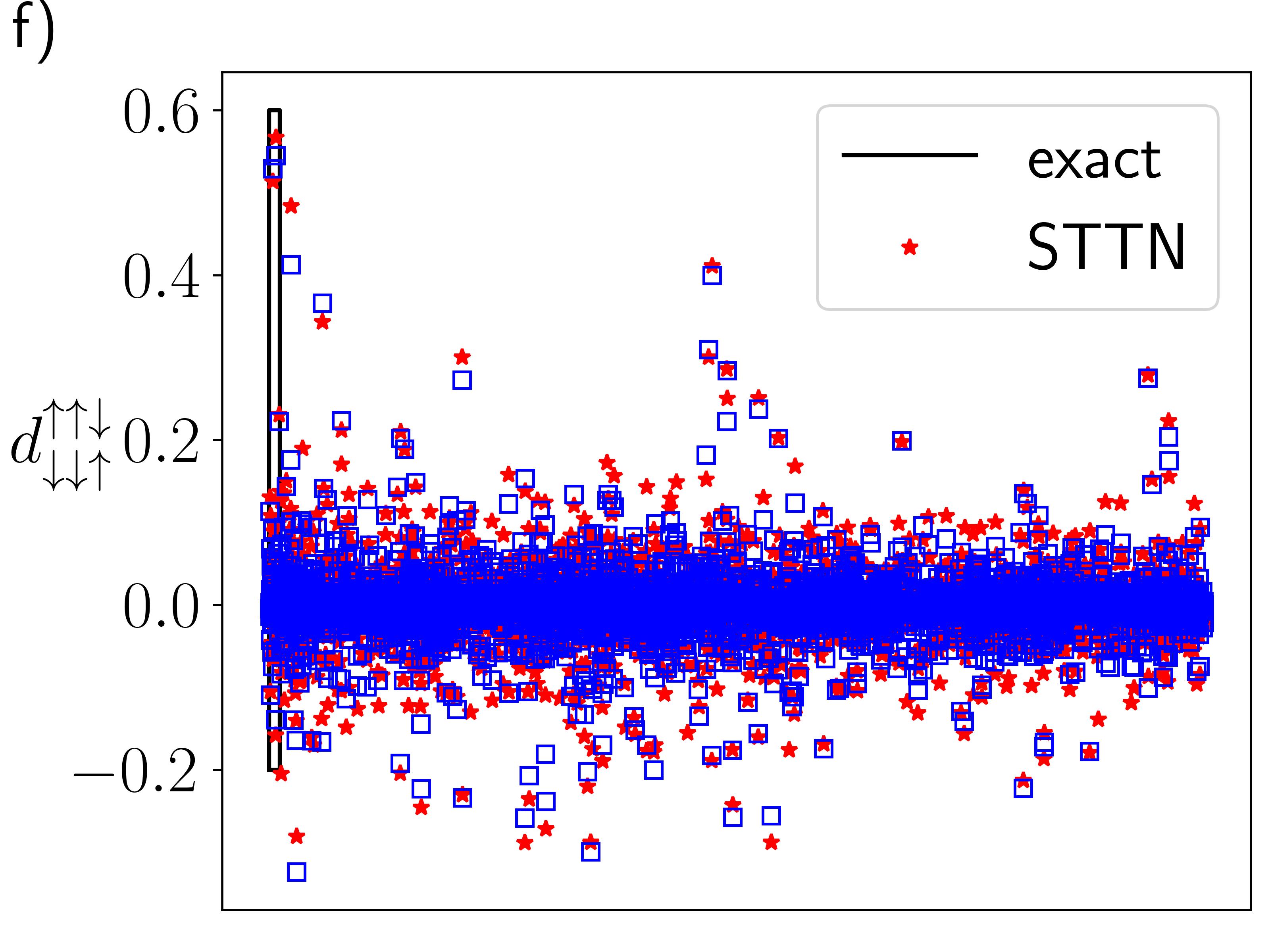}\\
\includegraphics[width=0.51\columnwidth]{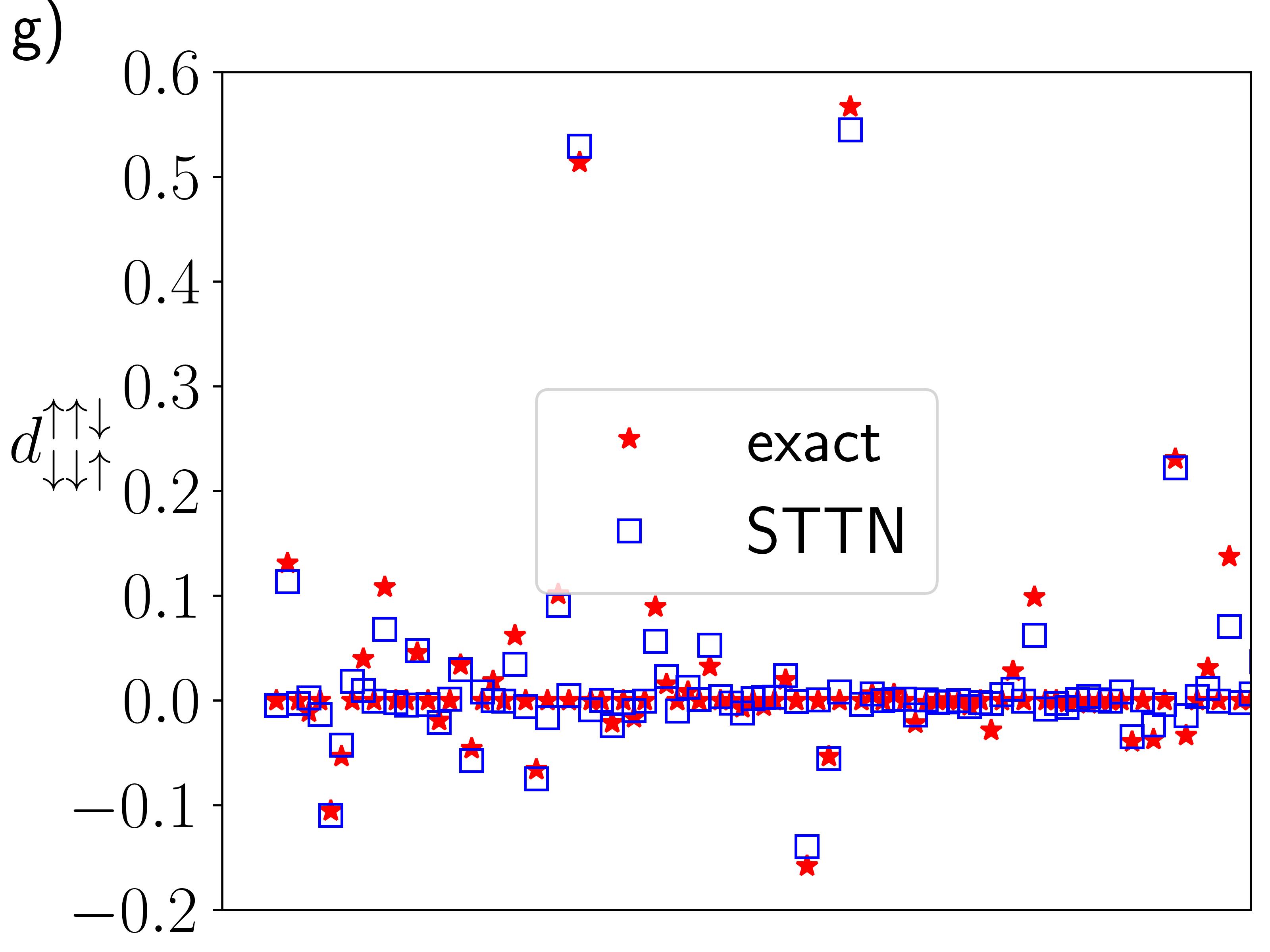}\
\caption{\label{fig:12_sites_HF_T_ampl_single_channel} For the 12 sites cluster, exact $T$-amplitudes and STTN fit using only one channel in each decomposition, for a) the single-excitation tensor, b) the up-up double-excitation tensor, c) the up-down double-excitation tensor, d) the region delimited by the black rectangle in c), e) the up-up-up triple-excitation tensor, f) the up-up-down triple-excitation tensor and g) the region delimited by the black rectangle in f). The horizontal axes correspond to arbitrary indices in the tensors. The tensor dimensions used in the STTN are listed in table \ref{tab:tensor_dim_T_SC}.}
\end{figure}
\begin{figure}[h]
\includegraphics[width=0.51\columnwidth]{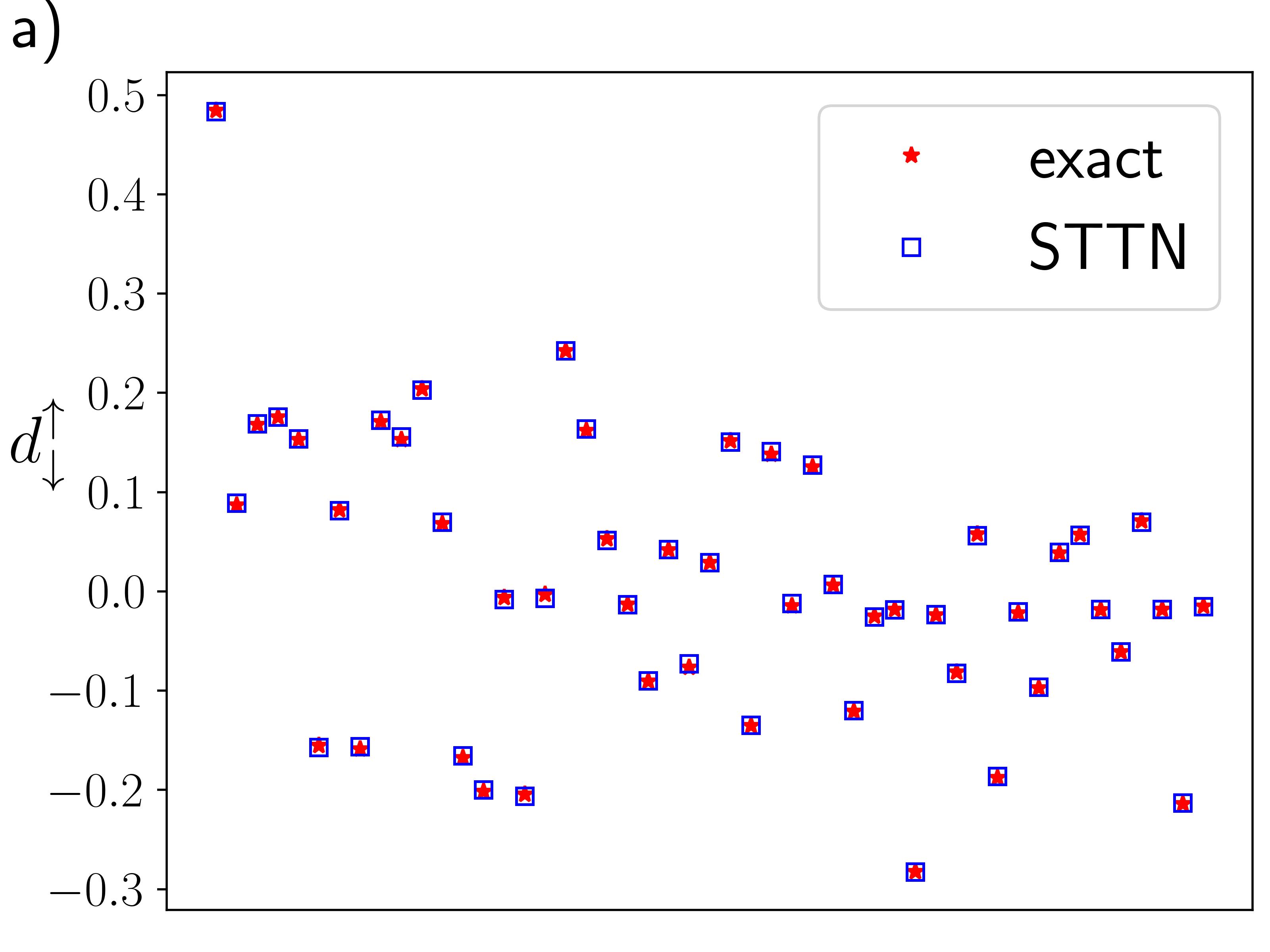}\includegraphics[width=0.51\columnwidth]{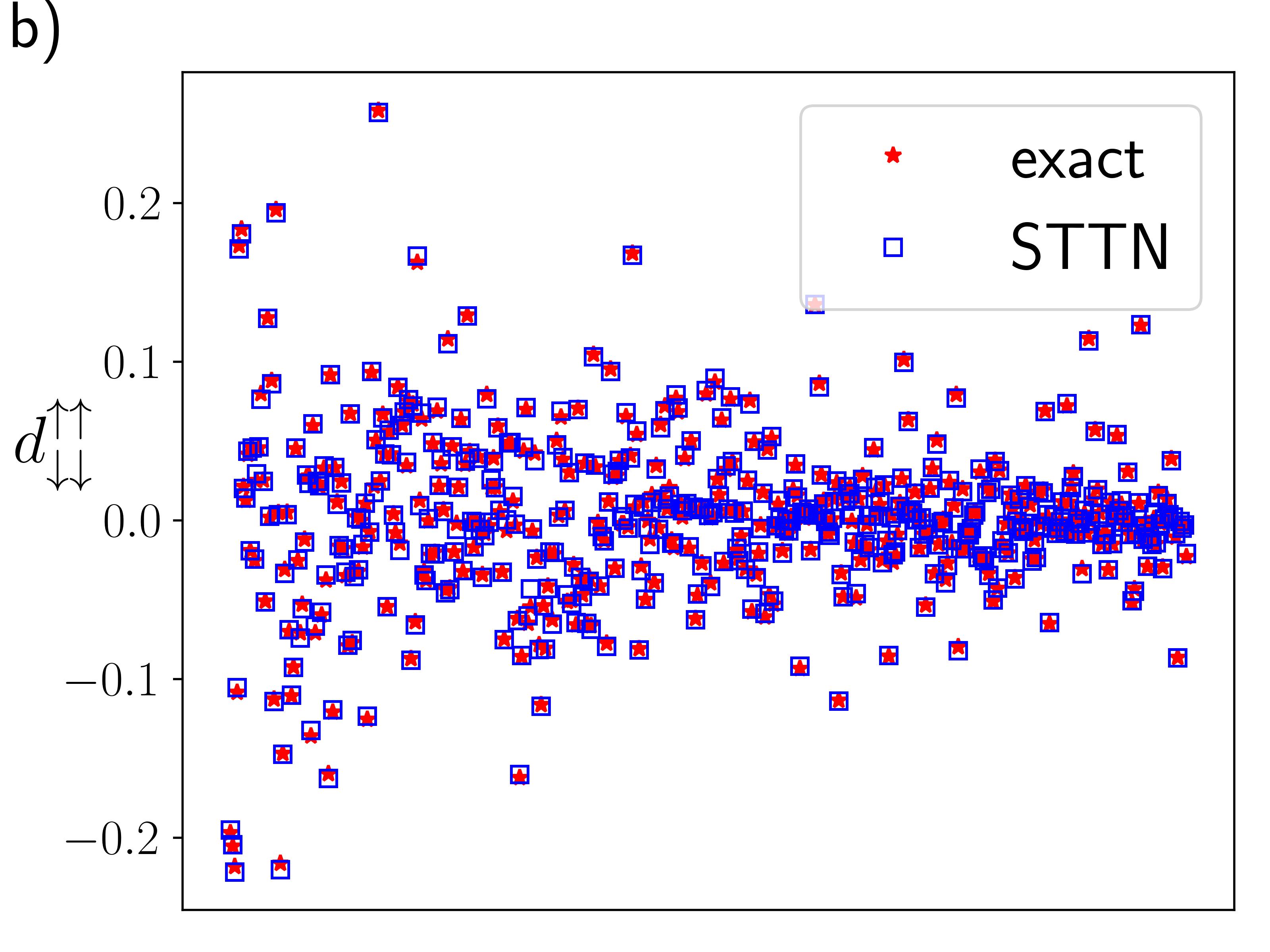}\\
\includegraphics[width=0.51\columnwidth]{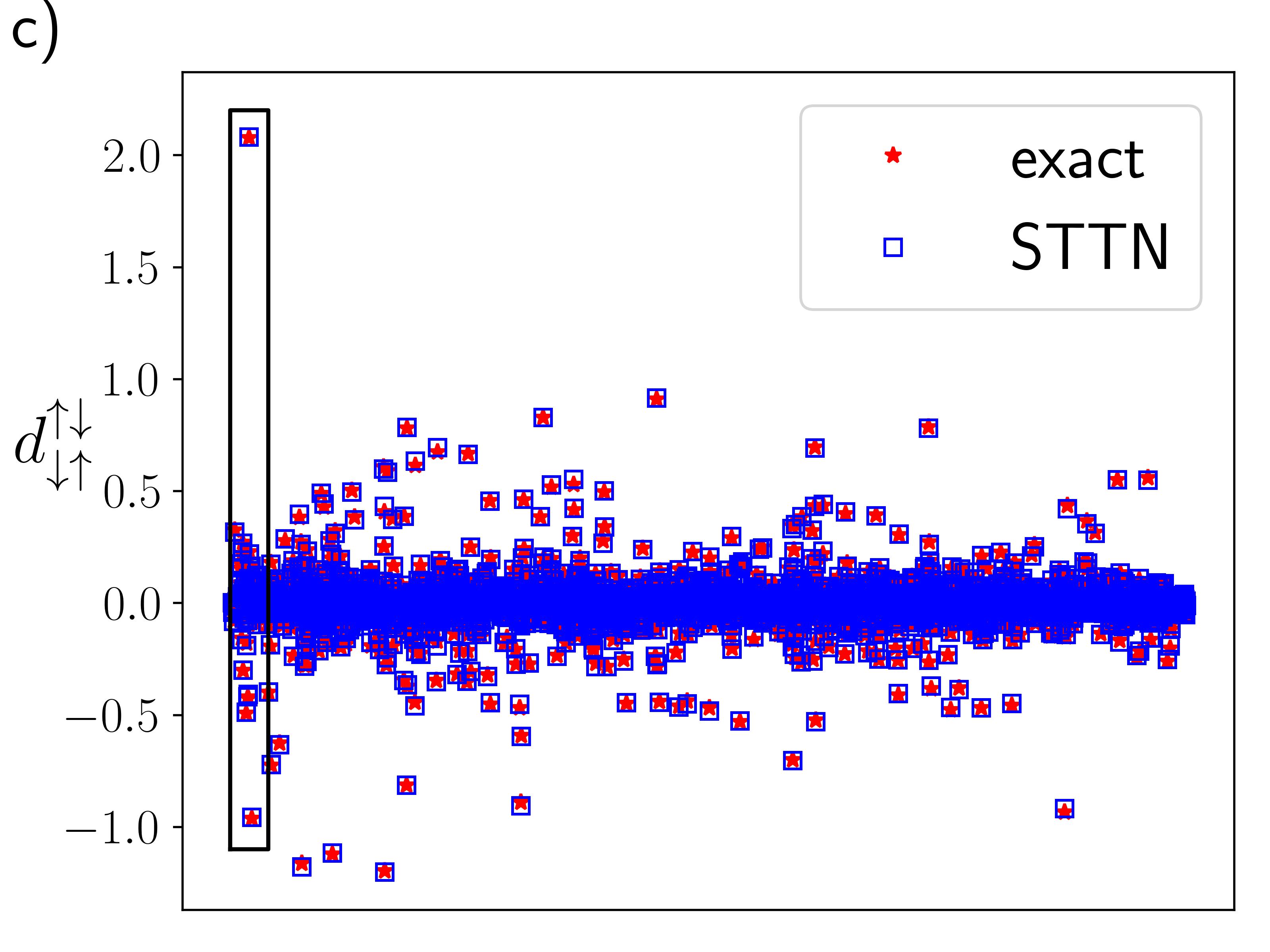}\includegraphics[width=0.51\columnwidth]{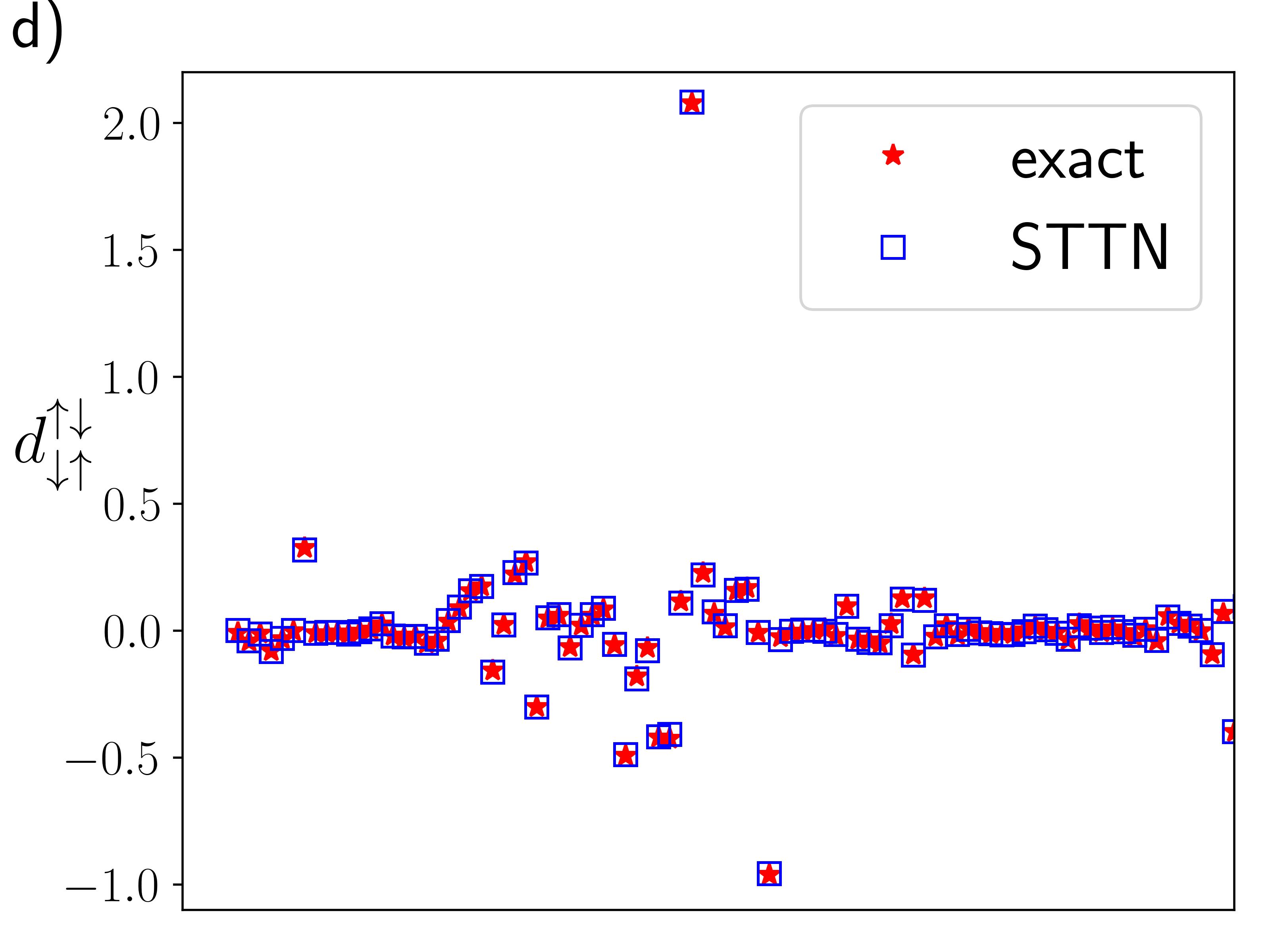}\\
\includegraphics[width=0.51\columnwidth]{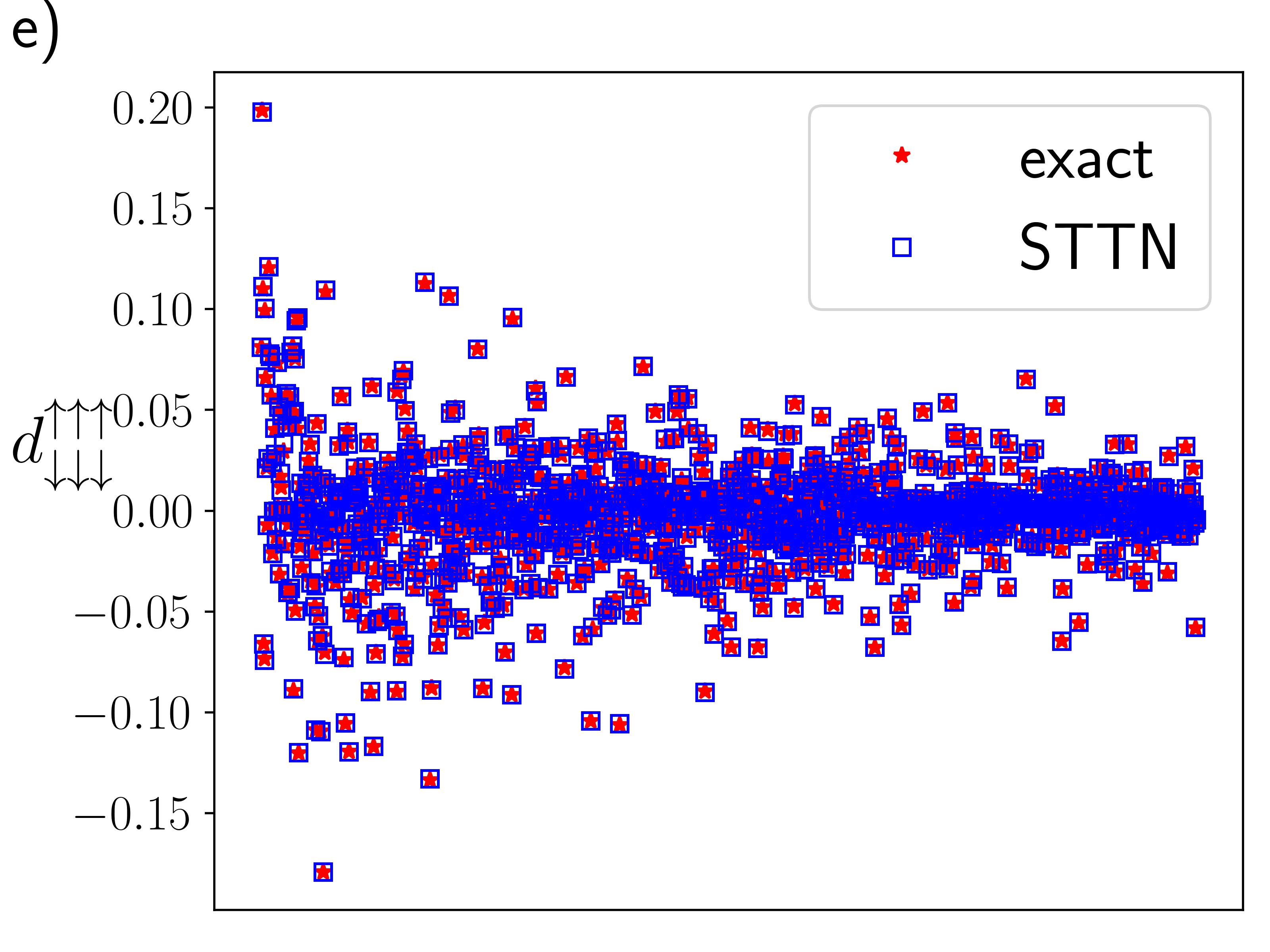}\includegraphics[width=0.51\columnwidth]{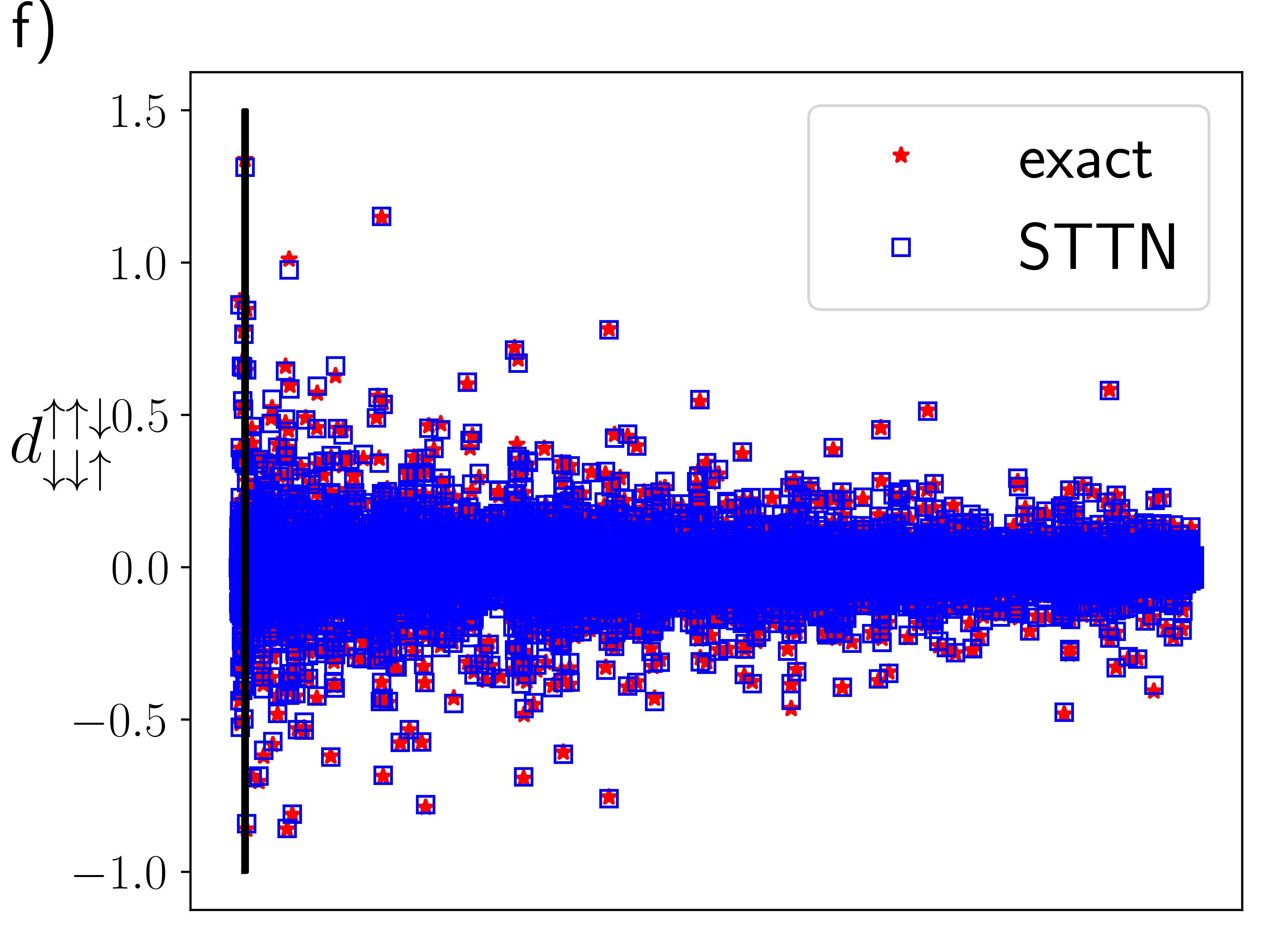}\\
\includegraphics[width=0.51\columnwidth]{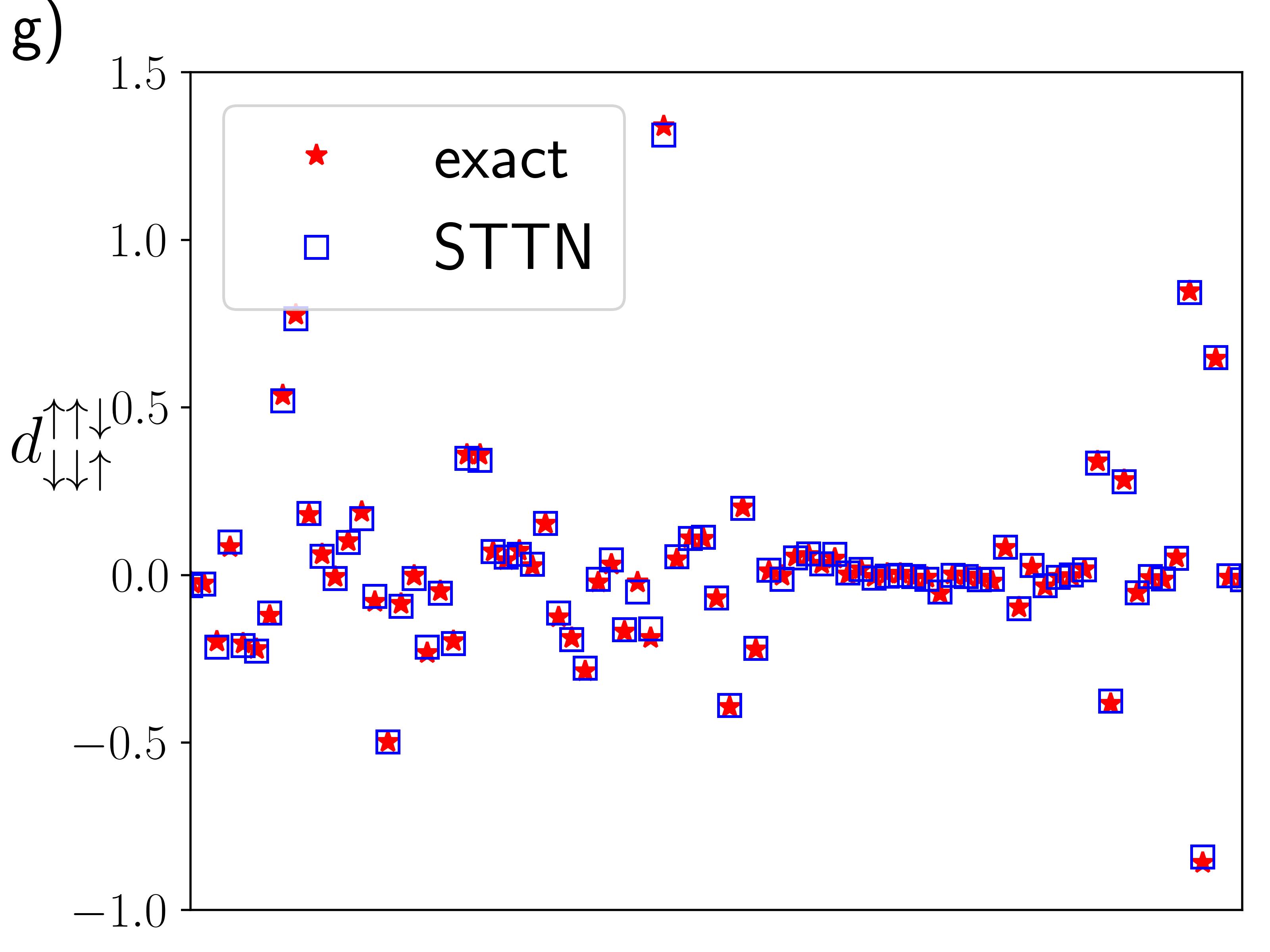}
\caption{\label{fig:14_sites_HF_T_ampl} For the 14 sites cluster, exact $T$-amplitudes and STTN fit for a) the single-excitation tensor, b) the up-up double-excitation tensor, c) the up-down double-excitation tensor, d) the region delimited by the black rectangle in c), e) the up-up-up triple-excitation tensor, f) the up-up-down triple-excitation tensor and g) the region delimited by the black rectangle in f). The horizontal axes correspond to arbitrary tensor indices. The tensor dimensions used in the STTN are listed in table \ref{tab:tensor_dim_T}.}
\end{figure}

Let us first consider the 10 sites cluster. In Fig. \ref{fig:10_sites_HF_T_ampl} are shown the elements of sample $T$-tensors and the corresponding STTN fit. Note that, due to the nature of the STTN structure, all the tensors are obtained using a single fit. All channels defined in the preceding section were included in that fit, using 857 free parameters to represent the 6075 $T$-amplitudes with three excitations or less. The STTN tensor dimensions are listed in the first column of table \ref{tab:tensor_dim_T}. Although the number of free parameters is much smaller than the number of tensor elements, we observe that the exact tensors are well reproduced by the STTN fit. If we use the energy equation \eqref{eq:DE_CC_T} with the STTN $T$-amplitudes, we obtain an error on the energy of $1.4\%$. The results for the STTN fit to the wave function coefficients are provided in Fig. \ref{fig:10_sites_HF_coef} of appendix \ref{sec:results_STTN_wv_cfs}. In that case the fit is slightly less accurate and the error on the energy is $1.6\%$.

For the 12 sites cluster, we have used 1804 free parameters to represents all the 18819 $T$-amplitudes up to three excitations, also using all the available channels in each decomposition. The results are shown in Fig. \ref{fig:12_sites_HF_T_ampl}. Here we also observe a very good agreement between the exact tensors and the STTN fit, which is confirmed by the small error on the energy of about $0.4\%$ obtained when using the energy equation with the STTN results. The STTN tensor dimensions for that fit are listed on the second column of table \ref{tab:tensor_dim_T}. The STTN fit directly to the wave function coefficients is shown in Fig. \ref{fig:12_sites_HF_coef} of appendix \ref{sec:results_STTN_wv_cfs}, and yield to a similar error on the energy.

To test the effect of combining different channels in a given tensor decompositions, we have also tried to fit the $T$-amplitudes and wave function coefficients using only one channel in each decomposition, but with a similar number of free parameters as with multiple channels. Figure \ref{fig:12_sites_HF_T_ampl_single_channel} shows the results for that type of fit to the $T$-amplitudes for the 12 sites cluster. In that case, the channels we have used are derived from the $p\bar{h}$ pairing, namely the $p\bar{h}p\bar{h}$ channel for $d^{\uparrow\uparrow}_{\downarrow\downarrow}$, the $p\bar{h}\bar{p}h$ channel for $d^{\uparrow\downarrow}_{\downarrow\uparrow}$, the $p\bar{h}p\bar{h}p\bar{h}$ channel for $d^{\uparrow\uparrow\uparrow}_{\downarrow\downarrow\downarrow}$ and the $p\bar{h}{,}p\bar{h}\bar{p}h$ channel for $d^{\uparrow\uparrow\downarrow}_{\downarrow\downarrow\uparrow}$. We have tried different combinations of STTN tensor dimensions for that fit and obtained the best fit using the the dimensions listed in table \ref{tab:tensor_dim_T_SC}, which yield 1802 free parameters. In Fig. \ref{fig:12_sites_HF_T_ampl_single_channel}, we observe a similar quality of fit as with multiple channels for the parallel-spin double- and triple excitation tensors (Figs. \ref{fig:12_sites_HF_T_ampl_single_channel}b and \ref{fig:12_sites_HF_T_ampl_single_channel}e), but a much lower quality of fit  for the $d^{\uparrow\downarrow}_{\downarrow\uparrow}$ (Figs. \ref{fig:12_sites_HF_T_ampl_single_channel}c and \ref{fig:12_sites_HF_T_ampl_single_channel}d) and $d^{\uparrow\uparrow\downarrow}_{\downarrow\downarrow\uparrow}$ (Figs. \ref{fig:12_sites_HF_T_ampl_single_channel}f and \ref{fig:12_sites_HF_T_ampl_single_channel}g) tensors. For the Hubbard model, those are the most strongly correlated tensors because of the strong correlation between excited particles(holes) with opposite spins. We also obtain a $\chi^2$ about 10 times larger and an error on the energy 20 times larger ($8\%$) than with multiple channels. Note that we have also tried the fit using the $\bar{p}h{,}p\bar{h}p\bar{h}$ channel instead of the $p\bar{h}{,}p\bar{h}\bar{p}h$ channel for the $d^{\uparrow\uparrow\downarrow}_{\downarrow\downarrow\uparrow}$ tensor, which produced a much worse fit quality. The same test performed on the 10 sites and 14 sites clusters yield similar results, namely that the $d^{\uparrow\downarrow}_{\downarrow\uparrow}$ and $d^{\uparrow\uparrow\downarrow}_{\downarrow\downarrow\uparrow}$ are not well fitted using a single channel, and the $\chi^2$ and error on the energy are much larger than with multiple channels for both systems (not shown).

For the STTN fit to the $T$-amplitudes for the 14 sites cluster, we have not included the $pp\bar{h}\bar{h}$, $p\bar{h}{,}pp\bar{h}\bar{h}$ and $\bar{p}h{,}pp\bar{h}\bar{h}$ channels in the decompositions. Therefore, the decompositions for the $d^{\uparrow\uparrow}_{\downarrow\downarrow}$ and $d^{\uparrow\uparrow\uparrow}_{\downarrow\downarrow\downarrow}$ tensors contain only one channel, and the decomposition for $d^{\uparrow\uparrow\downarrow}_{\downarrow\downarrow\uparrow}$ contain four channels instead of five. For that fit we have used 3365 free parameters to fit all 49049 amplitudes up to three excitations. On Fig. \ref{fig:14_sites_HF_T_ampl} showing the results of that fit, we can barely see the difference between the STTN fit and the exact amplitudes, which is consistent with the very small error of $0.1\%$ on the energy obtained when using the STTN results in the energy equation. The STTN tensor dimensions for that fit are listed on the third column of table \ref{tab:tensor_dim_T}. The quality of the STTN fit to the wave function coefficients, shown in Fig. \ref{fig:14_sites_HF_coef} of appendix \ref{sec:results_STTN_wv_cfs}, is also quite good, though not as good as for the $T$-amplitudes, with an error on the energy of $0.2\%$. Note that, for the 10 and 12 sites clusters, we have also tried using the same channels, discarding the $pp\bar{h}\bar{h}$, $p\bar{h}{,}pp\bar{h}\bar{h}$ and $\bar{p}h{,}pp\bar{h}\bar{h}$ channels, but the quality of the fits that included all the channels was better.

\begin{figure}
\includegraphics[width=\columnwidth]{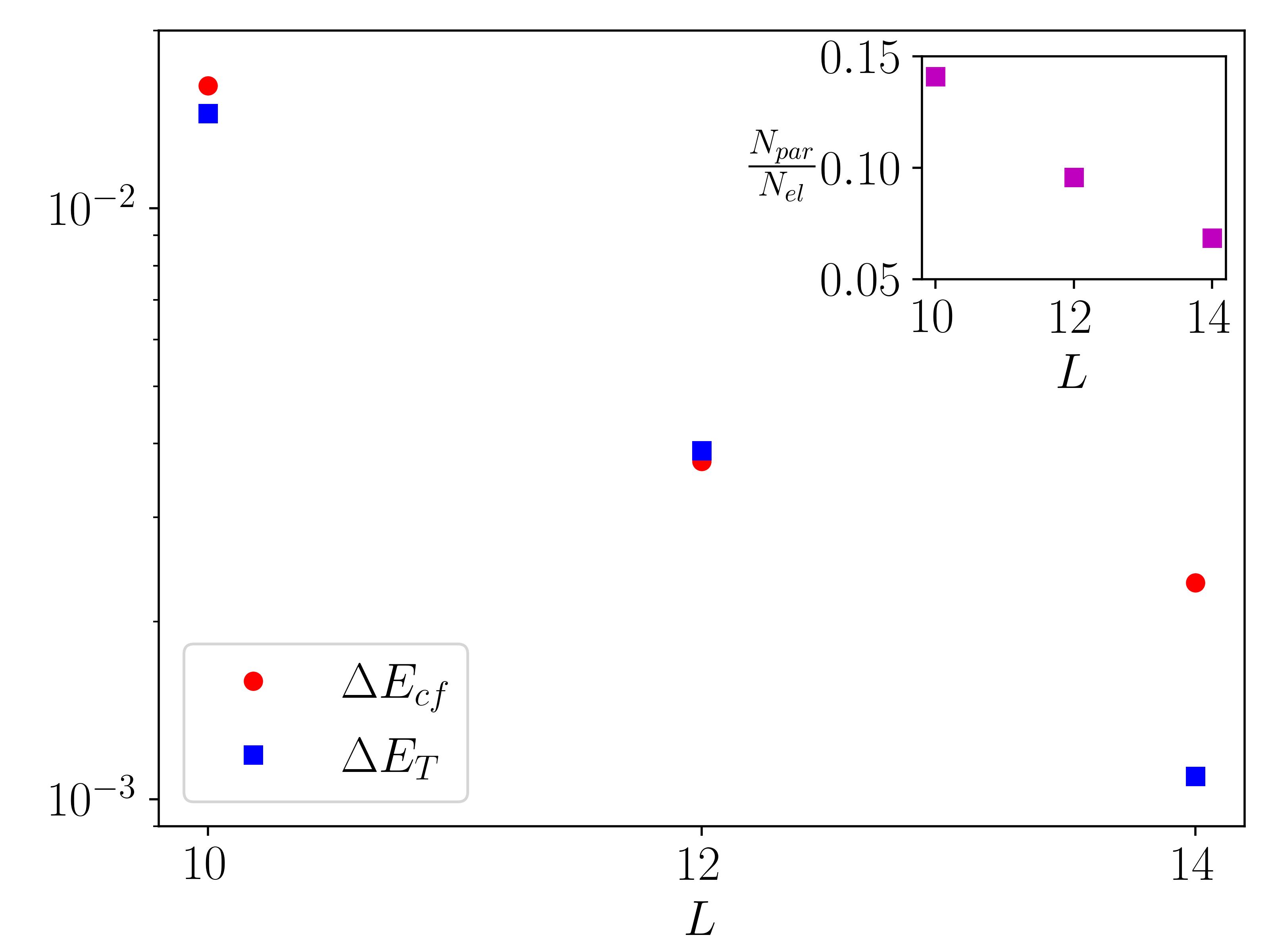}
\caption{\label{fig:DE_vs_L} Relative errors on the energy when the STTN are used to fit the wave function coefficients ($\Delta E_{cf}$) and the $T$-amplitudes ($\Delta E_{T}$), using multiple channels, as functions of system size $L$. Inset: ratio of the number of STTN free parameters to the number of tensor elements to fit as a function of system size.}
\end{figure}

Finally, on Fig. \ref{fig:DE_vs_L}, which shows the error on the energy as a function of system size for the two types of fits ($T$-amplitudes and wave function coefficients), we can see that the accuracy of the STTN fit improves as the system size increases, with a sharper improvement for the $T$-amplitudes. This is a quite remarkable result, especially given that, at the same time, the ratio of the number of STTN free parameters to the number of tensor elements to fit decreases as the size of the system increases (see inset of Fig. \ref{fig:DE_vs_L}).

\section{Discussion}\label{sec:discussion}

We can summarize ours results by three key findings: 1) STTN have the capacity to accurately express single-reference expansions of strongly correlated ground states using numbers of free parameters much smaller than the number of tensor elements represented, 2) the inclusion of multiple channels in the decomposition of a strongly correlated tensor is much more accurate than using a single channel, and 3) the accuracy of the STTN fit improves, despite the reduction in the ratio of the number of STTN free parameters to the number of tensor elements to fit, as the system size increases. If we define $N_{\leq k}$ as the number of free parameters required to represent the ground state wave function coefficients up to $k$ excitations, our results indicate that, for the systems considered, the true complexity of $N_{\leq k}$ is not (nearly) exponential in $k$, as would suggest the behavior of the number of Slater determinants as a function of $k$ (see Fig. \ref{fig:Npar_vs_Nx}). Although we do not have enough information to deduce the true scaling of $N_{\leq k}$, it appears to increase slowly enough so that the strong correlation extensions of SRCC based on tensor decompositions, as proposed in Ref. \onlinecite{DBergeron_TCC_2020}, are applicable in practice and that STTN are a good choice of representation to implement them.

In those extensions however, if $n$ is the number of excitations at which the series of CC equations is stopped (see Eq. \eqref{eq:CC_equations}), the tensor decompositions must be able to express all $T$-amplitudes (wave function coefficients) up to $n+2$ excitations, using a total number of free parameters, say $N^{TD}_{\leq n+2}$, smaller than the number of CC amplitude equations, i.e. the number of Slater determinants up to $n$ excitations, say $N^{SD}_{\leq n}$ (see section III of Ref \onlinecite{DBergeron_TCC_2020}). In this work, we have used $N^{TD}_{\leq n+1}<N^{SD}_{\leq n}$ instead of $N^{TD}_{\leq n+2}<N^{SD}_{\leq n}$. However, if the accuracy of the STTN improves as the size of the system increases and, as mentioned at the end of section \ref{sec:STTN}, the STTN decompositions for higher order tensors require only small additional numbers of free parameters, then, even for $n=2$, it becomes possible to reach the condition $N^{TD}_{\leq n+2}<N^{SD}_{\leq n}$ if the system is large enough. Using $n=3$ could also be a practical way to reach that condition, although computationally more costly for large systems.

We have also noticed that it seems slightly easier for STTN to represent $T$-amplitudes than wave function coefficients. If we carefully compare the magnitudes of $T$-amplitudes with the corresponding coefficients, we observe that the amplitudes decrease more slowly than the coefficients as the number of excitation increases. The difference of accuracy between the fits to amplitudes and coefficients could therefore be related to the fact that the different $T$-tensors to simultaneously fit are more similar in magnitude than the coefficients tensors. This is consistent with the fact the STTN fits are better when using spin-orbitals producing a reference energy slightly larger than the true HF ground state energy since, for the true HF ground state spin-orbitals, the difference of magnitude between the different $T$-tensors, for instance, is larger.

\section{Conclusion}\label{sec:conclusion}

We have shown that it is possible to accurately represent single-reference expansions of ground states of small two-dimensional Hubbard clusters in the strong correlation regime, up to three particle-hole excitations, using superpositions of tree-tensor networks (STTN), with a number of free parameters much smaller than the number of tensor elements that are represented. Our results also show that summing different channels is crucial to properly decompose the most strongly correlated tensors, and also suggest that the accuracy of the STTN improves as the system size increases. Those findings suggest that the strong correlation extension of CC proposed in Ref \onlinecite{DBergeron_TCC_2020}, and based on ``entanglement truncation'' rather than the standard truncation scheme of CC, are viable and could produce accurate results if implemented using STTN.

\section{Acknowledgements}

Thanks to Calcul Qu\'ebec and the \href{https://alliancecan.ca/en}{Digital Research Alliance of Canada} for the computing facilities. Special thanks to Andr\'e-Marie Tremblay for moral support.

\appendix

\section{Coupled cluster equations}\label{sec:CC_eqns}

If \eqref{eq:CC_ansatz} is an eigenstate of the Hamiltonian $\hat{H}$, we have $\hat{H}\ket{\psi}=E\ket{\psi}=e^{T}E\ket{\phi}$, where $E$ is the ground state energy. Therefore, assuming $\ket{\phi}$ is normalized, since any state $\ket{\phi_{j_1,j_2\ldots j_l}^{i_1i_2\ldots i_l}}$ is orthogonal to $\ket{\phi}$ for $l\neq 0$, if we project $\hat{H}\ket{\psi}$ on $\bra{\phi}e^{-T}$ or $\bra{\phi_{j_1,j_2\ldots j_l}^{i_1i_2\ldots i_l}}e^{-T}$, we obtain the equations
\begin{equation}\label{eq:CC_equations}
\begin{split}
\bra{\phi}e^{-T}\hat{H}e^{T}\ket{\phi}&=E\,,\\
\bra{\phi_{j_1,j_2\ldots j_l}^{i_1i_2\ldots i_l}}e^{-T}\hat{H}e^{T}\ket{\phi}&=0\,,\quad 1\leq l\leq n\,,
\end{split}
\end{equation}
which, after substitution of $T$ by Eq. \eqref{eq:def_T_CC} and $\hat{H}$ by its second quantization expression, yields a set of nonlinear equations defining the energy $E$ and the $T$-amplitudes $d_{j_1,j_2\ldots j_l}^{i_1i_2\ldots i_l}$ for $1\leq l\leq n$.

\section{The CC energy equation}\label{sec:DE_CC}

If we define the antisymmetrized Coulomb interaction as $V_{ijkl}=V_{ijkl}^c-V_{ijlk}^c$, where $V^c$ is the coulomb interaction, the general many-body Hamiltonian has the form
\begin{equation}\label{eq:H_K_V}
\hat{H}=\sum_{ij} t_{ij} a_i^\dagger a_j + \frac{1}{4}\sum_{ijkl} V_{ijkl} a_i^\dagger a_j^\dagger a_k a_l\,.
\end{equation}
Now, if we define the Fock matrix as
\begin{equation}
t^{\phi}_{ij}=t_{ij}+\sum_k V_{ikkj}n_k^{\phi}\,,
\end{equation}
where $n_k^{\phi}=\bra{\phi}a^\dagger_k a_k\ket{\phi}$, $\ket{\phi}$ being the reference Slater determinant,
using Eq. \eqref{eq:CC_equations}, with the definitions \eqref{eq:CC_ansatz} and \eqref{eq:def_T_CC}, the CC correlation energy equation can be derived as
\begin{equation}\label{eq:DE_CC_T}
\begin{split}
\Delta E&=E-\bra{\phi}\hat{H}\ket{\phi}\\
&=\sum_{ ij} t^{\phi}_{ij} d_{j}^{i}-\frac{1}{4}\sum_{i_1i_2, j_1j_2} d^{i_1i_2}_{j_1j_2} V_{j_2j_1i_2i_1}\\
&\qquad-\frac{1}{2}\sum_{i_1i_2, j_1j_2} d^{i_1}_{j_1}d^{i_2}_{j_2} V_{j_2j_1i_2i_1}\,,
\end{split}
\end{equation}
or, using \eqref{eq:wv_cfs_vs_T_ampl},
\begin{equation}\label{eq:DE_CC_cfs}
\Delta E=\sum_{ij} t^{\phi}_{ij}c^i_j-\frac{1}{4}\sum_{i_1i_2j_1j_2} c^{i_1i_2}_{j_1j_2}V_{j_2j_1i_2i_1}\,.
\end{equation}

\section{STTN fits to wave function coefficients}\label{sec:results_STTN_wv_cfs}

\begin{figure}[h]
\includegraphics[width=0.51\columnwidth]{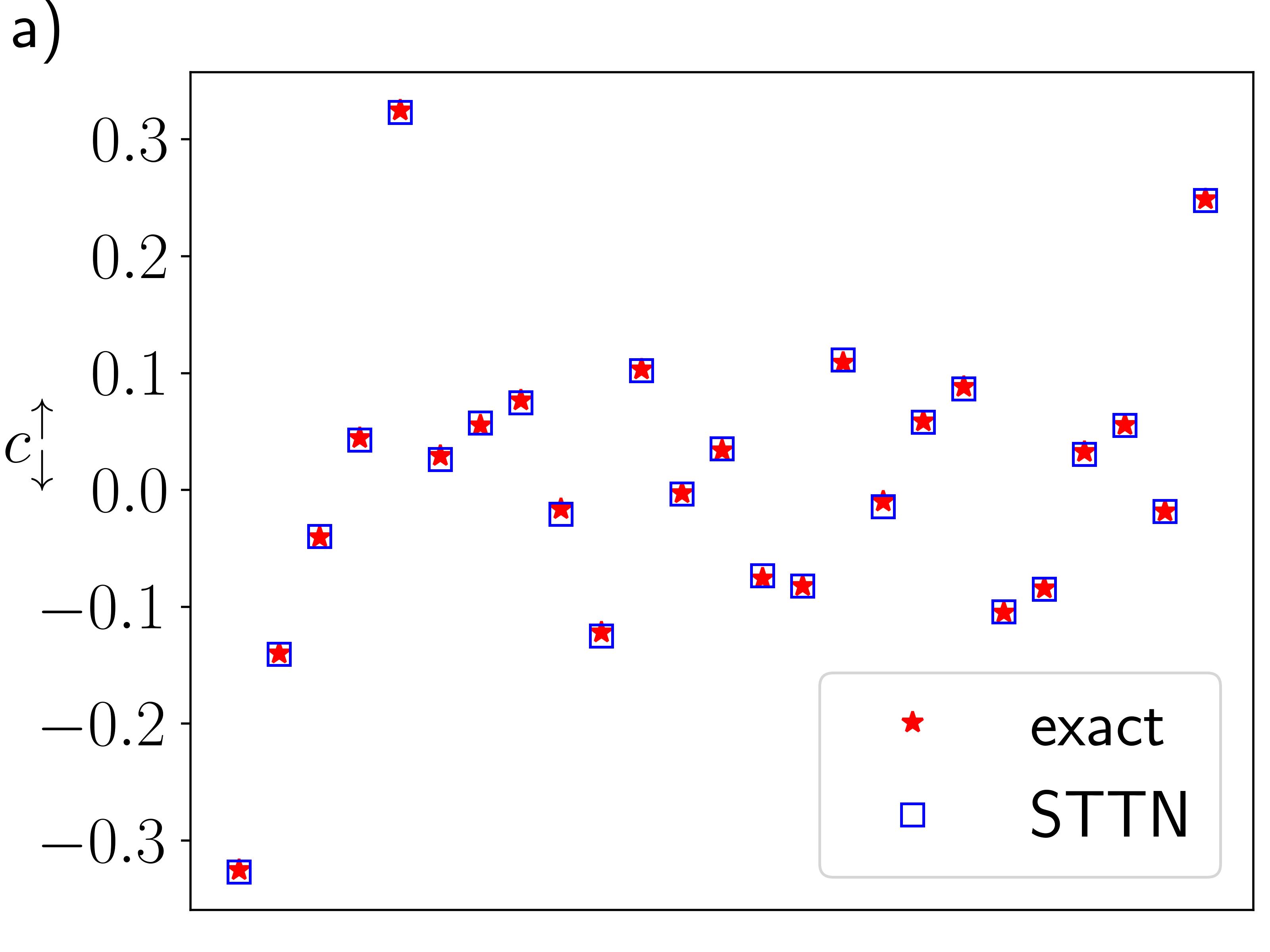}\includegraphics[width=0.51\columnwidth]{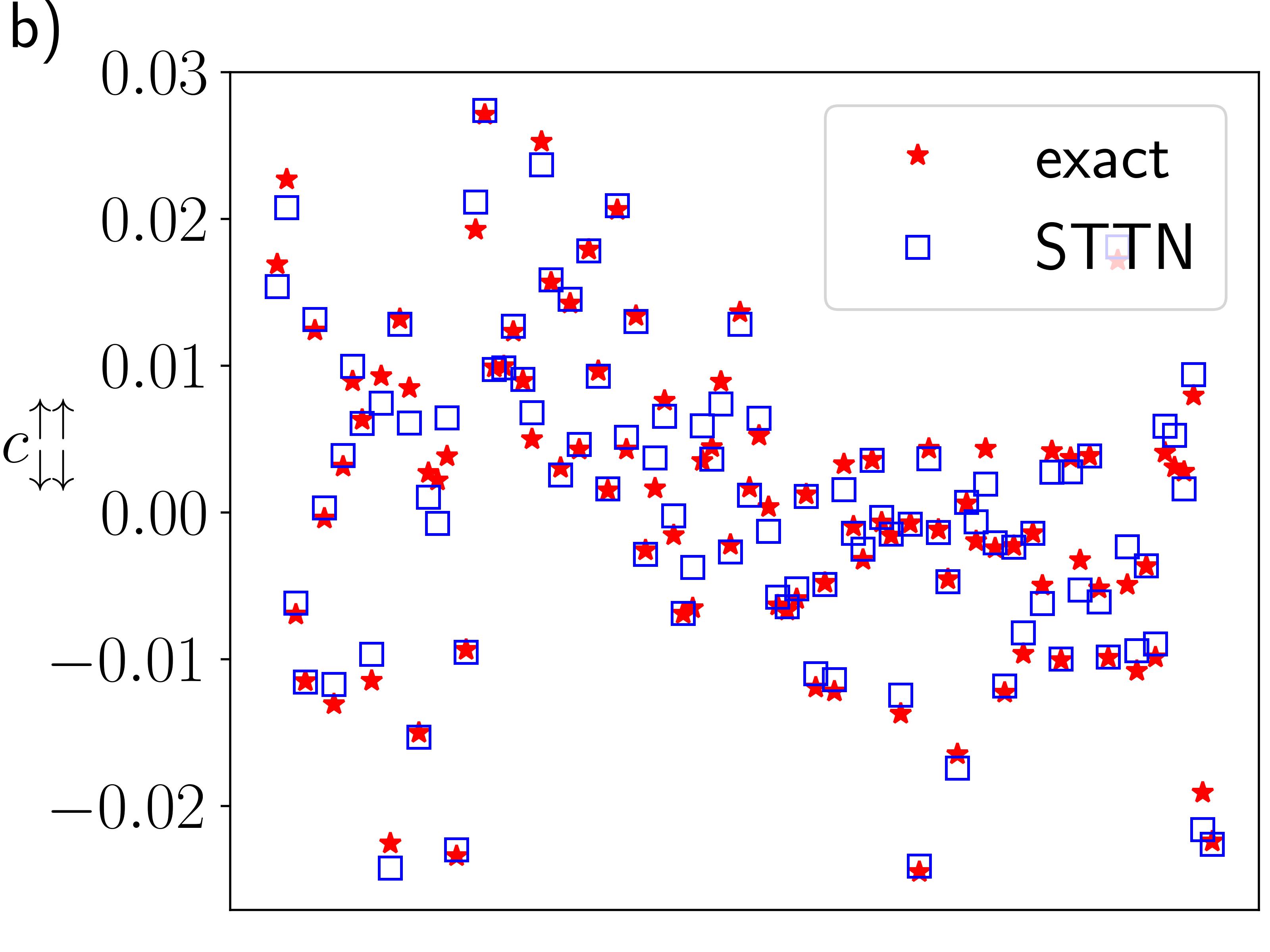}\\
\includegraphics[width=0.51\columnwidth]{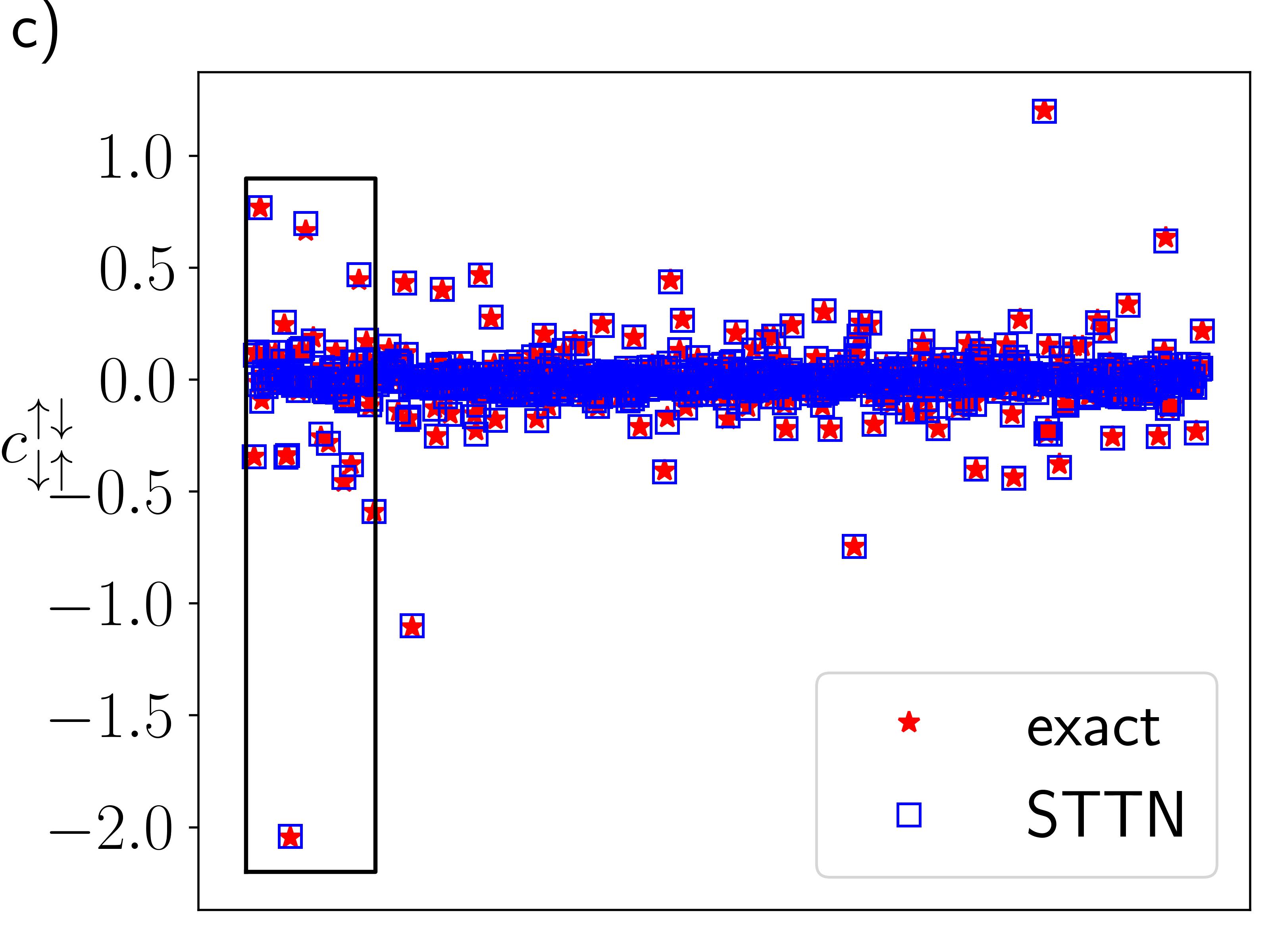}\includegraphics[width=0.51\columnwidth]{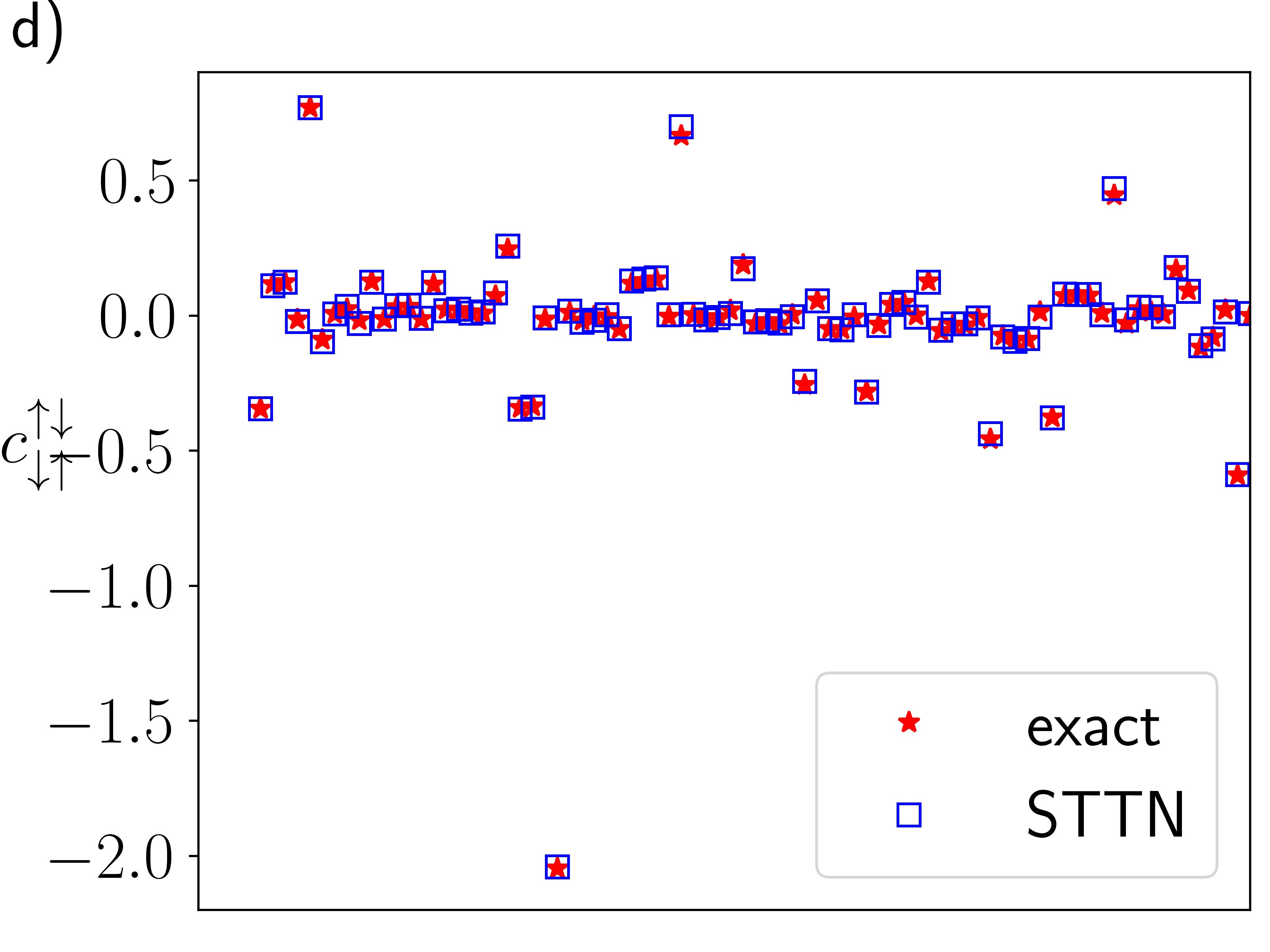}\\
\includegraphics[width=0.51\columnwidth]{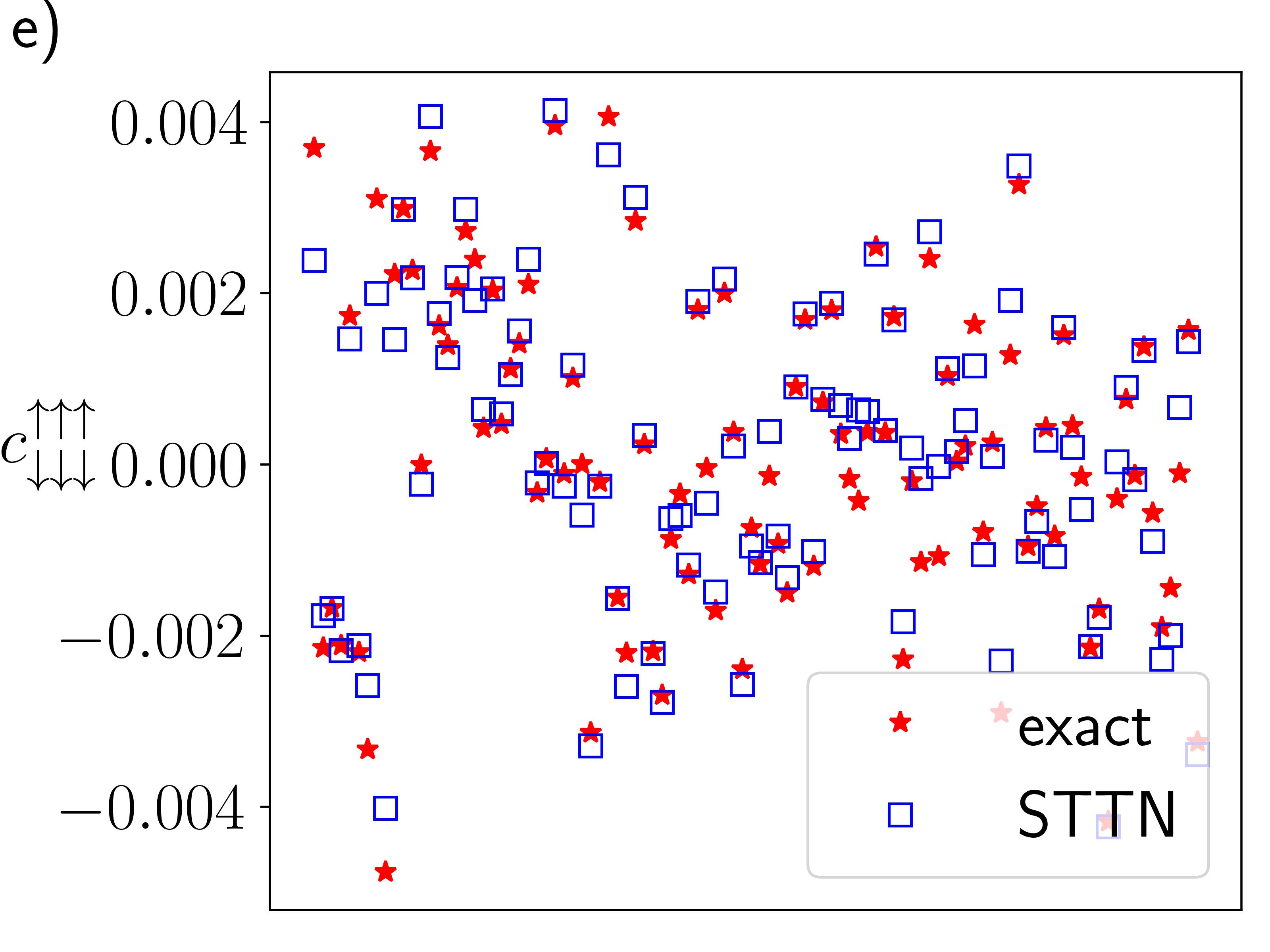}\includegraphics[width=0.51\columnwidth]{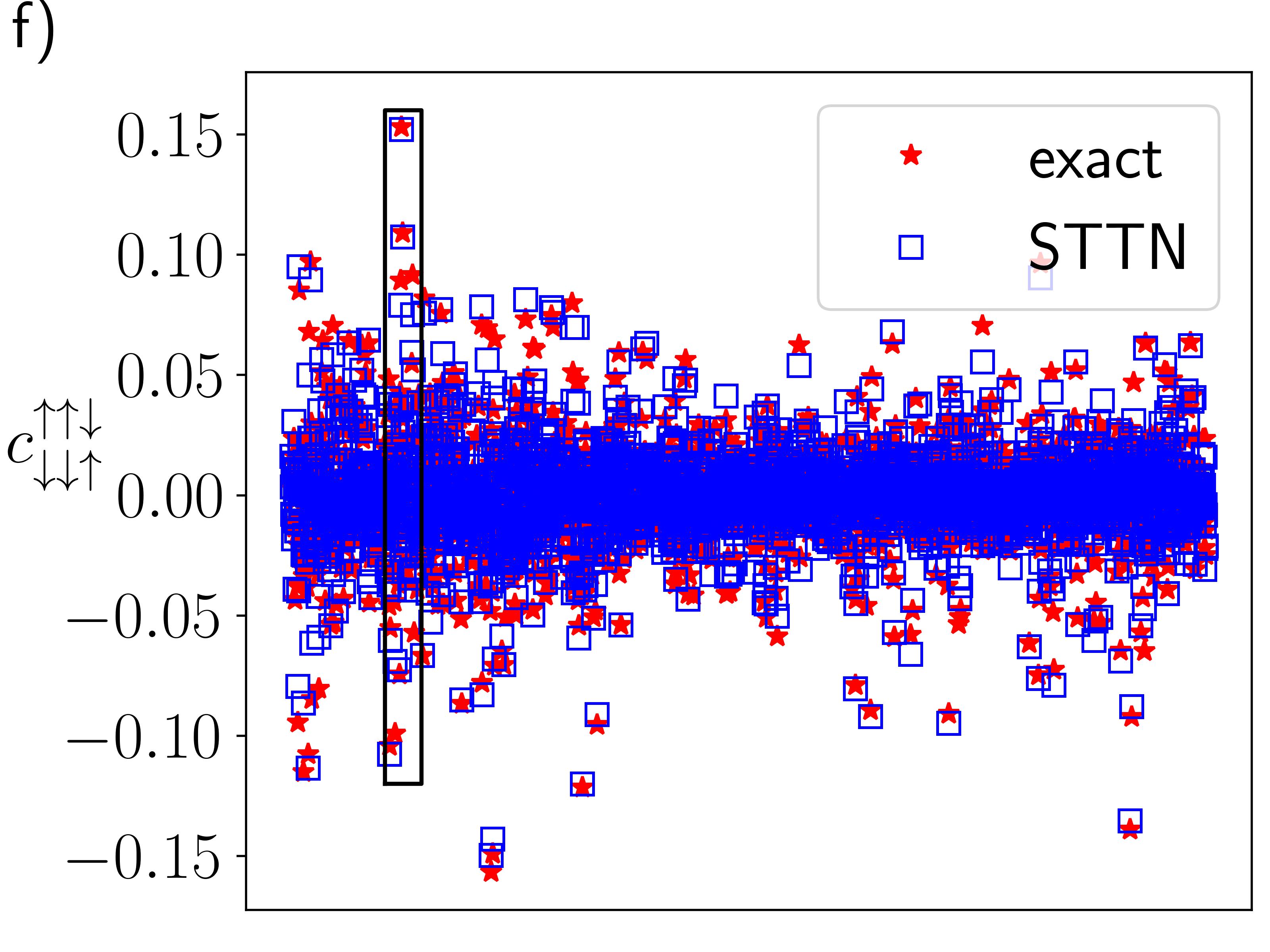}\\
\includegraphics[width=0.51\columnwidth]{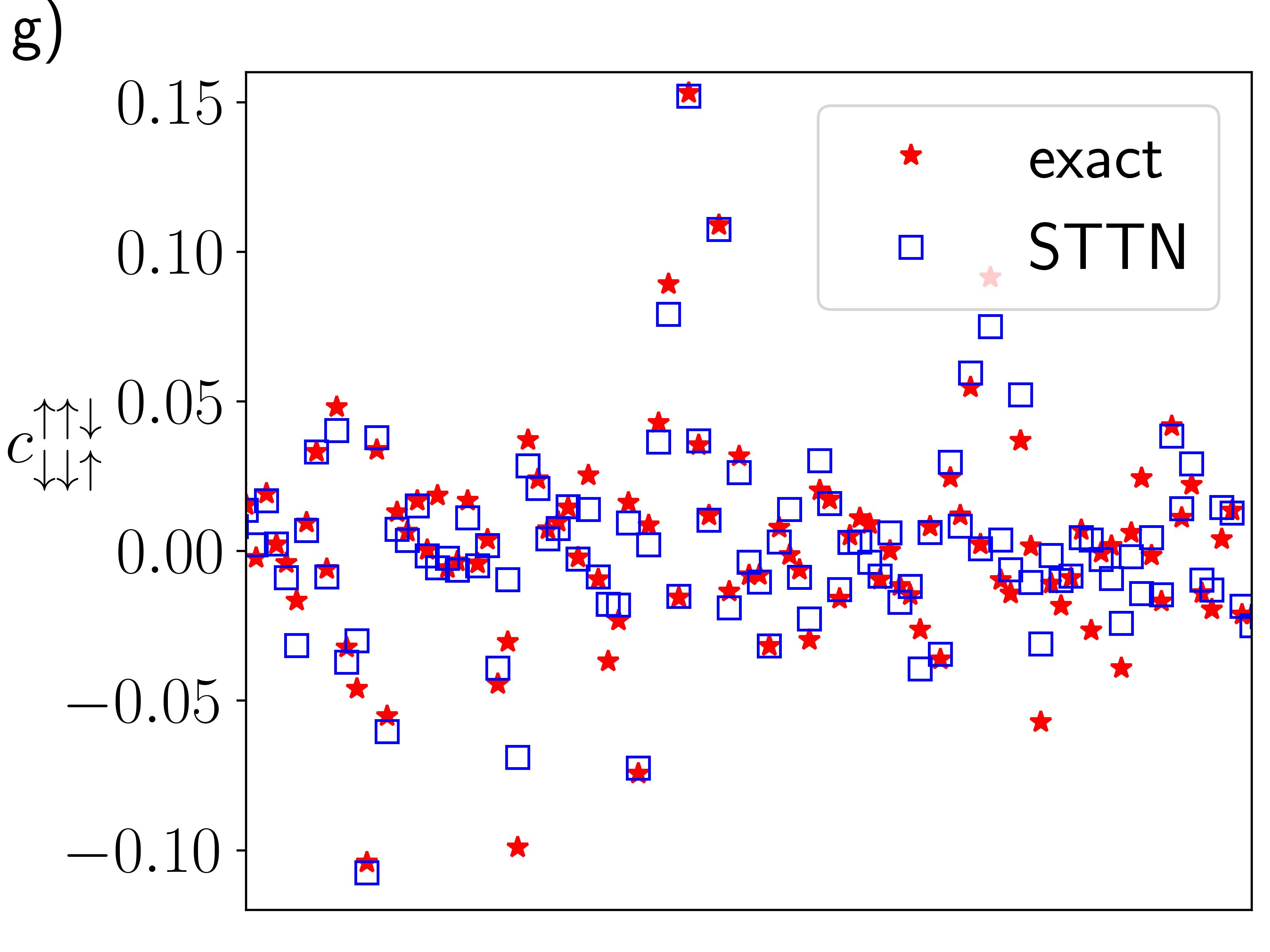}
\caption{\label{fig:10_sites_HF_coef} For the 10 sites cluster, exact wave function coefficients and STTN fit for a) the single-excitation tensor, b) the up-up double-excitation tensor, c) the up-down double-excitation tensor, d) the region delimited by the black rectangle in c), e) the up-up-up triple-excitation tensor, f) the up-up-down triple-excitation tensor and g) the region delimited by the black rectangle in f). The horizontal axes correspond to arbitrary tensor indices. The tensor dimensions used in the STTN are listed in table \ref{tab:tensor_dim_T}.}
\end{figure}

\begin{figure}[h]
\includegraphics[width=0.51\columnwidth]{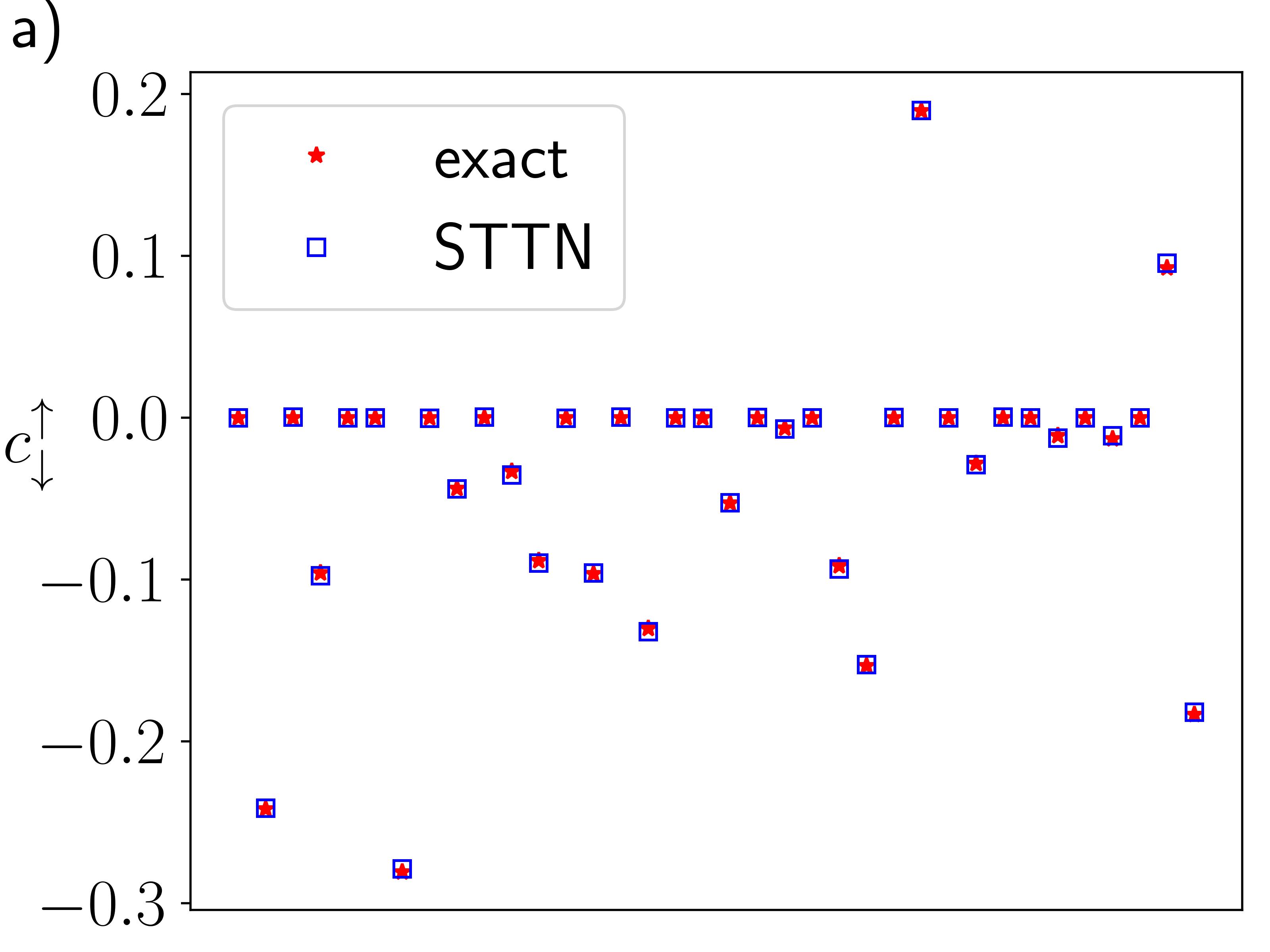}\includegraphics[width=0.51\columnwidth]{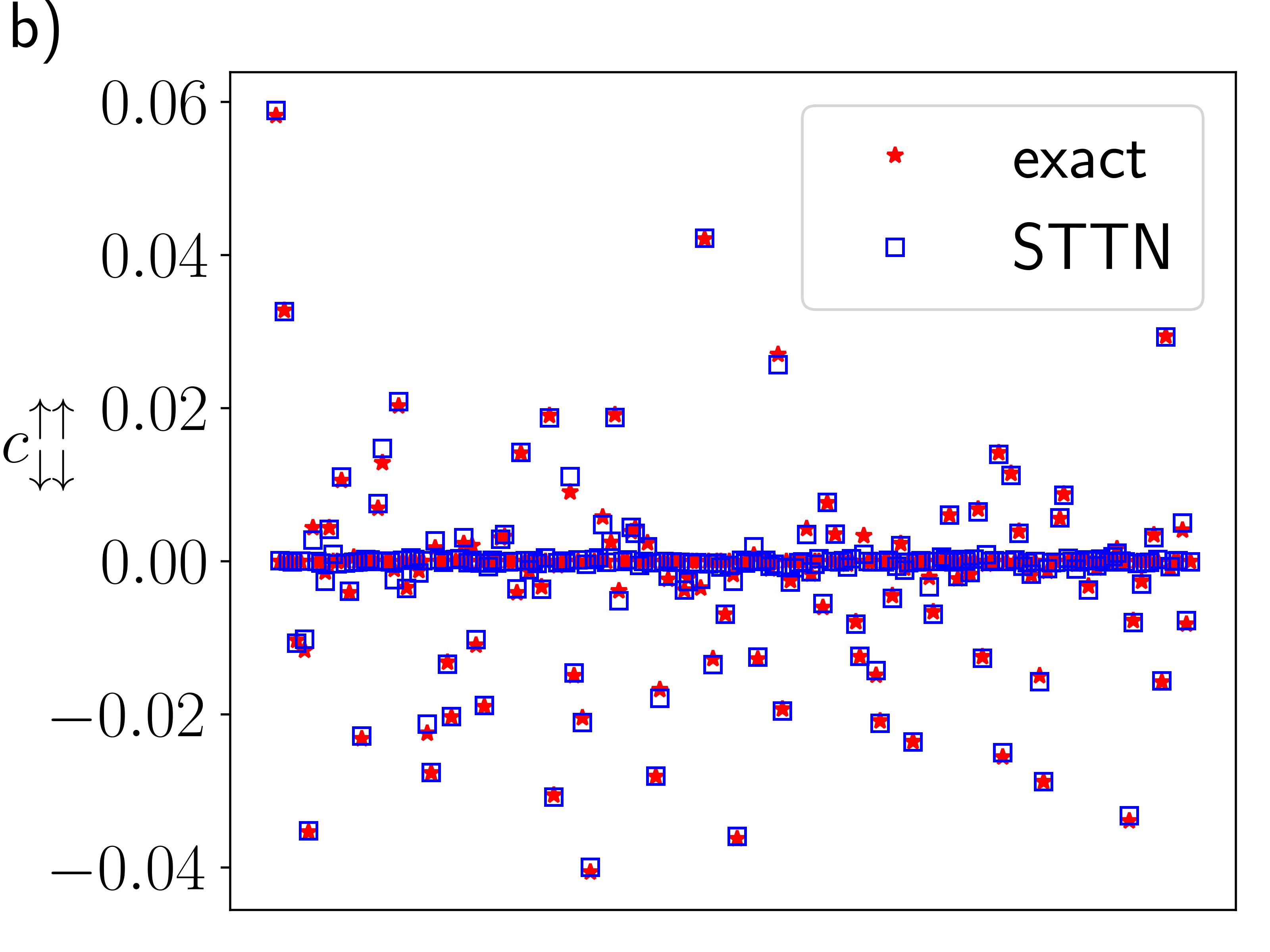}\\
\includegraphics[width=0.51\columnwidth]{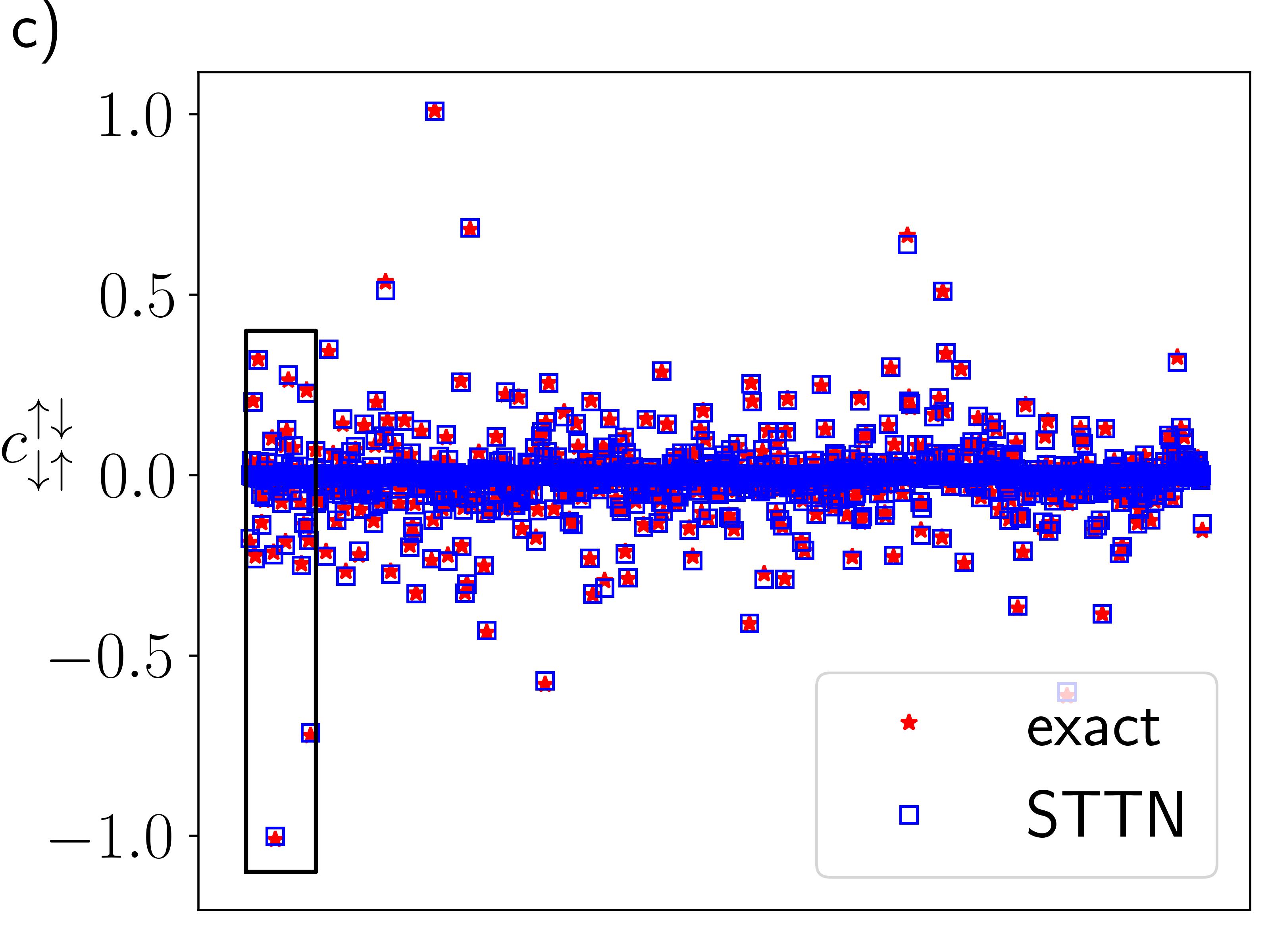}\includegraphics[width=0.51\columnwidth]{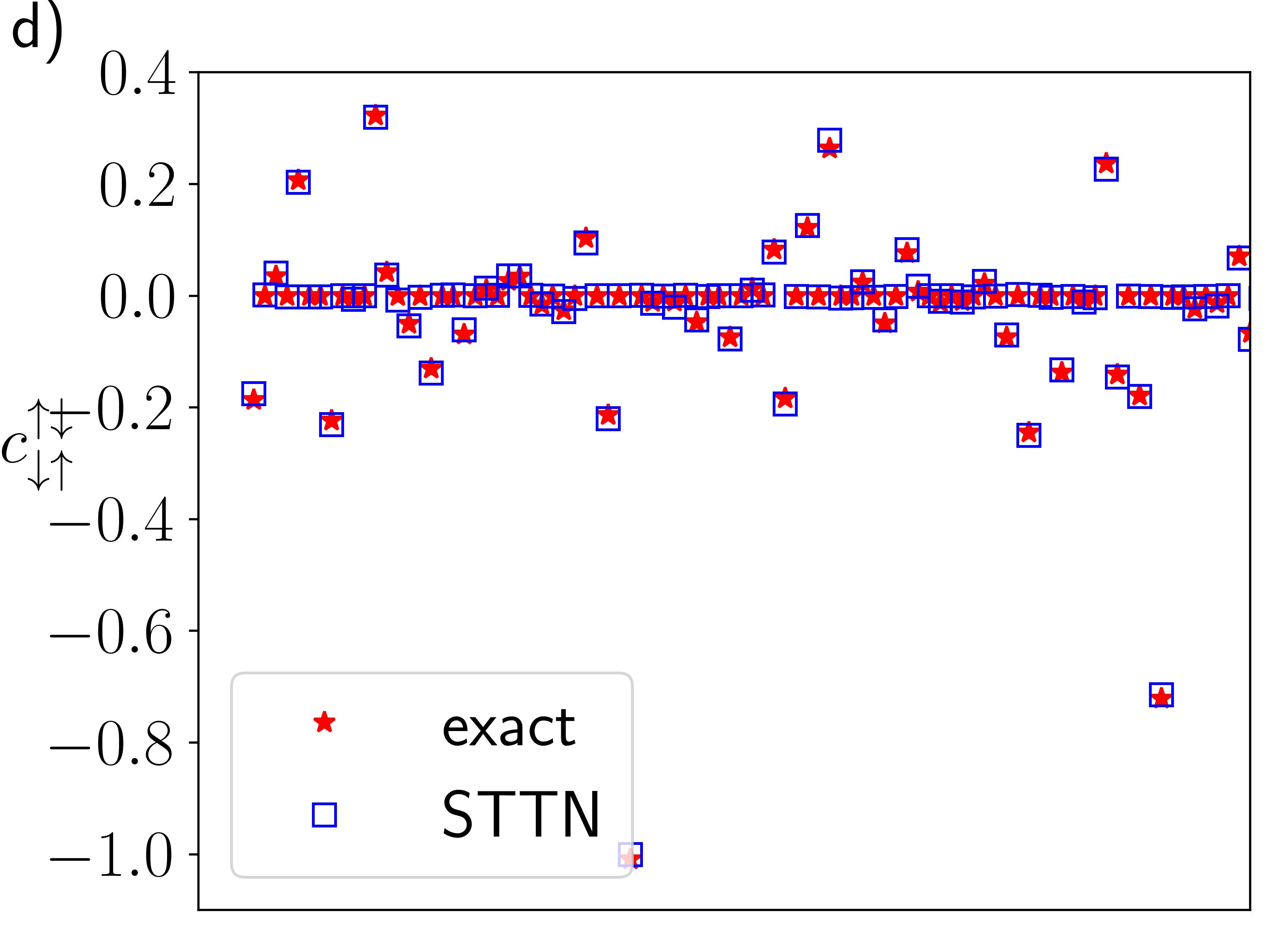}\\
\includegraphics[width=0.51\columnwidth]{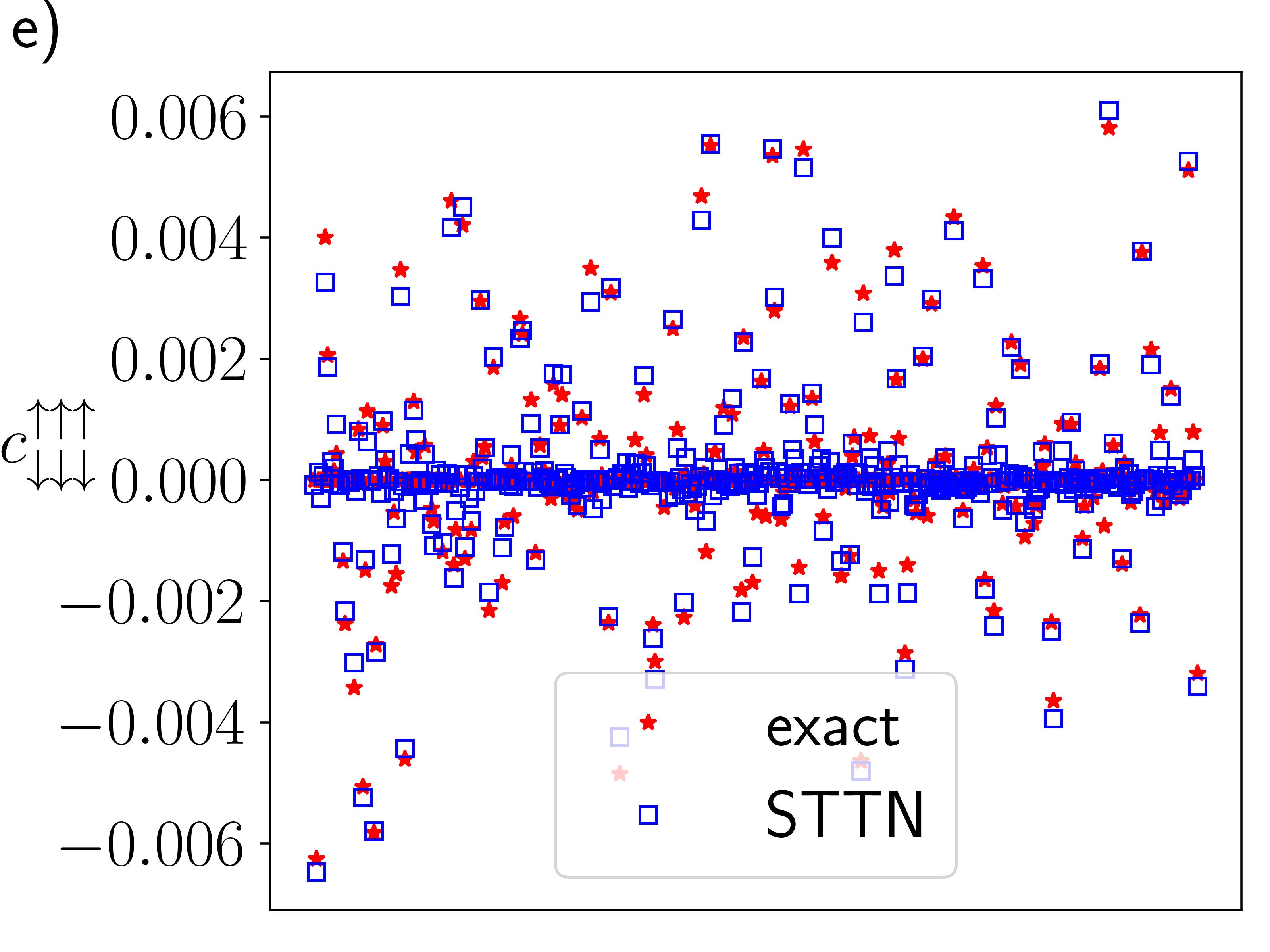}\includegraphics[width=0.51\columnwidth]{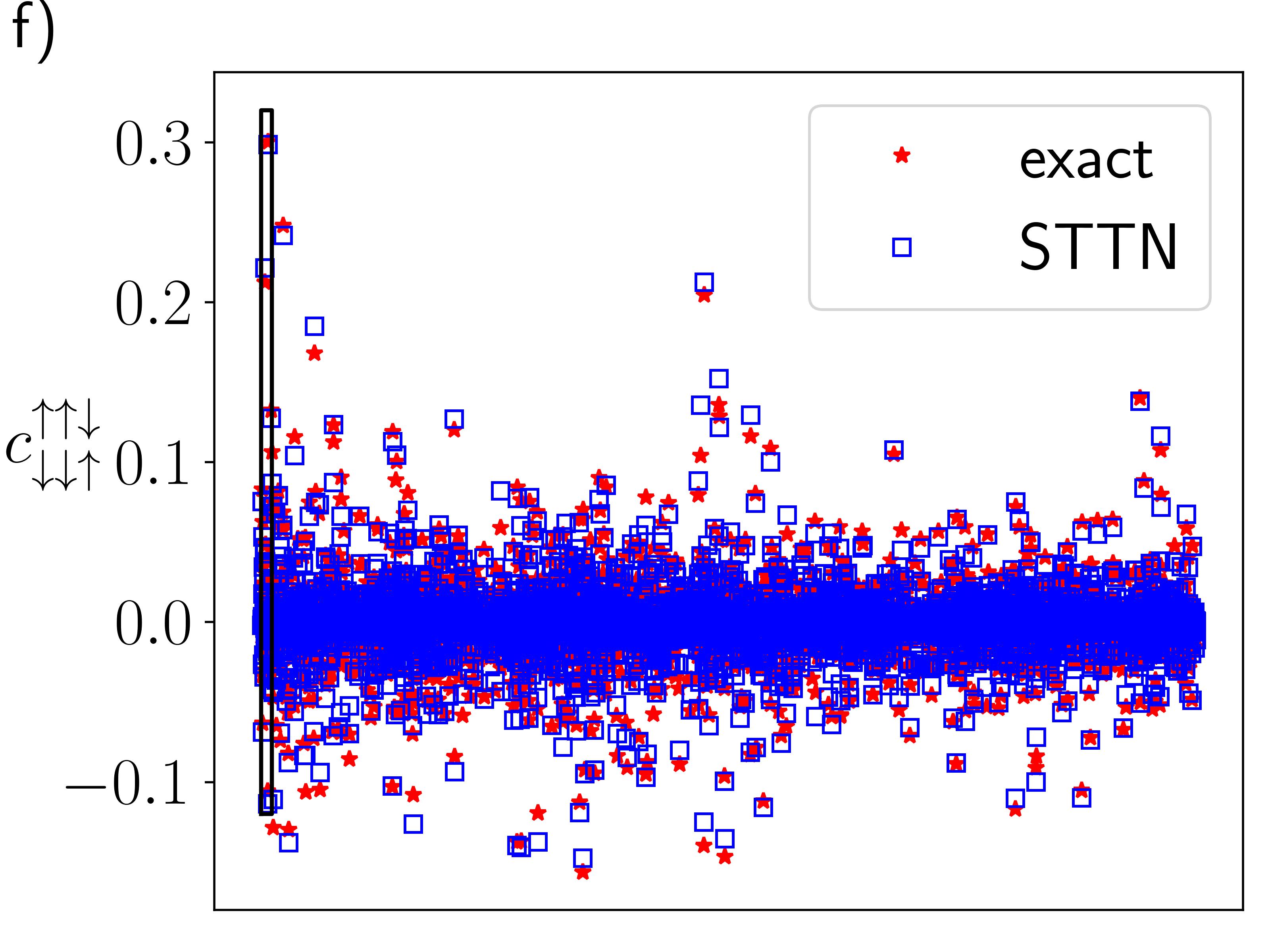}\\
\includegraphics[width=0.51\columnwidth]{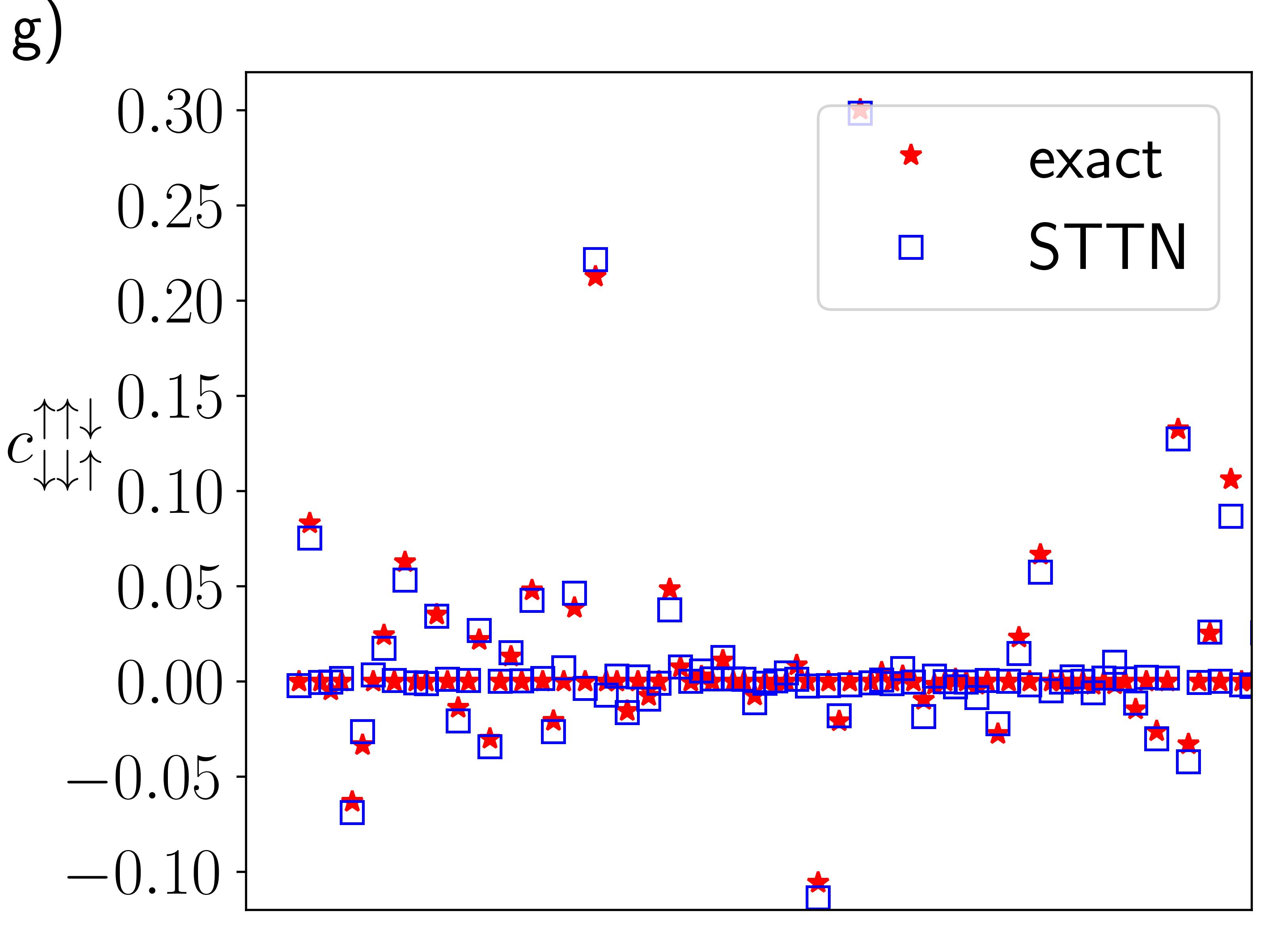}
\caption{\label{fig:12_sites_HF_coef} For the 12 sites cluster, exact wave function coefficients and STTN fit for a) the single-excitation tensor, b) the up-up double-excitation tensor, c) the up-down double-excitation tensor, d) the region delimited by the black rectangle in c), e) the up-up-up triple-excitation tensor, f) the up-up-down triple-excitation tensor and g) the region delimited by the black rectangle in f). The horizontal axes correspond to arbitrary tensor indices. The tensor dimensions used in the STTN are listed in table \ref{tab:tensor_dim_T}.}
\end{figure}

\begin{figure}[h]
\includegraphics[width=0.51\columnwidth]{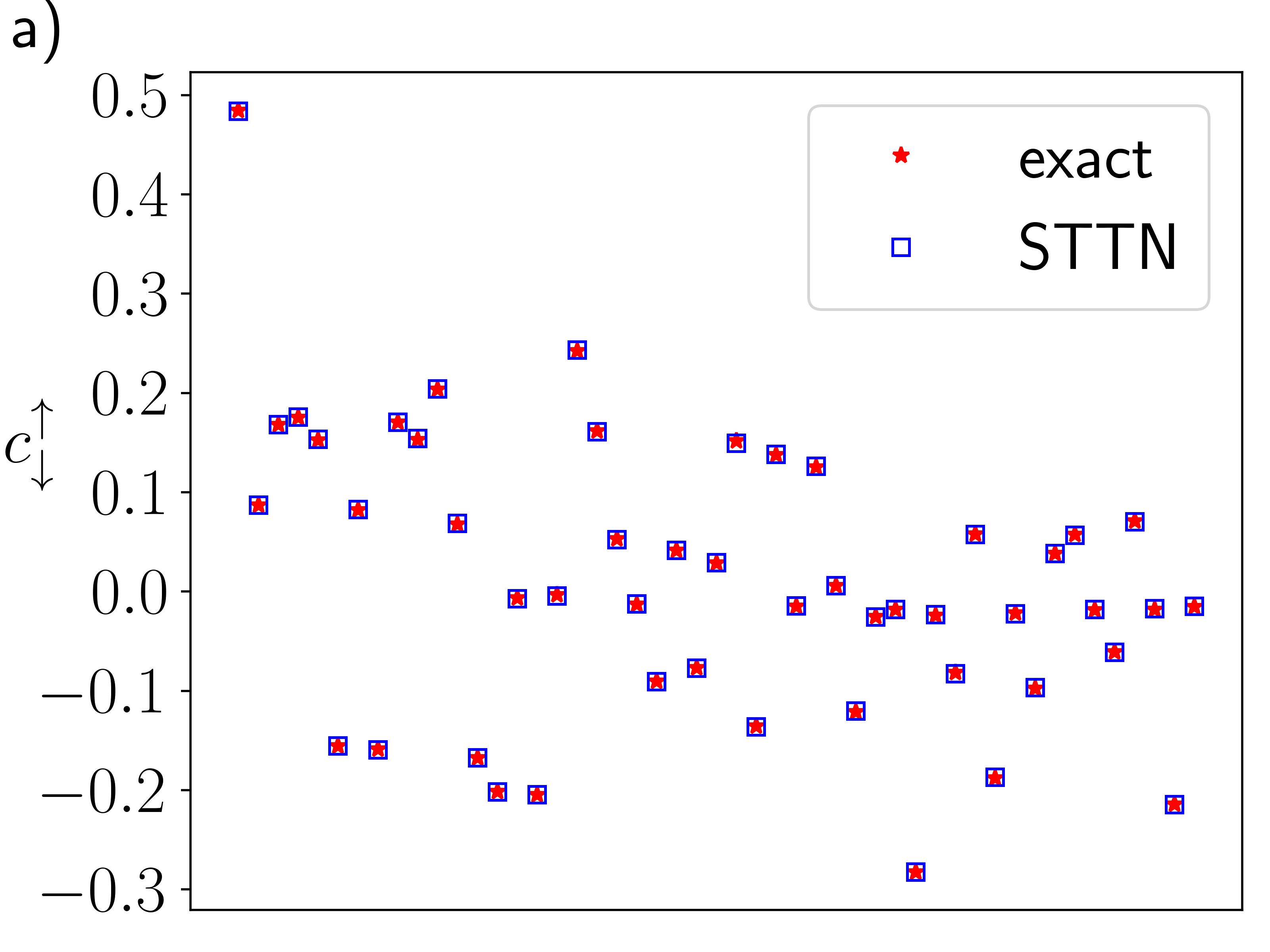}\includegraphics[width=0.51\columnwidth]{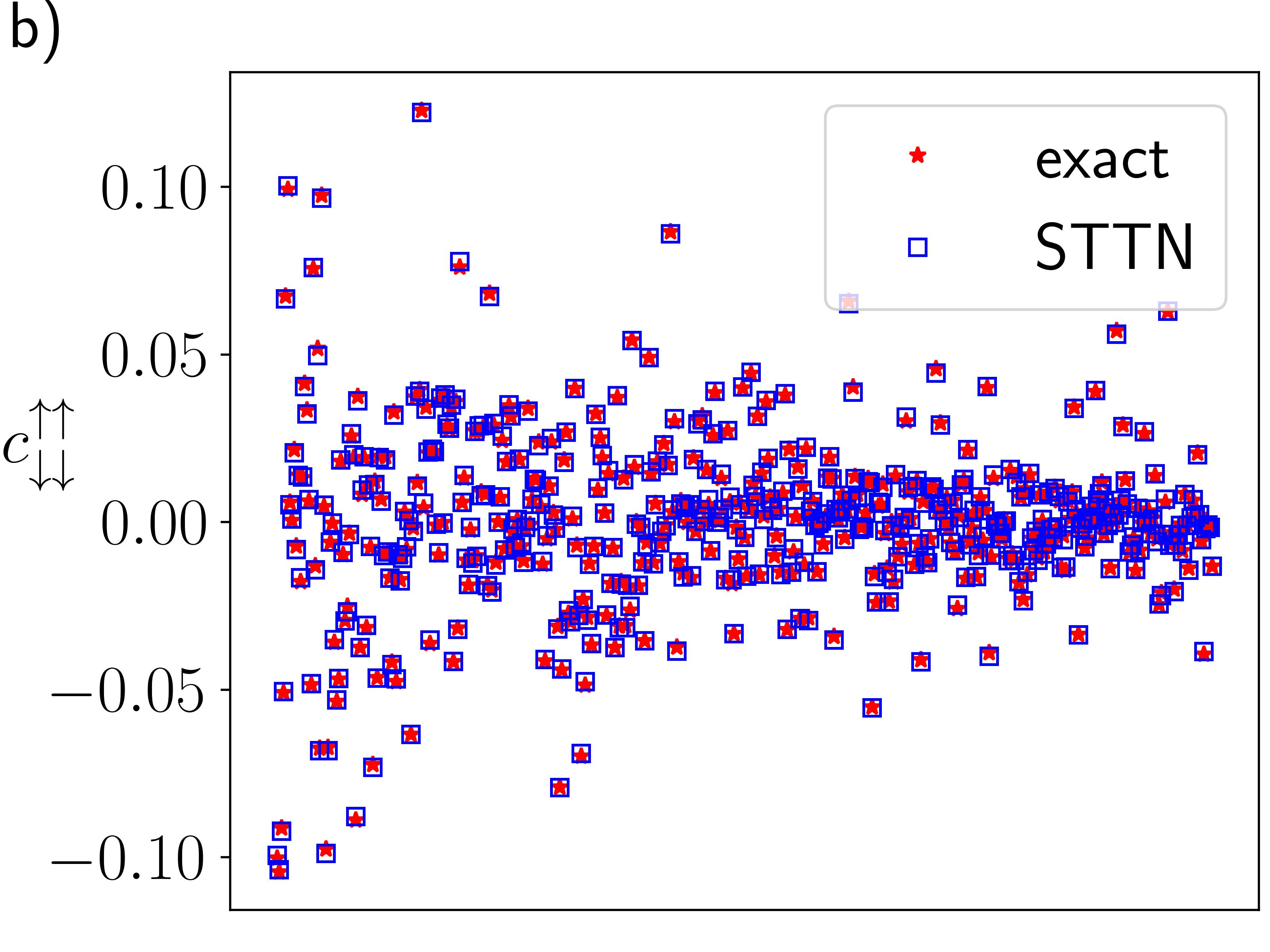}\\
\includegraphics[width=0.51\columnwidth]{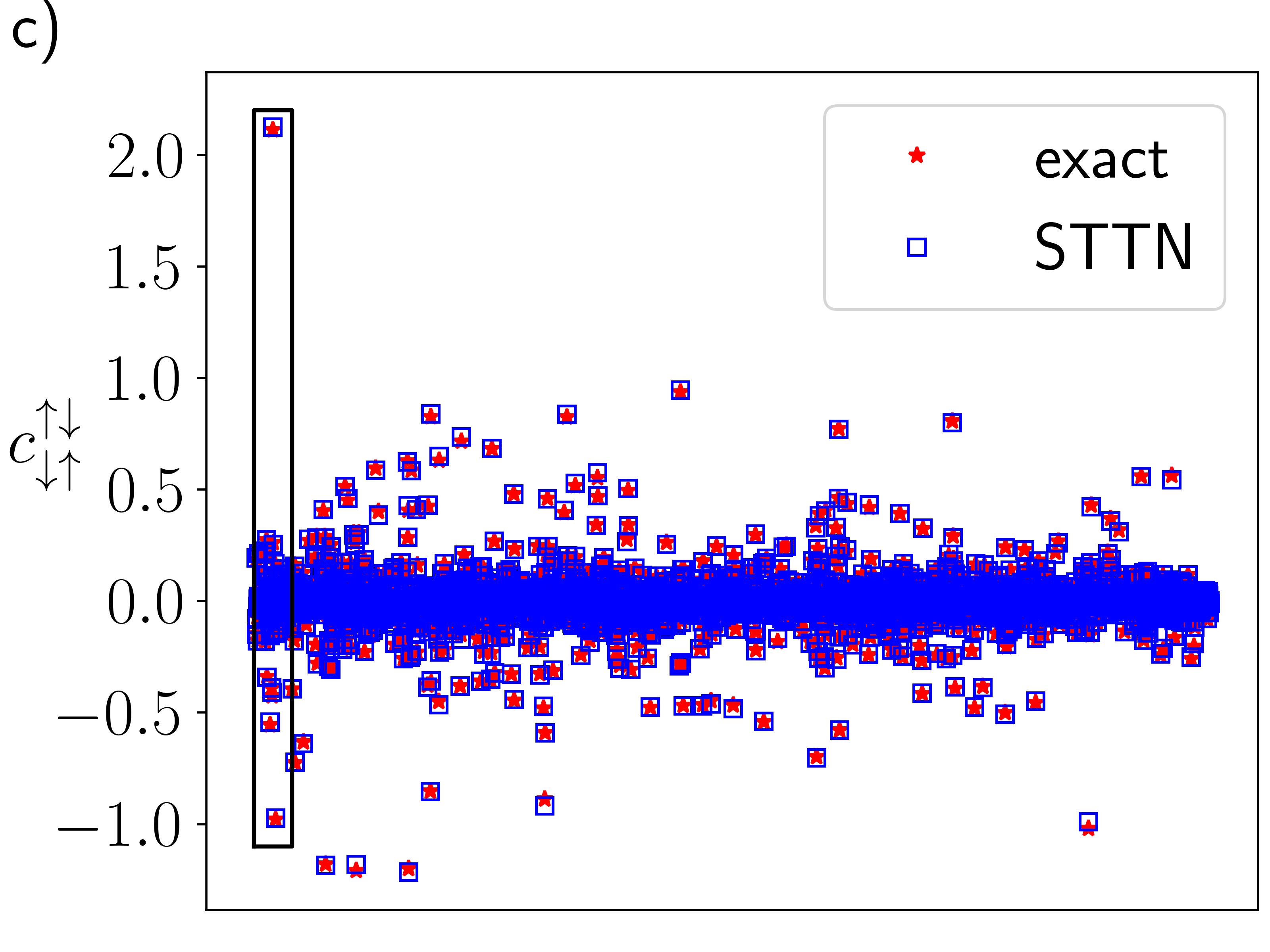}\includegraphics[width=0.51\columnwidth]{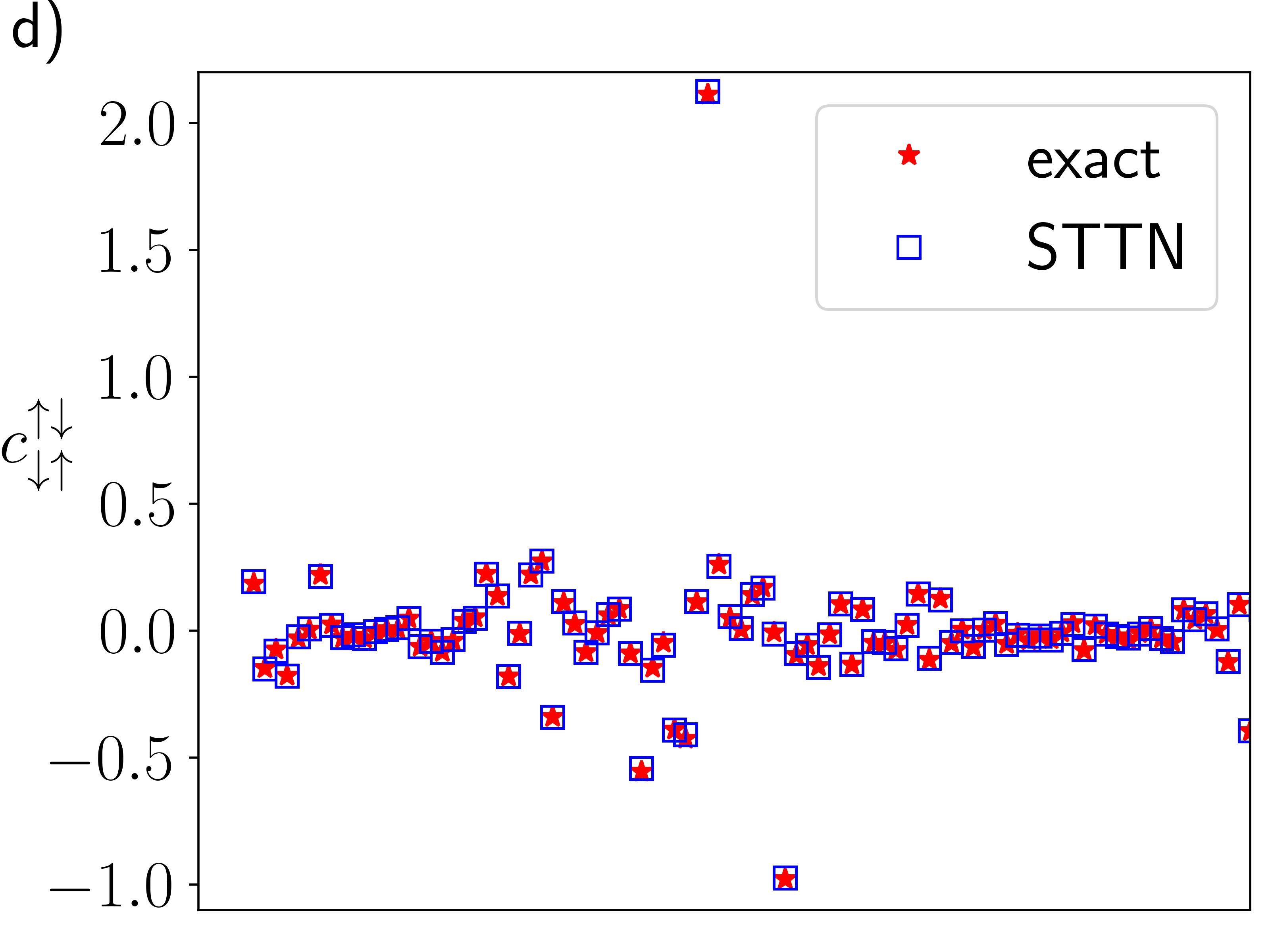}\\
\includegraphics[width=0.51\columnwidth]{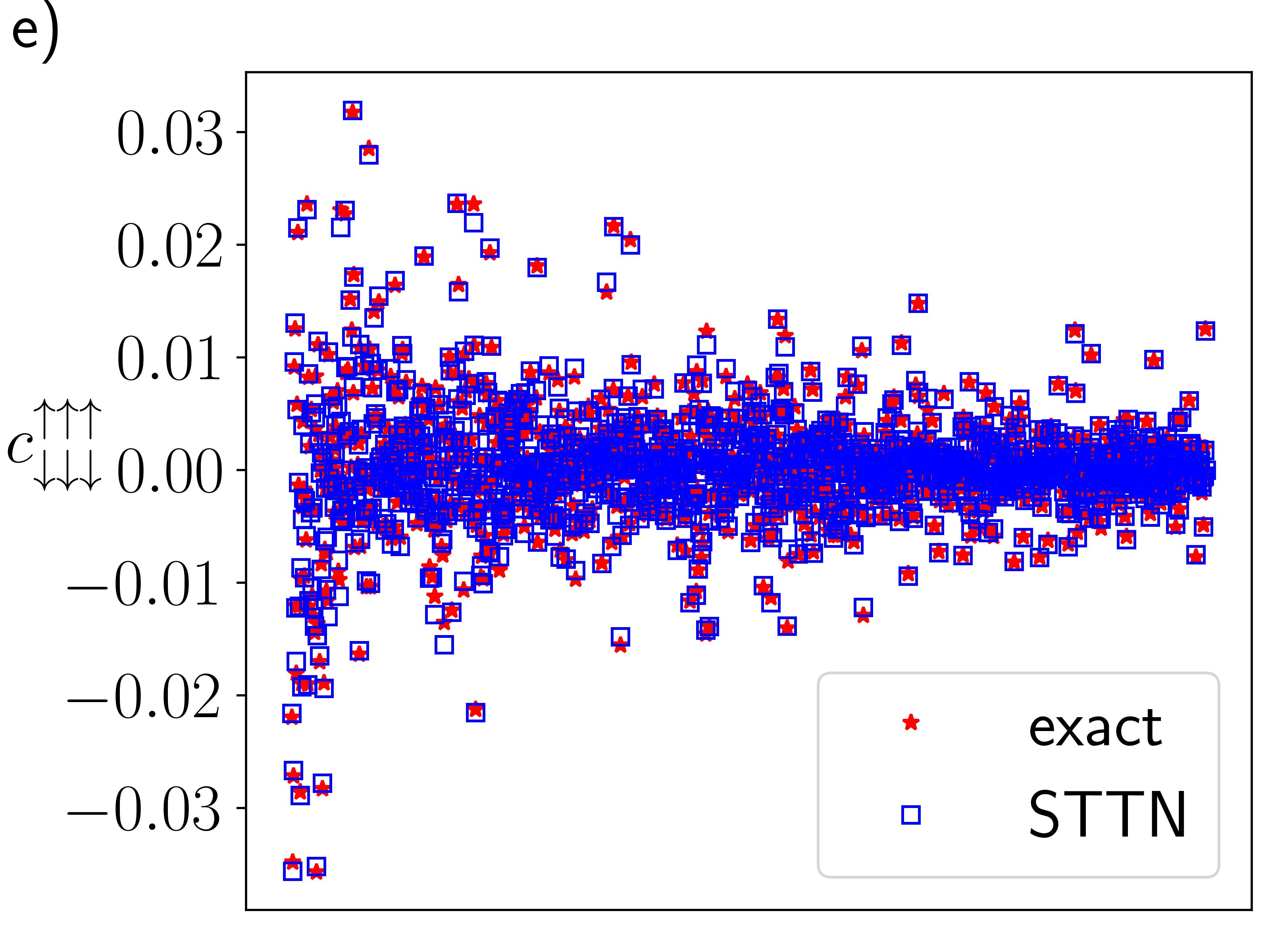}\includegraphics[width=0.51\columnwidth]{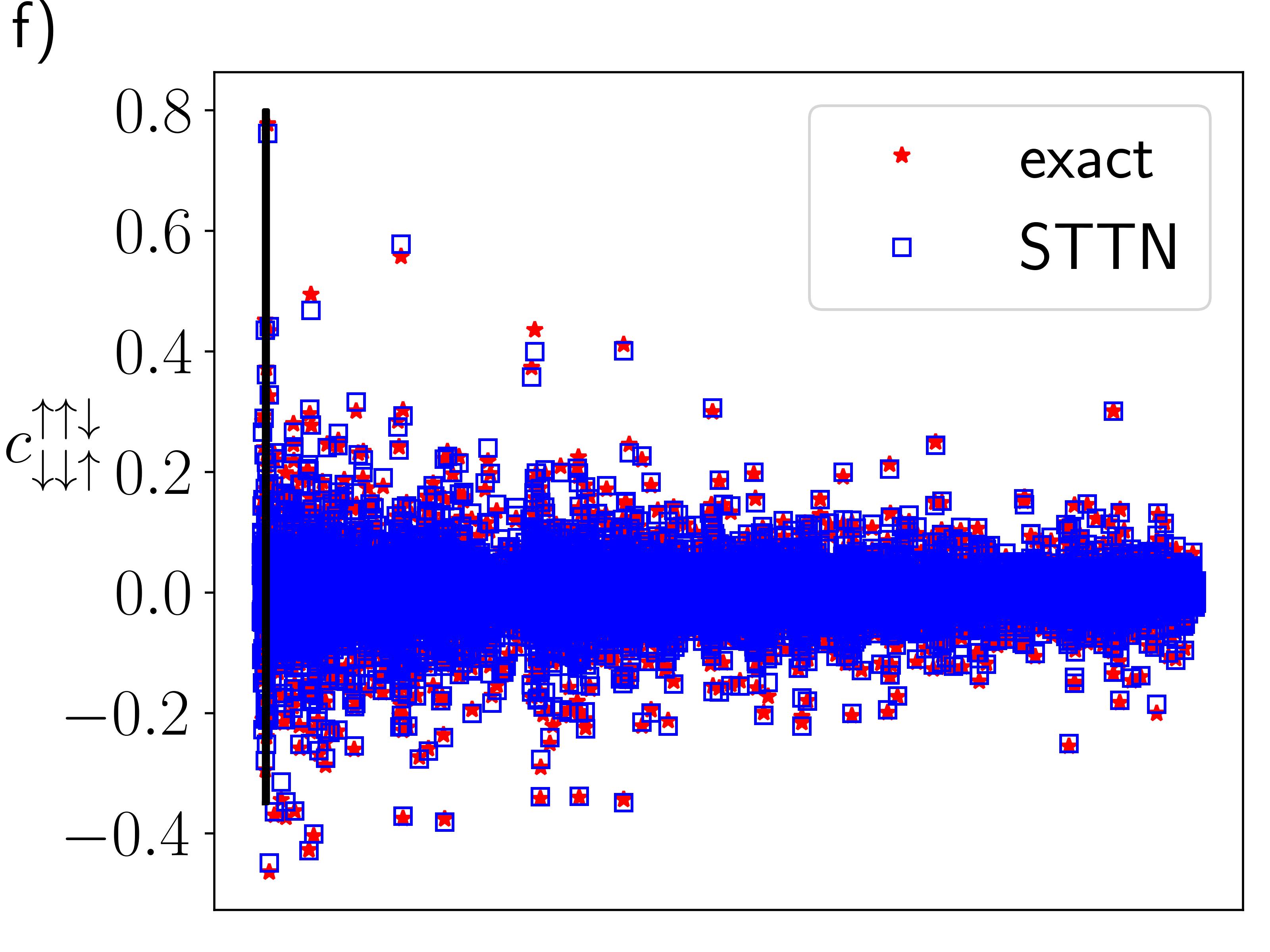}\\
\includegraphics[width=0.51\columnwidth]{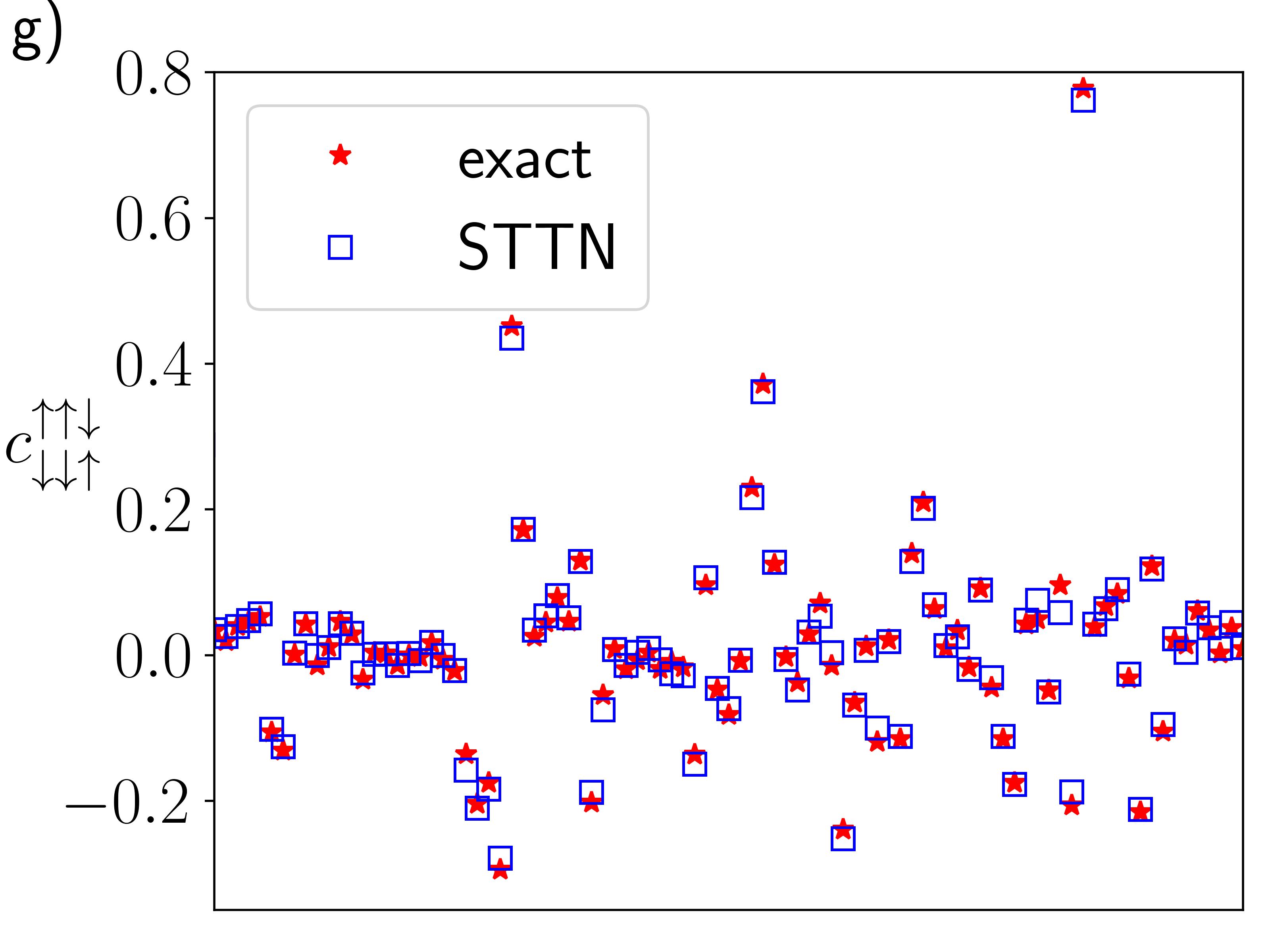}
\caption{\label{fig:14_sites_HF_coef} For the 14 sites cluster, exact wave function coefficients and STTN fit for a) the single-excitation tensor, b) the up-up double-excitation tensor, c) the up-down double-excitation tensor, d) the region delimited by the black rectangle in c), e) the up-up-up triple-excitation tensor, f) the up-up-down triple-excitation tensor and g) the region delimited by the black rectangle in f). The horizontal axes correspond to arbitrary tensor indices. The tensor dimensions used in the STTN are listed in table \ref{tab:tensor_dim_T}.}
\end{figure}

Figures \ref{fig:10_sites_HF_coef}, \ref{fig:12_sites_HF_coef} and \ref{fig:14_sites_HF_coef} of this appendix show the STTN fits to wave function coefficients directly, for the 10 sites, 12 sites and 14 sites clusters, respectively. The STTN tensor dimensions used in the fits are listed in table \ref{tab:tensor_dim_T}.

\section{Tensor dimensions}\label{sec:tensor_dim}

Table \ref{tab:tensor_dim_T} lists the STTN tensor dimensions for the three clusters and table \ref{tab:tensor_dim_T_SC} lists the tensor dimensions for the 12 sites cluster when only one channel is used in each decomposition.
\begin{table}[h]
\begin{tabular}{|c|c|c|c|}
\hline \backslashbox{dim}{L} & 10 & 12 & 14\\
\hline $s_p=s_h$ & 5 & 6 & 7\\
\hline $s_{p\bar{h}}=s_{\bar{p}h}$ & 4 & 5 & 9\\
\hline $s_{ph}=s_{\bar{p}\bar{h}}$ & 3 & 5 & 8\\
\hline $s_{p\bar{p}}=s_{h\bar{h}}$ & 3 & 5 & 8\\
\hline $s_{pp}=s_{hh}$ & 3 & 5 & 0\\
\hline $s_{p\bar{h}p\bar{h}}=s_{\bar{p}h\bar{p}h}$ & 2 & 3 & 3\\
\hline $s_{p\bar{h}\bar{p}h}$ & 3 & 3 & 3\\
\hline $s_{ph\bar{p}\bar{h}}$ & 2 & 2 & 3\\
\hline $s_{p\bar{p}\bar{h}h}$ & 2 & 2 & 2\\
\hline $s_{pp\bar{h}\bar{h}}$ & 2 & 2 & 0\\
\hline
\end{tabular}
\caption{STTN tensor dimensions for the three clusters.}
\label{tab:tensor_dim_T}
\end{table}
\begin{table}[h]
\begin{tabular}{|c|c|c|c|}
\hline \backslashbox{dim}{L} & 12\\
\hline $s_p=s_h$ & 6\\
\hline $s_{p\bar{h}}=s_{\bar{p}h}$ & 11\\
\hline $s_{ph}=s_{\bar{p}\bar{h}}$ & 0\\
\hline $s_{p\bar{p}}=s_{h\bar{h}}$ & 0\\
\hline $s_{pp}=s_{hh}$ & 0\\
\hline $s_{p\bar{h}p\bar{h}}=s_{\bar{p}h\bar{p}h}$ & 3\\
\hline $s_{p\bar{h}\bar{p}h}$ & 4\\
\hline $s_{ph\bar{p}\bar{h}}$ & 0\\
\hline $s_{p\bar{p}\bar{h}h}$ & 0\\
\hline $s_{pp\bar{h}\bar{h}}$ & 0\\
\hline
\end{tabular}
\caption{Tensor dimensions for the 12 sites clusters, when only one channel is used in each tensor decomposition.}
\label{tab:tensor_dim_T_SC}
\end{table}

\clearpage


%

\end{document}